\documentclass[aps,twocolumn,superscriptaddress,preprintnumbers,pre]{revtex4}
\usepackage{xr-hyper}
\usepackage{mathrsfs, hyperref}
\usepackage{amssymb, amsbsy, amsmath, latexsym, dsfont, array, layout, graphics,mathrsfs,braket,amsfonts,amsthm,
  amssymb,graphicx,subfigure, youngtab,color,bm,mathtools,braket,verbatim,url,cleveref,natbib,hypernat,extarrows,inputenc,multirow,bbm,dsfont}
  
\usepackage{xcolor}
\colorlet{RED}{red}
\usepackage{graphicx,psfrag}
\usepackage{amsfonts}
\usepackage[figuresright]{rotating}  
\usepackage{amssymb}
\usepackage{amsmath}
\usepackage{subfigure}
\usepackage{multirow}
\usepackage{tabularx}
\usepackage[resetlabels]{multibib}
\usepackage{amssymb, amsbsy, amsmath, latexsym, dsfont, array, layout, graphics,mathrsfs,braket,amsfonts,amsthm,
  amssymb,graphicx,subfigure,dcolumn, youngtab,color,bm,mathtools,braket,verbatim,url,cleveref,natbib,hypernat,extarrows,inputenc,multirow}
\usepackage{xcolor}
\graphicspath{ {./images/} }

\usepackage{mathrsfs, hyperref}
\usepackage{amssymb, amsbsy, amsmath, latexsym, dsfont, array, layout, graphics,mathrsfs,braket,amsfonts,amsthm,
  amssymb,graphicx,subfigure, youngtab,color,bm,mathtools,braket,verbatim,url,cleveref,natbib,hypernat,extarrows,inputenc,multirow,bbm}%,multicol
\usepackage{graphicx}% Include figure files
\usepackage{dcolumn}% Align table columns on decimal point
\usepackage{bm}% bold math
\usepackage{xcolor}
\usepackage[normalem]{ulem}
\usepackage[T1]{fontenc}
\graphicspath{ {./images/} }
\allowdisplaybreaks 

\newcommand{\splitatcommas}[1]{% 
\begingroup 
\begingroup\lccode`~=`, \lowercase{\endgroup 
\edef~{\mathchar\the\mathcode`, \penalty0 \noexpand\hspace{0pt plus 1em}}% 
}\mathcode`,="8000 #1% 
\endgroup 
} 

\begin{document}
\title{Beyond the Lowest Landau Level: Unlocking More Robust Fractional States Using Flat Chern Bands with Higher Vortexability}

\author{Yitong Zhang}\thanks{These three authors contributed equally}
\affiliation{%
Department of Physics, University of Michigan, Ann Arbor, MI 48109, USA
}
\author{Siddhartha Sarkar}\thanks{These three authors contributed equally}
\affiliation{%
Max Planck Institute for the Physics of Complex Systems, N\"othnitzer Stra\ss e 38, 01187 Dresden, Germany
}

\author{Xiaohan Wan}\thanks{These three authors contributed equally}
\affiliation{%
Department of Physics, University of Michigan, Ann Arbor, MI 48109, USA
}
\affiliation{
Department of Physics, University of Washington, Seattle, WA 98195, USA
}

\author{Daniel E. Parker}
\email{danielericparker@ucsd.edu}
\affiliation{%
Department of Physics, University of California at San Diego, La Jolla, CA 92093, USA}

\author{Shi-Zeng Lin}
\email{szl@lanl.gov}
\affiliation{%
Theoretical Division, T-4 and CNLS, Los Alamos National Laboratory, Los Alamos, New Mexico 87545, USA
}
\affiliation{%
 Center for Integrated Nanotechnologies (CINT), Los Alamos National Laboratory, Los Alamos, New Mexico 87545, USA
}
\author{Kai Sun}
\email{sunkai@umich.edu}
\affiliation{%
 Department of Physics, University of Michigan, Ann Arbor, MI 48109, USA
}
\begin{abstract}
Enhancing the many-body gap of a fractional state is crucial for realizing robust fractional excitations. For fractional Chern insulators, existing studies suggest that making flat Chern bands closely resemble the lowest Landau level (LLL) seems to maximize the excitation gap, providing an apparently optimal platform. In this work, we demonstrate that \emph{deforming away} from the LLL limit can, in fact, produce substantially larger FQH gaps. Using moir\'e flat bands with strongly non-Landau-level wavefunctions, we show that the gap can exceed that of the LLL by more than two orders of magnitude for short-range interactions and by factors of two to three for long-range interactions. This enhancement is generic across Abelian FCI states and follows a universal enhancement factor within each hierarchy. Using the Landau level framework, we identify the amplification of pseudopotentials as the microscopic origin of the observed enhancement. This finding demonstrates that pseudopotential engineering can substantially strengthen fractional topological phases.
We further examined non-Abelian states and found that, within finite-size resolution, this wavefunction construction method can also be used to manipulate and enhance the gap for certain interaction parameters.
\end{abstract}
\maketitle

\noindent\textit{Introduction.}--In the study of fractional quantum Hall (FQH) states~\cite{stormer1999fractional,laughlin1999nobel} and fractional Chern insulators (FCIs)~\cite{tang2011high,sun2011nearly,neupert2011fractional,sheng2011fractional,Regnault2011fractional,xiao2011interface,bernevig2012emergent,wu2012fractional,parameswaran2013fractional,roy2014band,wu2015fractional,repellin2020chern,ledwith2020fractional,simon2020contrasting,liu2021gate,mera2021engineering,li2021spontaneous,Devakul2021Magic,ledwith2021strong,wang2021exact,xie2021fractional,cai2023signatures,zeng2023thermodynamic,park2023observation,xu2023observation,ledwith2023vortexability,wu2024quantum,lu2024fractional,xie2025tunable}, a key open question is how to enhance the many-body gap of fractional states so that their exotic properties remain robust at higher temperatures and resilient against disorder. For FCIs, a common guiding principle in the literature have been Landau level mimicry: using flat Chern band whose properties mimic those of the lowest Landau level (LLL).
Model studies show that adiabatically deforming a flat Chern band toward the LLL—by making its Berry curvature more uniform and approaching the ideal geometric condition can enhance the excitation gap~\cite{wu2012fractional,parameswaran2013fractional,roy2014band,wu2015fractional,ledwith2020fractional,ledwith2021strong,wang2021exact,mera2021engineering,ledwith2023vortexability}.  Likewise, for non-Abelian fractional states, numerical studies indicate that the first Landau level provides a more favorable environment than known moir\'e flat bands~\cite{reddy2024non}. The mimicry paradigm can be summed up in the following hypothesis: while flat Chern bands can emulate Landau levels, for the same interaction strength, they typically exhibit smaller excitation gaps than the LLL for Abelian fractional states.

In this Letter, we demonstrate the opposite. Using moir\'e flat bands whose wavefunctions deviate substantially from the Landau level (LL) form, we show that deliberately deforming a Chern band away from ideal quantum geometry and the LLL limit can produce significantly larger FQH gaps. For short-range interactions, the enhancement over the LLL can reach two orders of magnitude and, for long-range interactions, is typically a factor of two to three. This enhancement is generic across Abelian FCI states and, within each hierarchy, follows a universal enhancement factor.

We first demonstrate this effect using a moir\'e model of a two-dimensional material with a quadratic band touching subjected to a periodic strain potential. Although illustrated in this specific setting, the underlying principles are general and apply broadly to other moir\'e systems. Within this platform, we construct exact flat Chern bands that resemble both the LLL (with ideal quantum geometry) and higher Landau levels, employing the theoretical framework of higher vortexability~\cite{fujimoto2024higher,liu2025theory}. By continuously interpolating between these two regimes, we find that the FCI gap reaches its maximum away from the limit of ideal quantum geometry. At fixed interaction strength, the maximal FCI gap surpasses not only the FQH gap of the LLL but also the gap in flat bands formed from interpolating Landau level wavefunctions.

To identify the underlying mechanism, we turn to the Landau level framework. It is worth noting that in LL systems, extensive effort has been devoted to studying how single-particle wavefunctions shape the pseudopotentials and thus the many-body gaps—a strategy known as pseudopotential engineering~\cite{zhang1986excitation,morf2002excitation,peterson2008finite,peterson2008orbital,peterson2010quantum,storni2010fractional,papic2011tunablePRB,papic2011tunablePRL,papic2012numerical}. The gap enhancement here stems from the same principle: our construction amplifies the pseudopotentials that set the energy scale of the fractional states. Because flat Chern bands offer a much larger design space, it enables even greater gap enhancement beyond LLs. To further underscore the central role of pseudopotentials, we explicitly compare two model systems with identical Chern numbers and Berry curvature distributions. Despite these similarities, the two models yield markedly different FCI gaps—demonstrating that distinct pseudopotentials encoded in their wavefunctions, rather than topology alone, determine the robustness of fractional phases. 

We have also examined non-Abelian (Moore–Read) states and find that for certain interaction parameters, their excitation gap can likewise be enhanced by hybridized wavefunctions.

\begin{figure}[t]
    \centering
    \includegraphics[width=0.48 \textwidth]{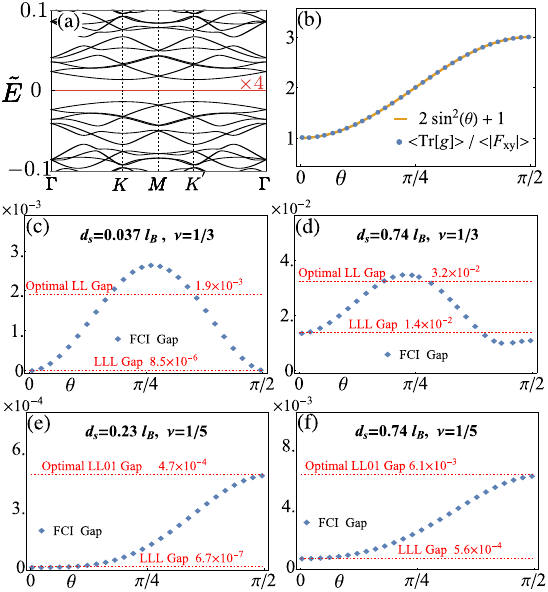}
\caption{\textbf{Many-body gap enhancement in moir\'e model.}
(a) Single particle band structure of the moir\'e Hamiltonian Eq.~\eqref{eq:moireHamiltonian} with four-fold degenerate exact flat bands. The vertical axis in (a) is normalized energy $\tilde{E}=E/(4 \pi/ \sqrt{3} a)^2$.
(b)~The ratio between the average trace of the quantum metric $\langle\mathrm{Tr}[g(\mathbf{k})]\rangle$ and the berry curvature $\braket{|F_{xy}(\mathbf{k})|}$ for the higher vortexable band $\tilde{\Psi}_{\mathbf{k},2}$ given in the main text. (c–f) Many-body gap of FCI states (blue dots) as a function of $\theta$. The gap increases markedly as $\theta$ deviates from the limit of ideal quantum geometry ($\theta=0$). 
The lower and upper horizontal red dashed lines in (c–f) mark, respectively, the gap of the corresponding FQH states in the LLL and the maximal many-body gap achieved in the LL-hybridization model with the same interaction strength.
Notably, in all cases the moir\'e model produces a gap larger than the LLL, and even exceeds the best LL-hybridization model values for $\nu=1/3$. The gap size here is measured in units of the Coulomb scale, $U^{\rm LL}_{\mathrm{int}}=e^2/\varepsilon l_B$, where $l_B=3^{1/4}a/(4\pi)^{1/2}$ is the magnetic length, $a$ is moir\'e lattice constant, and $\varepsilon$ is the dielectric constant.
The parameters used in ED are (c) $\nu=1/3$, $d_s/l_B=0.037$, $N_s=21$; 
(d) $\nu=1/3$, $d_s/l_B=0.74$, $N_s=21$; (e) $\nu=1/5$, $d_s/l_B=0.23$, $N_s=25$; 
(f) $\nu=1/5$, $d_s/l_B=0.74$, $N_s=25$.
}
\label{fig:moire}
\end{figure}

\noindent\textit{Moir\'e system.}--We consider the following moir\'e continuum Hamiltonian with chiral symmetry:
\begin{equation}\label{eq:moireHamiltonian}
\begin{split}
    \mathcal{H}(\mathbf{r}) &= \begin{pmatrix}0 & \mathcal{D}^\dagger(\mathbf{r})\\\mathcal{D}(\mathbf{r}) & 0\end{pmatrix},\\ \mathcal{D}(\mathbf{r}) &= \begin{pmatrix}
        -4\overline{\partial_z}^2+\tilde{A}(\mathbf{r}) & 2 i \gamma\overline{\partial_z}\\0 & -4\overline{\partial_z}^2+\tilde{A}(\mathbf{r})
    \end{pmatrix},
\end{split}
\end{equation}
where $z=x+iy$ is the complex coordinate, overline denotes complex conjugation, and $\tilde{A}(\mathbf{r}) = A_x(\mathbf{r})+iA_y(\mathbf{r})$ with $A_x=u_{xx}-u_{yy}$ and $A_y=u_{xy}$ representing moir\'e-periodic shear strain fields whose form is $\tilde{A}(\mathbf{r})= -\frac{\alpha}{2} \sum_{n=1}^{3}(\frac{\eta}{2} e^{i(1-n)\phi} \cos\left(2\mathbf{G}_n \cdot \mathbf{r}\right)-\beta e^{i(2-n)\phi} \cos\left((\mathbf{G}_n-\mathbf{G}_{n+1}) \cdot \mathbf{r}\right) + e^{i(1-n)\phi} \cos\left(\mathbf{G}_n \cdot \mathbf{r} \right))$, where $\phi=2\pi /3$, 
$\mathbf{G}_{1} = G(0,1)$ and $\mathbf{G}_{2,3} = G(\mp\sqrt{3}/2,-1/2)$ being the reciprocal lattice vectors, and $G = \frac{4\pi}{\sqrt{3}a}$ with moir\'e lattice constant $a$.
This Hamiltonian describes a homo-bilayer system, where the diagonal terms of $\mathcal{D}$ and $\mathcal{D}^\dagger$ represent the intralayer coupling, and the off-diagonal term encodes a momentum-dependent tunneling between the two layers with tunneling amplitude $\gamma$. Within each layer, the system hosts a chiral symmetric quadratic band touching (${\partial_z}^2$ and $\overline{\partial_z}^2$), and both layers are subjected to the same periodic moir\'e strain potential $\tilde{A}(\mathbf{r})$~\cite{wan2023topological,sarkar2023symmetry}.
As shown in Fig.~\ref{fig:moire}(a), 
for $(\alpha,\beta,\eta,\gamma)\approx(4.38G^2,0.5,-0.9,100G)$
the moir\'e field has a $p6mm$ space group symmetry, and the system hosts four exactly flat bands: two polarized on one sublattice with total Chern number $C=+2$, and the other two polarized on the opposite sublattice with $C=-2$. In the following, we focus on the two flat bands with $C=+2$.

As we show analytically in the Supplemental Material (SM~\cite{SM2025}), 
one of the two flat band wavefunctions takes the form $\Psi_{\mathbf{k},1}(\mathbf{r}) = \{\psi^\text{LLL}_\mathbf{k}(\mathbf{r}),0,0,0\}h(\mathbf{r})$, where $\psi^\text{LLL}_\mathbf{k}(\mathbf{r})$ denotes the  LLL wavefunction on a torus and $h(\mathbf{r})$ is a $\mathbf{k}$ independent scalar function such that  $|h(\mathbf{r})|$ is moir\'e periodic~\cite{ledwith2020fractional,wang2021exact}. 
Because the $\mathbf{k}$-dependence of this band resides entirely in the $\psi^\text{LLL}_\mathbf{k}$ term, this flat band shares the same Chern number $C = 1$ and ideal quantum geometry, $\text{Tr}[g(\mathbf{k})]=F_{xy}(\mathbf{k})$, as the LLL. Here, $g(\mathbf{k})$ denotes the quantum metric, and $F_{xy}(\mathbf{k})$ the Berry curvature.
The second flat band wavefunction can be written as $\Psi_{\mathbf{k},2}(\mathbf{r}) = \{l_B\psi^\text{LL1}_\mathbf{k}(\mathbf{r})/\sqrt{8},\gamma^{-1}\psi^\text{LLL}_\mathbf{k}(\mathbf{r}),0,0\}h(\mathbf{r})$, where $\psi^\text{LL1}_\mathbf{k}$ is the wavefunction of the first-LL ($n=1$) on a torus and $l_B =3^{1/4}a/(4\pi)^{1/2}$ is the magnetic length corresponding to one magnetic flux quantum per moir\'e unit cell. As $\gamma$ varies from $0$ to $\infty$, the wavefunction $\Psi_{\mathbf{k},2}$ continuously interpolates between a generalized LLL wavefunction and a generalized first-LL wavefunction. For convenience, we define $\tan \theta \equiv  l_B\gamma/\sqrt{8}$ such that $\theta = 0$ and $\pi/2$ correspond to $\gamma = 0$ and $\infty$, respectively; in this parametrization, $\Psi_{\mathbf{k},2}(\mathbf{r}) = \{\sin\theta \psi^\text{LL1}_\mathbf{k}(\mathbf{r}),\cos\theta\psi^\text{LLL}_\mathbf{k}(\mathbf{r}),0,0\}h(\mathbf{r})$.  

If $h(\mathbf{r}) = 1$, the wavefunctions above reduce to LL wavefunctions, and both the Berry curvature $F_{xy}(\mathbf{k})$ and the trace of the quantum metric $\mathrm{Tr}[g(\mathbf{k})]$ remain perfectly homogeneous in momentum space. 
For flat Chern bands, in contrast, the absence of net magnetic flux, together with the Bloch theorem, requires $h(\mathbf{r})$ to be spatially inhomogeneous. As shown in the SM~\cite{SM2025}, our model exhibits a strongly inhomogeneous $h(\mathbf{r})$, causing the wavefunctions $\Psi_{\mathbf{k},1}$ and $\Psi_{\mathbf{k},2}$ to deviate substantially from those of Landau level systems. Typically, such inhomogeneity in $h(\mathbf{r})$ would induce pronounced variations in the Berry curvature $F_{xy}(\mathbf{k})$ and the trace of the quantum metric $\mathrm{Tr}[g(\mathbf{k})]$ across momentum space~\cite{ledwith2020fractional,wang2021exact,shi2025effects,guerci2025fractionalization}. Here, however, we deliberately design the moir\'e potential so that, despite the spatial modulation of $h(\mathbf{r})$, both $F_{xy}(\mathbf{k})$ and $\mathrm{Tr}[g(\mathbf{k})]$ remain nearly homogeneous, with $\mathbf{k}$-space fluctuations below $5\%$ (see SM~\cite{SM2025}). 
In other words, these moir\'e flat bands deviate substantially from LLs in their real-space wavefunctions, yet closely mimic the geometric properties of LLs, i.e. the values of $F_{xy}(\mathbf{k})$ and $\mathrm{Tr}[g(\mathbf{k})]$. This is a key feature of our model that underlies its ability to produce larger many-body gaps than LLs

Due to the inhomogeneity of $h(\mathbf{r})$, the states $\Psi_{\mathbf{k},1}$ and $\Psi_{\mathbf{k},2}$ are not orthogonal. 
We therefore define the orthogonalized wavefunctions as 
$\tilde{\Psi}_{\mathbf{k},1} = \Psi_{\mathbf{k},1}$ and 
$\tilde{\Psi}_{\mathbf{k},2} = \Psi_{\mathbf{k},2} - 
\langle \Psi_{\mathbf{k},1} | \Psi_{\mathbf{k},2} \rangle 
\Psi_{\mathbf{k},1} / ||\Psi_{\mathbf{k},1}||^2$. 
Physically, these are the bands that are expected to be fully-filled and fully-empty at filling $\nu=1$ in the presence of quasi-long range interactions. 
As the parameter $\theta$ is varied, $\tilde{\Psi}_{\mathbf{k},2}$ retains Chern number $C = 1$, while the ratio 
$\langle \mathrm{Tr}[g(\mathbf{k})] \rangle / \langle F_{xy}(\mathbf{k}) \rangle=1+2\sin^2\theta$ as was shown in~\cite{liu2025theory}, evolving from 
$\lim_{\theta \to 0} \mathrm{Tr}[g(\mathbf{k})]/F_{xy}(\mathbf{k}) = 1$ 
to 
$\lim_{\theta \to \pi/2} \langle\mathrm{Tr}[g(\mathbf{k})]\rangle/\langle F_{xy}(\mathbf{k})\rangle = 3$, 
as shown in Fig.~\ref{fig:moire}(b). 
At $\theta = 0$, this band satisfies the ideal quantum geometry condition 
$\mathrm{Tr}[g(\mathbf{k})]/F_{xy}(\mathbf{k}) = 1$, 
and thus belongs to the family of ``vortexable'' flat bands~\cite{ledwith2023vortexability}, 
for which Laughlin-type wavefunctions can be constructed at fractional fillings, 
forming exact zero modes of the Haldane pseudopotentials~\cite{ledwith2020fractional,wang2021exact}. On the other hand, for $\theta>0$, where $\langle\mathrm{Tr}[g(\mathbf{k})]\rangle/\langle F_{xy}(\mathbf{k})\rangle > 1$, it belongs to the family of ``first (higher) vortexable'' bands~\cite{fujimoto2024higher,liu2025theory}, where non-Abelian FCIs states have been predicted.

We focus on the flat band spanned by
$\tilde{\Psi}_{\mathbf{k},2}$. To study the FCIs it can host, we consider projected repulsive interactions:
\begin{equation}
     H_\text{int} = \frac{1}{2A} \sum_\mathbf{q} V(\mathbf{q})\,:\rho(\mathbf{q})\,\rho(-\mathbf{q}):,
\end{equation}
where $A$ is the system area, colons denote normal ordering, and 
$V(\mathbf{q}) = 2\pi e^2 \tanh(d_s q)/(\varepsilon q)$ 
is the screened Coulomb interaction. 
Here $d_s$ is the separation between the screening electrodes. 
The projected density operator is 
$\rho(\mathbf{q}) = \sum_{\mathbf{k}} \lambda_{\mathbf{q}}(\mathbf{k}) c^\dagger_{\mathbf{k}} c_{\mathbf{k}+\mathbf{q}}$, 
with form factor 
$\lambda_{\mathbf{q}}(\mathbf{k}) = \langle \tilde{\Psi}_{\mathbf{k},2}(\mathbf{r}) | e^{-i\mathbf{q}\cdot\mathbf{r}} | \tilde{\Psi}_{\mathbf{k}+\mathbf{q},2}(\mathbf{r}) \rangle$.
We perform numerical exact diagonalization at filling fractions $\nu = 1/3$ and $\nu = 1/5$ for $\theta$ values between $0$ and $\pi/2$. At both fillings, and for the entire range of $\theta$, the ground state is found to be an FCI with many-body Chern number $C_\text{mb} = \nu$ (see End Matter Appendix A for the many-body spectra and particle entanglement spectra at a representative angle $\theta=\pi/4$). To quantify the stability of the FCI, Fig.~\ref{fig:moire}(c–f) shows the many-body excitation gap $\Delta_{\mathrm{mb}}$ as a function of $\theta$ for different screening lengths $d_s$.
Remarkably, the FCI gap is enhanced dramatically as $\theta$ increases, which tunes the band \textit{away} from ideal quantum geometry.
For comparison Fig.~\ref{fig:moire}(c–f) also shows the excitation gap for FQH states with the same interaction parameters in the LLL (lower dashed horizontal lines) and the maximum gap obtainable from the LL-hybridization model introduced later in the text (upper dashed horizontal lines). The gap of the present moir\'e model can exceed both benchmarks for suitable values of $\theta$.

For $\nu=1/3$ with short-range screening $d_s/l_B=0.037$, Fig.~\ref{fig:moire}(c), the gap develops a maximum near $\theta=\pi/4$, reaching values more than two orders of magnitude above the LLL baseline. By contrast, with longer-range interactions $d_s/l_B=0.74$, Fig.~\ref{fig:moire}(d), the enhancement is far milder. There the gap increases by about a factor of two, before decreasing.
For $\nu=1/5$, the largest gaps appear when the model approaches the generalized first Landau level ($\theta=\pi/2$), and again the effect is far more pronounced for short screening ($d_s/l_B=0.23$, Fig.~\ref{fig:moire}(e)) than for long-range interactions ($d_s/l_B=0.74$, Fig.~\ref{fig:moire}(f)).  

In the End Matter Fig.~\ref{fig:moire2}, we show many-body gap for another moir\'e model with much larger quantum geometric fluctuations. Remarkably, the qualitative features of gap enhancement remains the same even away from uniform quantum geometric distribution, although the overall gap sizes reduce in comparison with the model in Fig.~\ref{fig:moire}, suggesting that many-body gap enhancement as a function $\theta$ may be universal.

\noindent\textit{LL-hybridization model}--To understand the mechanism responsible for the many-body gap enhancement away from the vortexable limit, we construct a simplified ``toy model'' Hamiltonian that hybridizes two chosen Landau levels (LLs) labeled by $n_1$ and $n_2$. This type of mixed LL wavefunctions have been studied before in~\cite{papic2011tunablePRB,papic2011tunablePRL,papic2012numerical,yutushui2025numerical}.
Denoting the corresponding single-particle wavefunctions as $|\psi^{\text{LL}n_1}\rangle$ and $|\psi^{\text{LL}n_2}\rangle$, we define a two-component spinor wavefunction
\begin{equation}
\label{eq:hybrid_LL_wavefunction}
    \Psi_{n_1,n_2}(\theta)=\begin{pmatrix}\sin\theta\,|\psi^{\text{LL}n_2}\rangle\\[2pt]\cos\theta\,|\psi^{\text{LL}n_1}\rangle\end{pmatrix}, 
\qquad \theta\in[0,\pi/2],
\end{equation}
where the hybridization parameter $\theta$ continuously interpolates between the two LLs. Below, we will use LL$n_1n_2$ to label these models. For LL$01$, $\Psi_\theta$ coincides with $\Psi_{\mathbf{k},2}$ introduced early on, if $h(\mathbf{r})$ is set to unity. We can further introduce the following Hamiltonian
\begin{equation}
    \mathcal{H}_{n_2-n_1}(m) = \begin{pmatrix}
        m & (a^\dagger)^{n_2-n_1}\\(a)^{n_2-n_1} & -m
    \end{pmatrix},
\end{equation}
where $a$ and $a^\dagger$ are Landau level lowering and raising operators, respectively, and $m$ is a mass parameter. Without loss of generality, we take $n_2>n_1$. With the identification $\tan\theta = (\sqrt{E(n_1,n_2)^2+m^2}+m)/E(n_1,n_2)$, where $E(n_1,n_2)=\sqrt{n_2!/n_1!}$, one can verify that $\Psi_{n_1,n_2}(\theta)$ is an eigenstate of $\mathcal{H}_{n_2-n_1}(m)$ with eigenvalue $\sqrt{E(n_1,n_2)^2+m^2}$. The relation between $\theta$ and $m$ implies that $\theta = 0,\pi/4 \text{ and }\pi/2$ correspond to $m=-\infty, 0, \text{ and }\infty$. Notably, this Hamiltonian is the effective Hamiltonian of multilayer rhombohedral graphene with $n_2-n_1>1$ layers, subject to an out-of-plane magnetic field and a displacement field that controls the mass parameter $m$~\cite{vishwanathzero}. The special case $\mathcal{H}_{1}(m)$ describes graphene LLs in a magnetic field with a $C_{2z}$ breaking term $m$; this mass $m$ disappears at $\theta = \pi/4$, and the wavefunction $\Psi_{n,n+1}(\pi/4)$ corresponds to the $(n+1)$-th Landau level of graphene~\cite{semenoff1984condensed,shon1998quantum,tHoke2006fractional}.

\begin{figure}[t]
    \centering
    \includegraphics[width=0.48 \textwidth]{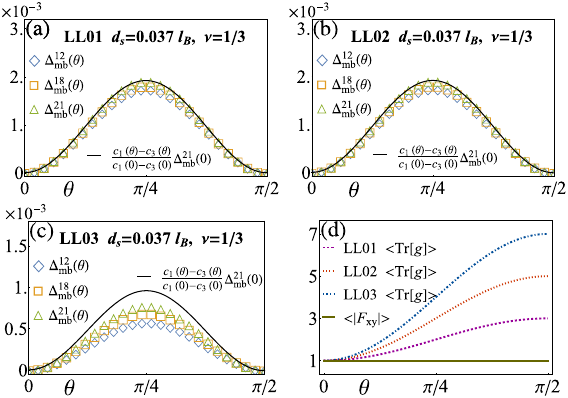}
    \caption{\textbf{Gap enhancement via Landau level hybridization and pseudopotential decomposition.} 
(a–c) Many-body gap and pseudopotentials as functions of the hybridization parameter $\theta$, for LL01, LL02, and LL03, respectively. As in Fig.~\ref{fig:moire}, the energy is measured in units of the Coulomb energy scale $U^{\rm LL}_{\rm int}$.  (d) shows the average trace of quantum metric and Berry curvature normalized by $l_B^2$ for the three models. Gap values are obtained for multiple finite clusters~\cite{SM2025}, using screened Coulomb interactions with screening length $d_s=0.037\,l_B$. The gap reaches its maximum near $\theta \!\approx\! \pi/4$, showing up to $\sim500$-fold enhancement for LL01 and LL02, with peak gap values of $\sim0.2\%$ of the Coulomb scale (compared to $0.0003\%$ for the LLL). For LL03, the enhancement is smaller--about $\sim200$-fold, yielding a maximum gap $\sim0.06\%$ of the Coulomb scale. 
}
\label{fig:GapEnhancement}
\end{figure}

This model exhibits the same gap enhancement phenomena as the moir\'e model. Explicitly, we consider the same screened Coulomb interaction, $V(q)= 2\pi e^2\tanh(q d_s)/(\varepsilon q)$, projected to the band Eq.~\eqref{eq:hybrid_LL_wavefunction} on a torus geometry. Fig.~\ref{fig:GapEnhancement} shows the Laughlin state gap at $\nu=1/3$ increases dramatically with $\theta$. For LL01 and LL02 hybridizations, the maximum is more than $500$ times the LLL gap, with a more modest ${\sim}200$ fold enhancement in the LL03 case. To understand this, we turn to pseudopotentials.

\noindent\textit{Haldane pseudopotentials}--
To analyze the impact of LL-hybridization on the FQH effect, it is convenient to employ the pseudopotential framework~\cite{haldane1983fractional,trugman1985exact,Haldane1987QHE}. In this approach, the effect of hybridization enters solely through the %\sout{projected} 
cyclotron part of the form factor,
\begin{equation}
F(q;\theta)
=\left[\cos^{2}\theta L_{n_1}\left(\frac{q^2l_B^2}{2}\right)+\sin^{2}\theta L_{n_2}\left(\frac{q^2l_B^2}{2}\right)\right]e^{-\frac{q^2l_B^2}{4}}.
\label{eq:form_factor}
\end{equation}
where $L_n(x)$ is the $n$\textsuperscript{th} Laguerre polynomial. The effective interaction after projection is $V_{\rm eff}(q;\theta)=V(q)[F(q;\theta)]^{2}$, 
Once the effective interaction is specified, the problem can be fully characterized by a discrete set of pseudopotential channels,
\begin{equation}
c_m(\theta)=\int\!\frac{d^{2}\mathbf q}{(2\pi)^{2}}\,V_{\rm eff}(q;\theta)\,v_m(q),\quad m=0,1,2,\cdots
\label{eq:c_m}
\end{equation}
where the basis function
%\begin{equation}
$v_m(q)=e^{-q^2l_B^2/2}L_m(q^2l_B^2)
\label{eq:v_m}
$
%\end{equation}
characterizes the momentum-space profile of a two-particle state with relative angular momentum $m$. 

We computed the pseudopotential coefficients $c_m$, both analytically and numerically (see SM~\cite{SM2025}). As $\theta$ varies, the key term $c_1$, which stabilizes the $\nu = 1/3$ Laughlin state, exhibits the same $\theta$-dependent trend as the many-body excitation gap. To quantify this, Fig.~\ref{fig:GapEnhancement} plots the pseudopotential difference $c_1(\theta) - c_3(\theta)$, normalized by the $\theta=0$ values of the many-body gap and pseudopotentials, $\Delta_{\mathrm{mb}}(0)/[c_1(0) - c_3(0)]$. The $\theta$-dependence of $c_1 - c_3$ closely tracks the enhancement of the many-body gap.
For LL01 and LL02 in the short-range limit ($d_s \ll l_B$), the pseudopotentials satisfy $c_1 \propto \sin^2(2\theta)$, while $c_3 = c_5 = 0$. The many-body gap follows the same $\theta$-dependence, peaking at $\theta = \pi/4$. The model LL03 shows a sizable $c_3$ contribution, and the reduced difference $c_1 - c_3$ correlates with a more moderate ${\sim}200$-fold gap enhancement, Fig.~\ref{fig:GapEnhancement}(c). Although the short-range limit provides analytical simplicity, it is not essential; both the pseudopotential and gap enhancements persist for finite screening lengths, as shown in SM~\cite{SM2025}. These results demonstrate that the observed gap enhancement originates from the $\theta$-dependent evolution of the pseudopotentials.

\begin{figure}[t]
  \centering
  \includegraphics[width=\linewidth]{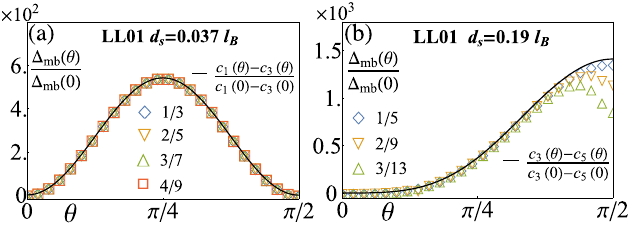}
  \caption{\textbf{Many-body gap ratio along Jain sequences.}
  Normalized many-body gap $\Delta_{\mathrm{mb}}(\theta)/\Delta_{\mathrm{mb}}(0)$ for the LL01 hybridization model.
  (a) \emph{$1/3$ sequence:} $\nu=1/3,\,2/5,\,3/7,\,4/9$ at short-distance screening $d_s/l_B=0.037$. 
  The enhancement exhibits a universal dome-like profile across the entire sequence, following the pseudopotential $\!c_1-c_3$  and peaking near $\theta=\pi/4$.
  (b) \emph{$1/5$ sequence:} $\nu=1/5,\,2/9,\,3/13$ at $d_s/l_B=0.19$. Here, the $\theta$ dependence tracks the pseudopotential $\!c_3-c_5$  throughout the sequence, with the maximum occurring near $\theta=\pi/2$.}
  \label{fig:jain}
\end{figure}
\noindent\textit{Jain Sequence}.--To further elucidate the connection between pseudopotentials and FQH gaps, in Fig.~\ref{fig:jain} we plot the gap ratio $\Delta_{\mathrm{mb}}(\theta)/\Delta_{\mathrm{mb}}(0)$ for FQH states within a given Jain sequence of filling $\nu = n/(2pn + 1)$, together with the pseudopotential ratio $[c_{2p-1}(\theta) - c_{2p+1}(\theta)] / [c_{2p-1}(0) - c_{2p+1}(0)]$. When normalized by their $\theta = 0$ values, the many-body gaps for all fractions within the same Jain sequence collapse onto a single universal curve, consistent with the corresponding pseudopotential ratio.
For the $1/3$ sequence, Fig.~\ref{fig:jain}(a), the gap enhancement follows $c_1-c_3$, achieving its maximum near $\theta \approx \pi/4$ since $c_1(\theta) \propto \sin^2(2\theta)$ for short-range interactions. %$c_1(\theta)\!\sim\!\sin^2(2\theta)$ at short range interaction.
For the $1/5$ sequence, Fig.~\ref{fig:jain}(b), the enhancement instead follows $c_3-c_5$, with maximum near
$\theta\!\approx\!\pi/2$ at finite screening length $d_s/l_B=0.19$.

\noindent{\it Non-Abelian states.}--This hybridization-based approach can also enhance the excitation gap of non-Abelian states~\cite{papic2011tunablePRB,papic2011tunablePRL,papic2012numerical}. As shown in Fig.~\ref{fig:em1} in the End Matter, using screened Coulomb interactions with a short screening length $d_s/l_B=0.5$, the energy gap of the Moore–Read state increases as we switch from the $n=1$ LL to hybridized wavefunctions composed of the lowest three LLs. Owing to their intrinsic complexity and stronger finite-size effects, non-Abelian states are considerably more intricate than their Abelian counterparts. Consequently, further studies will be needed to fully elucidate the connection between the excitation gap and the hybridized wavefunctions.

\noindent\textit{Discussion.}-- In addition to the hybridization between two LLs, in the SM~\cite{SM2025} we also consider mixing among multiple Landau levels. 
Among all cases studied, for filling factor $\nu = 1/3$ we find that the largest many-body gap occurs for the LL${01}$ hybridization with $\theta = \pi/4$ in the presence of short-range interactions, and at $\theta = 0.2\pi$ for long-range Coulomb interactions. Since the enhancement of many-body gap is for LL${01}$ and near $\theta=\pi/4$, it may be seen in the $n=1$ LL of graphene. For long-range interactions, the pseudopotential enhancement observed here agrees with earlier studies of Dirac-fermion Landau levels reported in Refs.~\cite{tHoke2006fractional,yutushui2025numerical}.

For the moir\'e model, although the wavefunctions deviate substantially from LLs due to the highly inhomogeneous function $h(\mathbf{r})$, we find that the hybridization between the generalized lowest and first LL exhibits the same trend of gap enhancement, i.e., moir\'e systems exhibit a similar pseudopotential enhancement. Moreover, the inhomogeneous form of $h(\mathbf{r})$ appears to further enhance the many-body gap beyond what can be achieved through LL-hybridization. The microscopic origin of this additional enhancement remains an interesting question for future study.

For comparison, we present in SM~\cite{SM2025}, Sec. VI, a single-component LL-hybridization model (that was previously used in Ref.~\cite{liu2025theory}) that shares identical Berry curvature $F_{xy}(\mathbf{k})$ and quantum metric trace $\mathrm{Tr}[g(\mathbf{k})]$ with the LL-hybridization models discussed above. However, unlike those models—which exhibit a many-body gap enhanced beyond that of the LLL—the comparison model shows a reduced gap smaller than the LLL. This contrast demonstrates that the single-particle quantities $F_{xy}(\mathbf{k})$ and $\mathrm{Tr}[g(\mathbf{k})]$ cannot fully determine the many-body gap.

Finally, we suggest a potential design principle to enhance gaps of fractionalized phases using higher vortexability. Higher vortexable bands, which appear within two band complexes of ideal bands, can often be tuned from lowest to first LL character. Both the moir\'e and LL model results here suggest that gaps are not necessarily maximized where zeroth or first Landau level mimicry is best, but instead along the one parameter family of higher vortexable bands. In addition to FCIs, this construction may also offer an platform for time-reversal-invariant fractional topological insulators (FTIs), as it suppresses the $s$-wave pseudopotential component $c_0$~\cite{SM2025}, a key ingredient for stabilizing FTIs~\cite{kwan2024abelianfractionaltopologicalinsulators}.

\bigskip
\let\oldaddcontentsline\addcontentsline% Store \addcontentsline
\renewcommand{\addcontentsline}[3]{}% Make \addcontentsline a no-op
\begin{acknowledgments}
\noindent \textit{Acknowledgements}.--The authors thank Siddharth Parameswaran for pointing out the LL hybridization model. This work was supported in part by Air Force Office of Scientific Research MURI FA9550-23-1-0334 and the Office of Naval Research MURI N00014-20-1-2479, and by the Gordon and Betty Moore Foundation Award N031710 (XW, YZ, KS). 
The work at LANL (SZL) was carried out under the auspices of the U.S. DOE NNSA under contract No. 89233218CNA000001 through the LDRD Program, and was supported by the Center for Nonlinear Studies at LANL, and was performed, in part, at the Center for Integrated Nanotechnologies, an Office of Science User Facility operated for the U.S. DOE Office of Science, under user proposals $\#2018BU0010$ and $\#2018BU0083$.
\end{acknowledgments}

\bibliographystyle{apsrev4-1}
\bibliography{ref}

\begin{thebibliography}{63}%
\makeatletter
\providecommand \@ifxundefined [1]{%
 \@ifx{#1\undefined}
}%
\providecommand \@ifnum [1]{%
 \ifnum #1\expandafter \@firstoftwo
 \else \expandafter \@secondoftwo
 \fi
}%
\providecommand \@ifx [1]{%
 \ifx #1\expandafter \@firstoftwo
 \else \expandafter \@secondoftwo
 \fi
}%
\providecommand \natexlab [1]{#1}%
\providecommand \enquote  [1]{``#1''}%
\providecommand \bibnamefont  [1]{#1}%
\providecommand \bibfnamefont [1]{#1}%
\providecommand \citenamefont [1]{#1}%
\providecommand \href@noop [0]{\@secondoftwo}%
\providecommand \href [0]{\begingroup \@sanitize@url \@href}%
\providecommand \@href[1]{\@@startlink{#1}\@@href}%
\providecommand \@@href[1]{\endgroup#1\@@endlink}%
\providecommand \@sanitize@url [0]{\catcode `\\12\catcode `\$12\catcode `\&12\catcode `\#12\catcode `\^12\catcode `\_12\catcode `\%12\relax}%
\providecommand \@@startlink[1]{}%
\providecommand \@@endlink[0]{}%
\providecommand \url  [0]{\begingroup\@sanitize@url \@url }%
\providecommand \@url [1]{\endgroup\@href {#1}{\urlprefix }}%
\providecommand \urlprefix  [0]{URL }%
\providecommand \Eprint [0]{\href }%
\providecommand \doibase [0]{http://dx.doi.org/}%
\providecommand \selectlanguage [0]{\@gobble}%
\providecommand \bibinfo  [0]{\@secondoftwo}%
\providecommand \bibfield  [0]{\@secondoftwo}%
\providecommand \translation [1]{[#1]}%
\providecommand \BibitemOpen [0]{}%
\providecommand \bibitemStop [0]{}%
\providecommand \bibitemNoStop [0]{.\EOS\space}%
\providecommand \EOS [0]{\spacefactor3000\relax}%
\providecommand \BibitemShut  [1]{\csname bibitem#1\endcsname}%
\let\auto@bib@innerbib\@empty
%</preamble>
\bibitem [{\citenamefont {Stormer}\ \emph {et~al.}(1999)\citenamefont {Stormer}, \citenamefont {Tsui},\ and\ \citenamefont {Gossard}}]{stormer1999fractional}%
  \BibitemOpen
  \bibfield  {author} {\bibinfo {author} {\bibfnamefont {H.~L.}\ \bibnamefont {Stormer}}, \bibinfo {author} {\bibfnamefont {D.~C.}\ \bibnamefont {Tsui}}, \ and\ \bibinfo {author} {\bibfnamefont {A.~C.}\ \bibnamefont {Gossard}},\ }\href@noop {} {\bibfield  {journal} {\bibinfo  {journal} {Reviews of Modern Physics}\ }\textbf {\bibinfo {volume} {71}},\ \bibinfo {pages} {S298} (\bibinfo {year} {1999})}\BibitemShut {NoStop}%
\bibitem [{\citenamefont {Laughlin}(1999)}]{laughlin1999nobel}%
  \BibitemOpen
  \bibfield  {author} {\bibinfo {author} {\bibfnamefont {R.~B.}\ \bibnamefont {Laughlin}},\ }\href@noop {} {\bibfield  {journal} {\bibinfo  {journal} {Reviews of Modern Physics}\ }\textbf {\bibinfo {volume} {71}},\ \bibinfo {pages} {863} (\bibinfo {year} {1999})}\BibitemShut {NoStop}%
\bibitem [{\citenamefont {Tang}\ \emph {et~al.}(2011)\citenamefont {Tang}, \citenamefont {Mei},\ and\ \citenamefont {Wen}}]{tang2011high}%
  \BibitemOpen
  \bibfield  {author} {\bibinfo {author} {\bibfnamefont {E.}~\bibnamefont {Tang}}, \bibinfo {author} {\bibfnamefont {J.-W.}\ \bibnamefont {Mei}}, \ and\ \bibinfo {author} {\bibfnamefont {X.-G.}\ \bibnamefont {Wen}},\ }\href@noop {} {\bibfield  {journal} {\bibinfo  {journal} {Phys. Rev. Lett.}\ }\textbf {\bibinfo {volume} {106}},\ \bibinfo {pages} {236802} (\bibinfo {year} {2011})}\BibitemShut {NoStop}%
\bibitem [{\citenamefont {Sun}\ \emph {et~al.}(2011)\citenamefont {Sun}, \citenamefont {Gu}, \citenamefont {Katsura},\ and\ \citenamefont {Das~Sarma}}]{sun2011nearly}%
  \BibitemOpen
  \bibfield  {author} {\bibinfo {author} {\bibfnamefont {K.}~\bibnamefont {Sun}}, \bibinfo {author} {\bibfnamefont {Z.}~\bibnamefont {Gu}}, \bibinfo {author} {\bibfnamefont {H.}~\bibnamefont {Katsura}}, \ and\ \bibinfo {author} {\bibfnamefont {S.}~\bibnamefont {Das~Sarma}},\ }\href@noop {} {\bibfield  {journal} {\bibinfo  {journal} {Phys. Rev. Lett.}\ }\textbf {\bibinfo {volume} {106}},\ \bibinfo {pages} {236803} (\bibinfo {year} {2011})}\BibitemShut {NoStop}%
\bibitem [{\citenamefont {Neupert}\ \emph {et~al.}(2011)\citenamefont {Neupert}, \citenamefont {Santos}, \citenamefont {Chamon},\ and\ \citenamefont {Mudry}}]{neupert2011fractional}%
  \BibitemOpen
  \bibfield  {author} {\bibinfo {author} {\bibfnamefont {T.}~\bibnamefont {Neupert}}, \bibinfo {author} {\bibfnamefont {L.}~\bibnamefont {Santos}}, \bibinfo {author} {\bibfnamefont {C.}~\bibnamefont {Chamon}}, \ and\ \bibinfo {author} {\bibfnamefont {C.}~\bibnamefont {Mudry}},\ }\href@noop {} {\bibfield  {journal} {\bibinfo  {journal} {Phys. Rev. Lett.}\ }\textbf {\bibinfo {volume} {106}},\ \bibinfo {pages} {236804} (\bibinfo {year} {2011})}\BibitemShut {NoStop}%
\bibitem [{\citenamefont {Sheng}\ \emph {et~al.}(2011)\citenamefont {Sheng}, \citenamefont {Gu}, \citenamefont {Sun},\ and\ \citenamefont {Sheng}}]{sheng2011fractional}%
  \BibitemOpen
  \bibfield  {author} {\bibinfo {author} {\bibfnamefont {D.~N.}\ \bibnamefont {Sheng}}, \bibinfo {author} {\bibfnamefont {Z.-C.}\ \bibnamefont {Gu}}, \bibinfo {author} {\bibfnamefont {K.}~\bibnamefont {Sun}}, \ and\ \bibinfo {author} {\bibfnamefont {L.}~\bibnamefont {Sheng}},\ }\href@noop {} {\bibfield  {journal} {\bibinfo  {journal} {Nat. Comm.}\ }\textbf {\bibinfo {volume} {2}},\ \bibinfo {pages} {389} (\bibinfo {year} {2011})}\BibitemShut {NoStop}%
\bibitem [{\citenamefont {Regnault}\ and\ \citenamefont {Bernevig}(2011)}]{Regnault2011fractional}%
  \BibitemOpen
  \bibfield  {author} {\bibinfo {author} {\bibfnamefont {N.}~\bibnamefont {Regnault}}\ and\ \bibinfo {author} {\bibfnamefont {B.~A.}\ \bibnamefont {Bernevig}},\ }\href@noop {} {\bibfield  {journal} {\bibinfo  {journal} {Phys. Rev. X}\ }\textbf {\bibinfo {volume} {1}},\ \bibinfo {pages} {021014} (\bibinfo {year} {2011})}\BibitemShut {NoStop}%
\bibitem [{\citenamefont {Xiao}\ \emph {et~al.}(2011)\citenamefont {Xiao}, \citenamefont {Zhu}, \citenamefont {Ran}, \citenamefont {Nagaosa},\ and\ \citenamefont {Okamoto}}]{xiao2011interface}%
  \BibitemOpen
  \bibfield  {author} {\bibinfo {author} {\bibfnamefont {D.}~\bibnamefont {Xiao}}, \bibinfo {author} {\bibfnamefont {W.}~\bibnamefont {Zhu}}, \bibinfo {author} {\bibfnamefont {Y.}~\bibnamefont {Ran}}, \bibinfo {author} {\bibfnamefont {N.}~\bibnamefont {Nagaosa}}, \ and\ \bibinfo {author} {\bibfnamefont {S.}~\bibnamefont {Okamoto}},\ }\href@noop {} {\bibfield  {journal} {\bibinfo  {journal} {Nature Communications}\ }\textbf {\bibinfo {volume} {2}} (\bibinfo {year} {2011})}\BibitemShut {NoStop}%
\bibitem [{\citenamefont {Bernevig}\ and\ \citenamefont {Regnault}(2012)}]{bernevig2012emergent}%
  \BibitemOpen
  \bibfield  {author} {\bibinfo {author} {\bibfnamefont {B.~A.}\ \bibnamefont {Bernevig}}\ and\ \bibinfo {author} {\bibfnamefont {N.}~\bibnamefont {Regnault}},\ }\href@noop {} {\bibfield  {journal} {\bibinfo  {journal} {Physical Review B—Condensed Matter and Materials Physics}\ }\textbf {\bibinfo {volume} {85}},\ \bibinfo {pages} {075128} (\bibinfo {year} {2012})}\BibitemShut {NoStop}%
\bibitem [{\citenamefont {Wu}\ \emph {et~al.}(2012)\citenamefont {Wu}, \citenamefont {Jain},\ and\ \citenamefont {Sun}}]{wu2012fractional}%
  \BibitemOpen
  \bibfield  {author} {\bibinfo {author} {\bibfnamefont {Y.-H.}\ \bibnamefont {Wu}}, \bibinfo {author} {\bibfnamefont {J.~K.}\ \bibnamefont {Jain}}, \ and\ \bibinfo {author} {\bibfnamefont {K.}~\bibnamefont {Sun}},\ }\href@noop {} {\bibfield  {journal} {\bibinfo  {journal} {Phys. Rev. B}\ }\textbf {\bibinfo {volume} {86}},\ \bibinfo {pages} {165129} (\bibinfo {year} {2012})}\BibitemShut {NoStop}%
\bibitem [{\citenamefont {Parameswaran}\ \emph {et~al.}(2013)\citenamefont {Parameswaran}, \citenamefont {Roy},\ and\ \citenamefont {Sondhi}}]{parameswaran2013fractional}%
  \BibitemOpen
  \bibfield  {author} {\bibinfo {author} {\bibfnamefont {S.~A.}\ \bibnamefont {Parameswaran}}, \bibinfo {author} {\bibfnamefont {R.}~\bibnamefont {Roy}}, \ and\ \bibinfo {author} {\bibfnamefont {S.~L.}\ \bibnamefont {Sondhi}},\ }\href@noop {} {\bibfield  {journal} {\bibinfo  {journal} {Comptes Rendus Physique}\ }\textbf {\bibinfo {volume} {14}},\ \bibinfo {pages} {816} (\bibinfo {year} {2013})}\BibitemShut {NoStop}%
\bibitem [{\citenamefont {Roy}(2014)}]{roy2014band}%
  \BibitemOpen
  \bibfield  {author} {\bibinfo {author} {\bibfnamefont {R.}~\bibnamefont {Roy}},\ }\href@noop {} {\bibfield  {journal} {\bibinfo  {journal} {Physical Review B}\ }\textbf {\bibinfo {volume} {90}},\ \bibinfo {pages} {165139} (\bibinfo {year} {2014})}\BibitemShut {NoStop}%
\bibitem [{\citenamefont {Wu}\ \emph {et~al.}(2015)\citenamefont {Wu}, \citenamefont {Jain},\ and\ \citenamefont {Sun}}]{wu2015fractional}%
  \BibitemOpen
  \bibfield  {author} {\bibinfo {author} {\bibfnamefont {Y.-H.}\ \bibnamefont {Wu}}, \bibinfo {author} {\bibfnamefont {J.~K.}\ \bibnamefont {Jain}}, \ and\ \bibinfo {author} {\bibfnamefont {K.}~\bibnamefont {Sun}},\ }\href@noop {} {\bibfield  {journal} {\bibinfo  {journal} {Phys. Rev. B}\ }\textbf {\bibinfo {volume} {91}},\ \bibinfo {pages} {041119} (\bibinfo {year} {2015})}\BibitemShut {NoStop}%
\bibitem [{\citenamefont {Repellin}\ and\ \citenamefont {Senthil}(2020)}]{repellin2020chern}%
  \BibitemOpen
  \bibfield  {author} {\bibinfo {author} {\bibfnamefont {C.}~\bibnamefont {Repellin}}\ and\ \bibinfo {author} {\bibfnamefont {T.}~\bibnamefont {Senthil}},\ }\href@noop {} {\bibfield  {journal} {\bibinfo  {journal} {Physical Review Research}\ }\textbf {\bibinfo {volume} {2}},\ \bibinfo {pages} {023238} (\bibinfo {year} {2020})}\BibitemShut {NoStop}%
\bibitem [{\citenamefont {Ledwith}\ \emph {et~al.}(2020)\citenamefont {Ledwith}, \citenamefont {Tarnopolsky}, \citenamefont {Khalaf},\ and\ \citenamefont {Vishwanath}}]{ledwith2020fractional}%
  \BibitemOpen
  \bibfield  {author} {\bibinfo {author} {\bibfnamefont {P.~J.}\ \bibnamefont {Ledwith}}, \bibinfo {author} {\bibfnamefont {G.}~\bibnamefont {Tarnopolsky}}, \bibinfo {author} {\bibfnamefont {E.}~\bibnamefont {Khalaf}}, \ and\ \bibinfo {author} {\bibfnamefont {A.}~\bibnamefont {Vishwanath}},\ }\href@noop {} {\bibfield  {journal} {\bibinfo  {journal} {Physical Review Research}\ }\textbf {\bibinfo {volume} {2}},\ \bibinfo {pages} {023237} (\bibinfo {year} {2020})}\BibitemShut {NoStop}%
\bibitem [{\citenamefont {Simon}\ and\ \citenamefont {Rudner}(2020)}]{simon2020contrasting}%
  \BibitemOpen
  \bibfield  {author} {\bibinfo {author} {\bibfnamefont {S.~H.}\ \bibnamefont {Simon}}\ and\ \bibinfo {author} {\bibfnamefont {M.~S.}\ \bibnamefont {Rudner}},\ }\href@noop {} {\bibfield  {journal} {\bibinfo  {journal} {Physical Review B}\ }\textbf {\bibinfo {volume} {102}},\ \bibinfo {pages} {165148} (\bibinfo {year} {2020})}\BibitemShut {NoStop}%
\bibitem [{\citenamefont {Liu}\ \emph {et~al.}(2021)\citenamefont {Liu}, \citenamefont {Abouelkomsan},\ and\ \citenamefont {Bergholtz}}]{liu2021gate}%
  \BibitemOpen
  \bibfield  {author} {\bibinfo {author} {\bibfnamefont {Z.}~\bibnamefont {Liu}}, \bibinfo {author} {\bibfnamefont {A.}~\bibnamefont {Abouelkomsan}}, \ and\ \bibinfo {author} {\bibfnamefont {E.~J.}\ \bibnamefont {Bergholtz}},\ }\href@noop {} {\bibfield  {journal} {\bibinfo  {journal} {Physical Review Letters}\ }\textbf {\bibinfo {volume} {126}},\ \bibinfo {pages} {026801} (\bibinfo {year} {2021})}\BibitemShut {NoStop}%
\bibitem [{\citenamefont {Mera}\ and\ \citenamefont {Ozawa}(2021)}]{mera2021engineering}%
  \BibitemOpen
  \bibfield  {author} {\bibinfo {author} {\bibfnamefont {B.}~\bibnamefont {Mera}}\ and\ \bibinfo {author} {\bibfnamefont {T.}~\bibnamefont {Ozawa}},\ }\href@noop {} {\bibfield  {journal} {\bibinfo  {journal} {Physical Review B}\ }\textbf {\bibinfo {volume} {104}},\ \bibinfo {pages} {115160} (\bibinfo {year} {2021})}\BibitemShut {NoStop}%
\bibitem [{\citenamefont {Li}\ \emph {et~al.}(2021)\citenamefont {Li}, \citenamefont {Kumar}, \citenamefont {Sun},\ and\ \citenamefont {Lin}}]{li2021spontaneous}%
  \BibitemOpen
  \bibfield  {author} {\bibinfo {author} {\bibfnamefont {H.}~\bibnamefont {Li}}, \bibinfo {author} {\bibfnamefont {U.}~\bibnamefont {Kumar}}, \bibinfo {author} {\bibfnamefont {K.}~\bibnamefont {Sun}}, \ and\ \bibinfo {author} {\bibfnamefont {S.-Z.}\ \bibnamefont {Lin}},\ }\href@noop {} {\bibfield  {journal} {\bibinfo  {journal} {Physical Review Research}\ }\textbf {\bibinfo {volume} {3}},\ \bibinfo {pages} {L032070} (\bibinfo {year} {2021})}\BibitemShut {NoStop}%
\bibitem [{\citenamefont {Devakul}\ \emph {et~al.}(2021)\citenamefont {Devakul}, \citenamefont {Crépel}, \citenamefont {Zhang},\ and\ \citenamefont {Fu}}]{Devakul2021Magic}%
  \BibitemOpen
  \bibfield  {author} {\bibinfo {author} {\bibfnamefont {T.}~\bibnamefont {Devakul}}, \bibinfo {author} {\bibfnamefont {V.}~\bibnamefont {Crépel}}, \bibinfo {author} {\bibfnamefont {Y.}~\bibnamefont {Zhang}}, \ and\ \bibinfo {author} {\bibfnamefont {L.}~\bibnamefont {Fu}},\ }\href@noop {} {\bibfield  {journal} {\bibinfo  {journal} {Nature Communications}\ }\textbf {\bibinfo {volume} {12}},\ \bibinfo {pages} {6730} (\bibinfo {year} {2021})}\BibitemShut {NoStop}%
\bibitem [{\citenamefont {Ledwith}\ \emph {et~al.}(2021)\citenamefont {Ledwith}, \citenamefont {Khalaf},\ and\ \citenamefont {Vishwanath}}]{ledwith2021strong}%
  \BibitemOpen
  \bibfield  {author} {\bibinfo {author} {\bibfnamefont {P.~J.}\ \bibnamefont {Ledwith}}, \bibinfo {author} {\bibfnamefont {E.}~\bibnamefont {Khalaf}}, \ and\ \bibinfo {author} {\bibfnamefont {A.}~\bibnamefont {Vishwanath}},\ }\href@noop {} {\bibfield  {journal} {\bibinfo  {journal} {Annals of Physics}\ }\textbf {\bibinfo {volume} {435}},\ \bibinfo {pages} {168646} (\bibinfo {year} {2021})}\BibitemShut {NoStop}%
\bibitem [{\citenamefont {Wang}\ \emph {et~al.}(2021)\citenamefont {Wang}, \citenamefont {Cano}, \citenamefont {Millis}, \citenamefont {Liu},\ and\ \citenamefont {Yang}}]{wang2021exact}%
  \BibitemOpen
  \bibfield  {author} {\bibinfo {author} {\bibfnamefont {J.}~\bibnamefont {Wang}}, \bibinfo {author} {\bibfnamefont {J.}~\bibnamefont {Cano}}, \bibinfo {author} {\bibfnamefont {A.~J.}\ \bibnamefont {Millis}}, \bibinfo {author} {\bibfnamefont {Z.}~\bibnamefont {Liu}}, \ and\ \bibinfo {author} {\bibfnamefont {B.}~\bibnamefont {Yang}},\ }\href@noop {} {\bibfield  {journal} {\bibinfo  {journal} {Physical review letters}\ }\textbf {\bibinfo {volume} {127}},\ \bibinfo {pages} {246403} (\bibinfo {year} {2021})}\BibitemShut {NoStop}%
\bibitem [{\citenamefont {Xie}\ \emph {et~al.}(2021)\citenamefont {Xie}, \citenamefont {Pierce}, \citenamefont {Park}, \citenamefont {Parker}, \citenamefont {Khalaf}, \citenamefont {Ledwith}, \citenamefont {Cao}, \citenamefont {Lee}, \citenamefont {Chen}, \citenamefont {Forrester} \emph {et~al.}}]{xie2021fractional}%
  \BibitemOpen
  \bibfield  {author} {\bibinfo {author} {\bibfnamefont {Y.}~\bibnamefont {Xie}}, \bibinfo {author} {\bibfnamefont {A.~T.}\ \bibnamefont {Pierce}}, \bibinfo {author} {\bibfnamefont {J.~M.}\ \bibnamefont {Park}}, \bibinfo {author} {\bibfnamefont {D.~E.}\ \bibnamefont {Parker}}, \bibinfo {author} {\bibfnamefont {E.}~\bibnamefont {Khalaf}}, \bibinfo {author} {\bibfnamefont {P.}~\bibnamefont {Ledwith}}, \bibinfo {author} {\bibfnamefont {Y.}~\bibnamefont {Cao}}, \bibinfo {author} {\bibfnamefont {S.~H.}\ \bibnamefont {Lee}}, \bibinfo {author} {\bibfnamefont {S.}~\bibnamefont {Chen}}, \bibinfo {author} {\bibfnamefont {P.~R.}\ \bibnamefont {Forrester}},  \emph {et~al.},\ }\href@noop {} {\bibfield  {journal} {\bibinfo  {journal} {Nature}\ }\textbf {\bibinfo {volume} {600}},\ \bibinfo {pages} {439} (\bibinfo {year} {2021})}\BibitemShut {NoStop}%
\bibitem [{\citenamefont {Cai}\ \emph {et~al.}(2023)\citenamefont {Cai}, \citenamefont {Anderson}, \citenamefont {Wang}, \citenamefont {Zhang}, \citenamefont {Liu}, \citenamefont {Holtzmann}, \citenamefont {Zhang}, \citenamefont {Fan}, \citenamefont {Taniguchi}, \citenamefont {Watanabe} \emph {et~al.}}]{cai2023signatures}%
  \BibitemOpen
  \bibfield  {author} {\bibinfo {author} {\bibfnamefont {J.}~\bibnamefont {Cai}}, \bibinfo {author} {\bibfnamefont {E.}~\bibnamefont {Anderson}}, \bibinfo {author} {\bibfnamefont {C.}~\bibnamefont {Wang}}, \bibinfo {author} {\bibfnamefont {X.}~\bibnamefont {Zhang}}, \bibinfo {author} {\bibfnamefont {X.}~\bibnamefont {Liu}}, \bibinfo {author} {\bibfnamefont {W.}~\bibnamefont {Holtzmann}}, \bibinfo {author} {\bibfnamefont {Y.}~\bibnamefont {Zhang}}, \bibinfo {author} {\bibfnamefont {F.}~\bibnamefont {Fan}}, \bibinfo {author} {\bibfnamefont {T.}~\bibnamefont {Taniguchi}}, \bibinfo {author} {\bibfnamefont {K.}~\bibnamefont {Watanabe}},  \emph {et~al.},\ }\href@noop {} {\bibfield  {journal} {\bibinfo  {journal} {Nature}\ }\textbf {\bibinfo {volume} {622}},\ \bibinfo {pages} {63} (\bibinfo {year} {2023})}\BibitemShut {NoStop}%
\bibitem [{\citenamefont {Zeng}\ \emph {et~al.}(2023)\citenamefont {Zeng}, \citenamefont {Xia}, \citenamefont {Kang}, \citenamefont {Zhu}, \citenamefont {Kn{\"u}ppel}, \citenamefont {Vaswani}, \citenamefont {Watanabe}, \citenamefont {Taniguchi}, \citenamefont {Mak},\ and\ \citenamefont {Shan}}]{zeng2023thermodynamic}%
  \BibitemOpen
  \bibfield  {author} {\bibinfo {author} {\bibfnamefont {Y.}~\bibnamefont {Zeng}}, \bibinfo {author} {\bibfnamefont {Z.}~\bibnamefont {Xia}}, \bibinfo {author} {\bibfnamefont {K.}~\bibnamefont {Kang}}, \bibinfo {author} {\bibfnamefont {J.}~\bibnamefont {Zhu}}, \bibinfo {author} {\bibfnamefont {P.}~\bibnamefont {Kn{\"u}ppel}}, \bibinfo {author} {\bibfnamefont {C.}~\bibnamefont {Vaswani}}, \bibinfo {author} {\bibfnamefont {K.}~\bibnamefont {Watanabe}}, \bibinfo {author} {\bibfnamefont {T.}~\bibnamefont {Taniguchi}}, \bibinfo {author} {\bibfnamefont {K.~F.}\ \bibnamefont {Mak}}, \ and\ \bibinfo {author} {\bibfnamefont {J.}~\bibnamefont {Shan}},\ }\href@noop {} {\bibfield  {journal} {\bibinfo  {journal} {Nature}\ }\textbf {\bibinfo {volume} {622}},\ \bibinfo {pages} {69} (\bibinfo {year} {2023})}\BibitemShut {NoStop}%
\bibitem [{\citenamefont {Park}\ \emph {et~al.}(2023)\citenamefont {Park}, \citenamefont {Cai}, \citenamefont {Anderson}, \citenamefont {Zhang}, \citenamefont {Zhu}, \citenamefont {Liu}, \citenamefont {Wang}, \citenamefont {Holtzmann}, \citenamefont {Hu}, \citenamefont {Liu} \emph {et~al.}}]{park2023observation}%
  \BibitemOpen
  \bibfield  {author} {\bibinfo {author} {\bibfnamefont {H.}~\bibnamefont {Park}}, \bibinfo {author} {\bibfnamefont {J.}~\bibnamefont {Cai}}, \bibinfo {author} {\bibfnamefont {E.}~\bibnamefont {Anderson}}, \bibinfo {author} {\bibfnamefont {Y.}~\bibnamefont {Zhang}}, \bibinfo {author} {\bibfnamefont {J.}~\bibnamefont {Zhu}}, \bibinfo {author} {\bibfnamefont {X.}~\bibnamefont {Liu}}, \bibinfo {author} {\bibfnamefont {C.}~\bibnamefont {Wang}}, \bibinfo {author} {\bibfnamefont {W.}~\bibnamefont {Holtzmann}}, \bibinfo {author} {\bibfnamefont {C.}~\bibnamefont {Hu}}, \bibinfo {author} {\bibfnamefont {Z.}~\bibnamefont {Liu}},  \emph {et~al.},\ }\href@noop {} {\bibfield  {journal} {\bibinfo  {journal} {Nature}\ }\textbf {\bibinfo {volume} {622}},\ \bibinfo {pages} {74} (\bibinfo {year} {2023})}\BibitemShut {NoStop}%
\bibitem [{\citenamefont {Xu}\ \emph {et~al.}(2023)\citenamefont {Xu}, \citenamefont {Sun}, \citenamefont {Jia}, \citenamefont {Liu}, \citenamefont {Xu}, \citenamefont {Li}, \citenamefont {Gu}, \citenamefont {Watanabe}, \citenamefont {Taniguchi}, \citenamefont {Tong} \emph {et~al.}}]{xu2023observation}%
  \BibitemOpen
  \bibfield  {author} {\bibinfo {author} {\bibfnamefont {F.}~\bibnamefont {Xu}}, \bibinfo {author} {\bibfnamefont {Z.}~\bibnamefont {Sun}}, \bibinfo {author} {\bibfnamefont {T.}~\bibnamefont {Jia}}, \bibinfo {author} {\bibfnamefont {C.}~\bibnamefont {Liu}}, \bibinfo {author} {\bibfnamefont {C.}~\bibnamefont {Xu}}, \bibinfo {author} {\bibfnamefont {C.}~\bibnamefont {Li}}, \bibinfo {author} {\bibfnamefont {Y.}~\bibnamefont {Gu}}, \bibinfo {author} {\bibfnamefont {K.}~\bibnamefont {Watanabe}}, \bibinfo {author} {\bibfnamefont {T.}~\bibnamefont {Taniguchi}}, \bibinfo {author} {\bibfnamefont {B.}~\bibnamefont {Tong}},  \emph {et~al.},\ }\href@noop {} {\bibfield  {journal} {\bibinfo  {journal} {Physical Review X}\ }\textbf {\bibinfo {volume} {13}},\ \bibinfo {pages} {031037} (\bibinfo {year} {2023})}\BibitemShut {NoStop}%
\bibitem [{\citenamefont {Ledwith}\ \emph {et~al.}(2023)\citenamefont {Ledwith}, \citenamefont {Vishwanath},\ and\ \citenamefont {Parker}}]{ledwith2023vortexability}%
  \BibitemOpen
  \bibfield  {author} {\bibinfo {author} {\bibfnamefont {P.~J.}\ \bibnamefont {Ledwith}}, \bibinfo {author} {\bibfnamefont {A.}~\bibnamefont {Vishwanath}}, \ and\ \bibinfo {author} {\bibfnamefont {D.~E.}\ \bibnamefont {Parker}},\ }\href@noop {} {\bibfield  {journal} {\bibinfo  {journal} {Physical Review B}\ }\textbf {\bibinfo {volume} {108}},\ \bibinfo {pages} {205144} (\bibinfo {year} {2023})}\BibitemShut {NoStop}%
\bibitem [{\citenamefont {Wu}\ \emph {et~al.}(2024)\citenamefont {Wu}, \citenamefont {Sarkar}, \citenamefont {Wan}, \citenamefont {Sun},\ and\ \citenamefont {Lin}}]{wu2024quantum}%
  \BibitemOpen
  \bibfield  {author} {\bibinfo {author} {\bibfnamefont {A.-K.}\ \bibnamefont {Wu}}, \bibinfo {author} {\bibfnamefont {S.}~\bibnamefont {Sarkar}}, \bibinfo {author} {\bibfnamefont {X.}~\bibnamefont {Wan}}, \bibinfo {author} {\bibfnamefont {K.}~\bibnamefont {Sun}}, \ and\ \bibinfo {author} {\bibfnamefont {S.-Z.}\ \bibnamefont {Lin}},\ }\href@noop {} {\bibfield  {journal} {\bibinfo  {journal} {Physical Review Research}\ }\textbf {\bibinfo {volume} {6}},\ \bibinfo {pages} {L032063} (\bibinfo {year} {2024})}\BibitemShut {NoStop}%
\bibitem [{\citenamefont {Lu}\ \emph {et~al.}(2024)\citenamefont {Lu}, \citenamefont {Han}, \citenamefont {Yao}, \citenamefont {Reddy}, \citenamefont {Yang}, \citenamefont {Seo}, \citenamefont {Watanabe}, \citenamefont {Taniguchi}, \citenamefont {Fu},\ and\ \citenamefont {Ju}}]{lu2024fractional}%
  \BibitemOpen
  \bibfield  {author} {\bibinfo {author} {\bibfnamefont {Z.}~\bibnamefont {Lu}}, \bibinfo {author} {\bibfnamefont {T.}~\bibnamefont {Han}}, \bibinfo {author} {\bibfnamefont {Y.}~\bibnamefont {Yao}}, \bibinfo {author} {\bibfnamefont {A.~P.}\ \bibnamefont {Reddy}}, \bibinfo {author} {\bibfnamefont {J.}~\bibnamefont {Yang}}, \bibinfo {author} {\bibfnamefont {J.}~\bibnamefont {Seo}}, \bibinfo {author} {\bibfnamefont {K.}~\bibnamefont {Watanabe}}, \bibinfo {author} {\bibfnamefont {T.}~\bibnamefont {Taniguchi}}, \bibinfo {author} {\bibfnamefont {L.}~\bibnamefont {Fu}}, \ and\ \bibinfo {author} {\bibfnamefont {L.}~\bibnamefont {Ju}},\ }\href@noop {} {\bibfield  {journal} {\bibinfo  {journal} {Nature}\ }\textbf {\bibinfo {volume} {626}},\ \bibinfo {pages} {759} (\bibinfo {year} {2024})}\BibitemShut {NoStop}%
\bibitem [{\citenamefont {Xie}\ \emph {et~al.}(2025)\citenamefont {Xie}, \citenamefont {Huo}, \citenamefont {Lu}, \citenamefont {Feng}, \citenamefont {Zhang}, \citenamefont {Wang}, \citenamefont {Yang}, \citenamefont {Watanabe}, \citenamefont {Taniguchi}, \citenamefont {Liu} \emph {et~al.}}]{xie2025tunable}%
  \BibitemOpen
  \bibfield  {author} {\bibinfo {author} {\bibfnamefont {J.}~\bibnamefont {Xie}}, \bibinfo {author} {\bibfnamefont {Z.}~\bibnamefont {Huo}}, \bibinfo {author} {\bibfnamefont {X.}~\bibnamefont {Lu}}, \bibinfo {author} {\bibfnamefont {Z.}~\bibnamefont {Feng}}, \bibinfo {author} {\bibfnamefont {Z.}~\bibnamefont {Zhang}}, \bibinfo {author} {\bibfnamefont {W.}~\bibnamefont {Wang}}, \bibinfo {author} {\bibfnamefont {Q.}~\bibnamefont {Yang}}, \bibinfo {author} {\bibfnamefont {K.}~\bibnamefont {Watanabe}}, \bibinfo {author} {\bibfnamefont {T.}~\bibnamefont {Taniguchi}}, \bibinfo {author} {\bibfnamefont {K.}~\bibnamefont {Liu}},  \emph {et~al.},\ }\href@noop {} {\bibfield  {journal} {\bibinfo  {journal} {Nature Materials}\ ,\ \bibinfo {pages} {1}} (\bibinfo {year} {2025})}\BibitemShut {NoStop}%
\bibitem [{\citenamefont {Reddy}\ \emph {et~al.}(2024)\citenamefont {Reddy}, \citenamefont {Paul}, \citenamefont {Abouelkomsan},\ and\ \citenamefont {Fu}}]{reddy2024non}%
  \BibitemOpen
  \bibfield  {author} {\bibinfo {author} {\bibfnamefont {A.~P.}\ \bibnamefont {Reddy}}, \bibinfo {author} {\bibfnamefont {N.}~\bibnamefont {Paul}}, \bibinfo {author} {\bibfnamefont {A.}~\bibnamefont {Abouelkomsan}}, \ and\ \bibinfo {author} {\bibfnamefont {L.}~\bibnamefont {Fu}},\ }\href@noop {} {\bibfield  {journal} {\bibinfo  {journal} {Physical review letters}\ }\textbf {\bibinfo {volume} {133}},\ \bibinfo {pages} {166503} (\bibinfo {year} {2024})}\BibitemShut {NoStop}%
\bibitem [{\citenamefont {Fujimoto}\ \emph {et~al.}(2025)\citenamefont {Fujimoto}, \citenamefont {Parker}, \citenamefont {Dong}, \citenamefont {Khalaf}, \citenamefont {Vishwanath},\ and\ \citenamefont {Ledwith}}]{fujimoto2024higher}%
  \BibitemOpen
  \bibfield  {author} {\bibinfo {author} {\bibfnamefont {M.}~\bibnamefont {Fujimoto}}, \bibinfo {author} {\bibfnamefont {D.~E.}\ \bibnamefont {Parker}}, \bibinfo {author} {\bibfnamefont {J.}~\bibnamefont {Dong}}, \bibinfo {author} {\bibfnamefont {E.}~\bibnamefont {Khalaf}}, \bibinfo {author} {\bibfnamefont {A.}~\bibnamefont {Vishwanath}}, \ and\ \bibinfo {author} {\bibfnamefont {P.}~\bibnamefont {Ledwith}},\ }\href@noop {} {\bibfield  {journal} {\bibinfo  {journal} {Phys. Rev. Lett.}\ }\textbf {\bibinfo {volume} {134}},\ \bibinfo {pages} {106502} (\bibinfo {year} {2025})}\BibitemShut {NoStop}%
\bibitem [{\citenamefont {Liu}\ \emph {et~al.}(2025)\citenamefont {Liu}, \citenamefont {Mera}, \citenamefont {Fujimoto}, \citenamefont {Ozawa},\ and\ \citenamefont {Wang}}]{liu2025theory}%
  \BibitemOpen
  \bibfield  {author} {\bibinfo {author} {\bibfnamefont {Z.}~\bibnamefont {Liu}}, \bibinfo {author} {\bibfnamefont {B.}~\bibnamefont {Mera}}, \bibinfo {author} {\bibfnamefont {M.}~\bibnamefont {Fujimoto}}, \bibinfo {author} {\bibfnamefont {T.}~\bibnamefont {Ozawa}}, \ and\ \bibinfo {author} {\bibfnamefont {J.}~\bibnamefont {Wang}},\ }\href@noop {} {\bibfield  {journal} {\bibinfo  {journal} {Physical Review X}\ }\textbf {\bibinfo {volume} {15}},\ \bibinfo {pages} {031019} (\bibinfo {year} {2025})}\BibitemShut {NoStop}%
\bibitem [{\citenamefont {Zhang}\ and\ \citenamefont {Sarma}(1986)}]{zhang1986excitation}%
  \BibitemOpen
  \bibfield  {author} {\bibinfo {author} {\bibfnamefont {F.-C.}\ \bibnamefont {Zhang}}\ and\ \bibinfo {author} {\bibfnamefont {S.~D.}\ \bibnamefont {Sarma}},\ }\href@noop {} {\bibfield  {journal} {\bibinfo  {journal} {Physical Review B}\ }\textbf {\bibinfo {volume} {33}},\ \bibinfo {pages} {2903} (\bibinfo {year} {1986})}\BibitemShut {NoStop}%
\bibitem [{\citenamefont {Morf}\ \emph {et~al.}(2002)\citenamefont {Morf}, \citenamefont {d’Ambrumenil},\ and\ \citenamefont {Sarma}}]{morf2002excitation}%
  \BibitemOpen
  \bibfield  {author} {\bibinfo {author} {\bibfnamefont {R.}~\bibnamefont {Morf}}, \bibinfo {author} {\bibfnamefont {N.}~\bibnamefont {d’Ambrumenil}}, \ and\ \bibinfo {author} {\bibfnamefont {S.~D.}\ \bibnamefont {Sarma}},\ }\href@noop {} {\bibfield  {journal} {\bibinfo  {journal} {Physical Review B}\ }\textbf {\bibinfo {volume} {66}},\ \bibinfo {pages} {075408} (\bibinfo {year} {2002})}\BibitemShut {NoStop}%
\bibitem [{\citenamefont {Peterson}\ \emph {et~al.}(2008{\natexlab{a}})\citenamefont {Peterson}, \citenamefont {Jolicoeur},\ and\ \citenamefont {Das~Sarma}}]{peterson2008finite}%
  \BibitemOpen
  \bibfield  {author} {\bibinfo {author} {\bibfnamefont {M.~R.}\ \bibnamefont {Peterson}}, \bibinfo {author} {\bibfnamefont {T.}~\bibnamefont {Jolicoeur}}, \ and\ \bibinfo {author} {\bibfnamefont {S.}~\bibnamefont {Das~Sarma}},\ }\href@noop {} {\bibfield  {journal} {\bibinfo  {journal} {Physical review letters}\ }\textbf {\bibinfo {volume} {101}},\ \bibinfo {pages} {016807} (\bibinfo {year} {2008}{\natexlab{a}})}\BibitemShut {NoStop}%
\bibitem [{\citenamefont {Peterson}\ \emph {et~al.}(2008{\natexlab{b}})\citenamefont {Peterson}, \citenamefont {Jolicoeur},\ and\ \citenamefont {Das~Sarma}}]{peterson2008orbital}%
  \BibitemOpen
  \bibfield  {author} {\bibinfo {author} {\bibfnamefont {M.~R.}\ \bibnamefont {Peterson}}, \bibinfo {author} {\bibfnamefont {T.}~\bibnamefont {Jolicoeur}}, \ and\ \bibinfo {author} {\bibfnamefont {S.}~\bibnamefont {Das~Sarma}},\ }\href@noop {} {\bibfield  {journal} {\bibinfo  {journal} {Physical Review B—Condensed Matter and Materials Physics}\ }\textbf {\bibinfo {volume} {78}},\ \bibinfo {pages} {155308} (\bibinfo {year} {2008}{\natexlab{b}})}\BibitemShut {NoStop}%
\bibitem [{\citenamefont {Peterson}\ and\ \citenamefont {Das~Sarma}(2010)}]{peterson2010quantum}%
  \BibitemOpen
  \bibfield  {author} {\bibinfo {author} {\bibfnamefont {M.~R.}\ \bibnamefont {Peterson}}\ and\ \bibinfo {author} {\bibfnamefont {S.}~\bibnamefont {Das~Sarma}},\ }\href@noop {} {\bibfield  {journal} {\bibinfo  {journal} {Physical Review B—Condensed Matter and Materials Physics}\ }\textbf {\bibinfo {volume} {81}},\ \bibinfo {pages} {165304} (\bibinfo {year} {2010})}\BibitemShut {NoStop}%
\bibitem [{\citenamefont {Storni}\ \emph {et~al.}(2010)\citenamefont {Storni}, \citenamefont {Morf},\ and\ \citenamefont {Das~Sarma}}]{storni2010fractional}%
  \BibitemOpen
  \bibfield  {author} {\bibinfo {author} {\bibfnamefont {M.}~\bibnamefont {Storni}}, \bibinfo {author} {\bibfnamefont {R.}~\bibnamefont {Morf}}, \ and\ \bibinfo {author} {\bibfnamefont {S.}~\bibnamefont {Das~Sarma}},\ }\href@noop {} {\bibfield  {journal} {\bibinfo  {journal} {Physical review letters}\ }\textbf {\bibinfo {volume} {104}},\ \bibinfo {pages} {076803} (\bibinfo {year} {2010})}\BibitemShut {NoStop}%
\bibitem [{\citenamefont {Papi{\'c}}\ \emph {et~al.}(2011{\natexlab{a}})\citenamefont {Papi{\'c}}, \citenamefont {Abanin}, \citenamefont {Barlas},\ and\ \citenamefont {Bhatt}}]{papic2011tunablePRB}%
  \BibitemOpen
  \bibfield  {author} {\bibinfo {author} {\bibfnamefont {Z.}~\bibnamefont {Papi{\'c}}}, \bibinfo {author} {\bibfnamefont {D.}~\bibnamefont {Abanin}}, \bibinfo {author} {\bibfnamefont {Y.}~\bibnamefont {Barlas}}, \ and\ \bibinfo {author} {\bibfnamefont {R.~N.}\ \bibnamefont {Bhatt}},\ }\href@noop {} {\bibfield  {journal} {\bibinfo  {journal} {Physical Review B—Condensed Matter and Materials Physics}\ }\textbf {\bibinfo {volume} {84}},\ \bibinfo {pages} {241306} (\bibinfo {year} {2011}{\natexlab{a}})}\BibitemShut {NoStop}%
\bibitem [{\citenamefont {Papi{\'c}}\ \emph {et~al.}(2011{\natexlab{b}})\citenamefont {Papi{\'c}}, \citenamefont {Thomale},\ and\ \citenamefont {Abanin}}]{papic2011tunablePRL}%
  \BibitemOpen
  \bibfield  {author} {\bibinfo {author} {\bibfnamefont {Z.}~\bibnamefont {Papi{\'c}}}, \bibinfo {author} {\bibfnamefont {R.}~\bibnamefont {Thomale}}, \ and\ \bibinfo {author} {\bibfnamefont {D.}~\bibnamefont {Abanin}},\ }\href@noop {} {\bibfield  {journal} {\bibinfo  {journal} {Physical Review Letters}\ }\textbf {\bibinfo {volume} {107}},\ \bibinfo {pages} {176602} (\bibinfo {year} {2011}{\natexlab{b}})}\BibitemShut {NoStop}%
\bibitem [{\citenamefont {Papi{\'c}}\ \emph {et~al.}(2012)\citenamefont {Papi{\'c}}, \citenamefont {Abanin}, \citenamefont {Barias},\ and\ \citenamefont {Bhatt}}]{papic2012numerical}%
  \BibitemOpen
  \bibfield  {author} {\bibinfo {author} {\bibfnamefont {Z.}~\bibnamefont {Papi{\'c}}}, \bibinfo {author} {\bibfnamefont {D.}~\bibnamefont {Abanin}}, \bibinfo {author} {\bibfnamefont {Y.}~\bibnamefont {Barias}}, \ and\ \bibinfo {author} {\bibfnamefont {R.~N.}\ \bibnamefont {Bhatt}},\ }in\ \href@noop {} {\emph {\bibinfo {booktitle} {Journal of Physics: Conference Series}}},\ Vol.\ \bibinfo {volume} {402}\ (\bibinfo {organization} {IOP Publishing},\ \bibinfo {year} {2012})\ p.\ \bibinfo {pages} {012020}\BibitemShut {NoStop}%
\bibitem [{\citenamefont {Wan}\ \emph {et~al.}(2023)\citenamefont {Wan}, \citenamefont {Sarkar}, \citenamefont {Lin},\ and\ \citenamefont {Sun}}]{wan2023topological}%
  \BibitemOpen
  \bibfield  {author} {\bibinfo {author} {\bibfnamefont {X.}~\bibnamefont {Wan}}, \bibinfo {author} {\bibfnamefont {S.}~\bibnamefont {Sarkar}}, \bibinfo {author} {\bibfnamefont {S.-Z.}\ \bibnamefont {Lin}}, \ and\ \bibinfo {author} {\bibfnamefont {K.}~\bibnamefont {Sun}},\ }\href@noop {} {\bibfield  {journal} {\bibinfo  {journal} {Physical Review Letters}\ }\textbf {\bibinfo {volume} {130}},\ \bibinfo {pages} {216401} (\bibinfo {year} {2023})}\BibitemShut {NoStop}%
\bibitem [{\citenamefont {Sarkar}\ \emph {et~al.}(2025)\citenamefont {Sarkar}, \citenamefont {Wan}, \citenamefont {Lin},\ and\ \citenamefont {Sun}}]{sarkar2023symmetry}%
  \BibitemOpen
  \bibfield  {author} {\bibinfo {author} {\bibfnamefont {S.}~\bibnamefont {Sarkar}}, \bibinfo {author} {\bibfnamefont {X.}~\bibnamefont {Wan}}, \bibinfo {author} {\bibfnamefont {S.-Z.}\ \bibnamefont {Lin}}, \ and\ \bibinfo {author} {\bibfnamefont {K.}~\bibnamefont {Sun}},\ }\href@noop {} {\bibfield  {journal} {\bibinfo  {journal} {Physical Review Letters}\ }\textbf {\bibinfo {volume} {135}},\ \bibinfo {pages} {016501} (\bibinfo {year} {2025})}\BibitemShut {NoStop}%
\bibitem [{SM2()}]{SM2025}%
  \BibitemOpen
  \href@noop {} {\enquote {\bibinfo {title} {See \uppercase{S}upplemental \uppercase{M}aterial ... for details},}\ }\BibitemShut {NoStop}%
\bibitem [{\citenamefont {Shi}\ \emph {et~al.}(2025)\citenamefont {Shi}, \citenamefont {Cano},\ and\ \citenamefont {Morales-Dur{\'a}n}}]{shi2025effects}%
  \BibitemOpen
  \bibfield  {author} {\bibinfo {author} {\bibfnamefont {J.}~\bibnamefont {Shi}}, \bibinfo {author} {\bibfnamefont {J.}~\bibnamefont {Cano}}, \ and\ \bibinfo {author} {\bibfnamefont {N.}~\bibnamefont {Morales-Dur{\'a}n}},\ }\href@noop {} {\bibfield  {journal} {\bibinfo  {journal} {arXiv preprint arXiv:2503.15900}\ } (\bibinfo {year} {2025})}\BibitemShut {NoStop}%
\bibitem [{\citenamefont {Guerci}\ \emph {et~al.}(2025)\citenamefont {Guerci}, \citenamefont {Abouelkomsan},\ and\ \citenamefont {Fu}}]{guerci2025fractionalization}%
  \BibitemOpen
  \bibfield  {author} {\bibinfo {author} {\bibfnamefont {D.}~\bibnamefont {Guerci}}, \bibinfo {author} {\bibfnamefont {A.}~\bibnamefont {Abouelkomsan}}, \ and\ \bibinfo {author} {\bibfnamefont {L.}~\bibnamefont {Fu}},\ }\href@noop {} {\bibfield  {journal} {\bibinfo  {journal} {arXiv preprint arXiv:2506.10938}\ } (\bibinfo {year} {2025})}\BibitemShut {NoStop}%
\bibitem [{\citenamefont {Yutushui}\ \emph {et~al.}(2025)\citenamefont {Yutushui}, \citenamefont {Dey},\ and\ \citenamefont {Mross}}]{yutushui2025numerical}%
  \BibitemOpen
  \bibfield  {author} {\bibinfo {author} {\bibfnamefont {M.}~\bibnamefont {Yutushui}}, \bibinfo {author} {\bibfnamefont {A.}~\bibnamefont {Dey}}, \ and\ \bibinfo {author} {\bibfnamefont {D.~F.}\ \bibnamefont {Mross}},\ }\href@noop {} {\bibfield  {journal} {\bibinfo  {journal} {arXiv preprint arXiv:2508.14162}\ } (\bibinfo {year} {2025})}\BibitemShut {NoStop}%
\bibitem [{\citenamefont {Vishwanath}(2023)}]{vishwanathzero}%
  \BibitemOpen
  \bibfield  {author} {\bibinfo {author} {\bibfnamefont {A.}~\bibnamefont {Vishwanath}},\ }\href@noop {} {\bibfield  {journal} {\bibinfo  {journal} {J. Club Condens. Matter Phys}\ } (\bibinfo {year} {2023})}\BibitemShut {NoStop}%
\bibitem [{\citenamefont {Semenoff}(1984)}]{semenoff1984condensed}%
  \BibitemOpen
  \bibfield  {author} {\bibinfo {author} {\bibfnamefont {G.~W.}\ \bibnamefont {Semenoff}},\ }\href@noop {} {\bibfield  {journal} {\bibinfo  {journal} {Physical Review Letters}\ }\textbf {\bibinfo {volume} {53}},\ \bibinfo {pages} {2449} (\bibinfo {year} {1984})}\BibitemShut {NoStop}%
\bibitem [{\citenamefont {Shon}\ and\ \citenamefont {Ando}(1998)}]{shon1998quantum}%
  \BibitemOpen
  \bibfield  {author} {\bibinfo {author} {\bibfnamefont {N.~H.}\ \bibnamefont {Shon}}\ and\ \bibinfo {author} {\bibfnamefont {T.}~\bibnamefont {Ando}},\ }\href@noop {} {\bibfield  {journal} {\bibinfo  {journal} {Journal of the Physical Society of Japan}\ }\textbf {\bibinfo {volume} {67}},\ \bibinfo {pages} {2421} (\bibinfo {year} {1998})}\BibitemShut {NoStop}%
\bibitem [{\citenamefont {T{\H{o}}ke}\ \emph {et~al.}(2006)\citenamefont {T{\H{o}}ke}, \citenamefont {Lammert}, \citenamefont {Crespi},\ and\ \citenamefont {Jain}}]{tHoke2006fractional}%
  \BibitemOpen
  \bibfield  {author} {\bibinfo {author} {\bibfnamefont {C.}~\bibnamefont {T{\H{o}}ke}}, \bibinfo {author} {\bibfnamefont {P.~E.}\ \bibnamefont {Lammert}}, \bibinfo {author} {\bibfnamefont {V.~H.}\ \bibnamefont {Crespi}}, \ and\ \bibinfo {author} {\bibfnamefont {J.~K.}\ \bibnamefont {Jain}},\ }\href@noop {} {\bibfield  {journal} {\bibinfo  {journal} {Physical Review B—Condensed Matter and Materials Physics}\ }\textbf {\bibinfo {volume} {74}},\ \bibinfo {pages} {235417} (\bibinfo {year} {2006})}\BibitemShut {NoStop}%
\bibitem [{\citenamefont {Haldane}(1983)}]{haldane1983fractional}%
  \BibitemOpen
  \bibfield  {author} {\bibinfo {author} {\bibfnamefont {F.~D.~M.}\ \bibnamefont {Haldane}},\ }\href@noop {} {\bibfield  {journal} {\bibinfo  {journal} {Physical Review Letters}\ }\textbf {\bibinfo {volume} {51}},\ \bibinfo {pages} {605} (\bibinfo {year} {1983})}\BibitemShut {NoStop}%
\bibitem [{\citenamefont {Trugman}\ and\ \citenamefont {Kivelson}(1985)}]{trugman1985exact}%
  \BibitemOpen
  \bibfield  {author} {\bibinfo {author} {\bibfnamefont {S.~A.}\ \bibnamefont {Trugman}}\ and\ \bibinfo {author} {\bibfnamefont {S.}~\bibnamefont {Kivelson}},\ }\href@noop {} {\bibfield  {journal} {\bibinfo  {journal} {Physical Review B}\ }\textbf {\bibinfo {volume} {31}},\ \bibinfo {pages} {5280} (\bibinfo {year} {1985})}\BibitemShut {NoStop}%
\bibitem [{\citenamefont {Haldane}(1987)}]{Haldane1987QHE}%
  \BibitemOpen
  \bibfield  {author} {\bibinfo {author} {\bibfnamefont {F.~D.~M.}\ \bibnamefont {Haldane}},\ }in\ \href@noop {} {\emph {\bibinfo {booktitle} {The Quantum Hall Effect}}},\ \bibinfo {editor} {edited by\ \bibinfo {editor} {\bibfnamefont {R.~E.}\ \bibnamefont {Prange}}\ and\ \bibinfo {editor} {\bibfnamefont {S.~M.}\ \bibnamefont {Girvin}}}\ (\bibinfo  {publisher} {Springer},\ \bibinfo {address} {New York},\ \bibinfo {year} {1987})\BibitemShut {NoStop}%
\bibitem [{\citenamefont {Kwan}\ \emph {et~al.}(2024)\citenamefont {Kwan}, \citenamefont {Wagner}, \citenamefont {Yu}, \citenamefont {Dagnino}, \citenamefont {Jiang}, \citenamefont {Xu}, \citenamefont {Bernevig}, \citenamefont {Neupert},\ and\ \citenamefont {Regnault}}]{kwan2024abelianfractionaltopologicalinsulators}%
  \BibitemOpen
  \bibfield  {author} {\bibinfo {author} {\bibfnamefont {Y.~H.}\ \bibnamefont {Kwan}}, \bibinfo {author} {\bibfnamefont {G.}~\bibnamefont {Wagner}}, \bibinfo {author} {\bibfnamefont {J.}~\bibnamefont {Yu}}, \bibinfo {author} {\bibfnamefont {A.~K.}\ \bibnamefont {Dagnino}}, \bibinfo {author} {\bibfnamefont {Y.}~\bibnamefont {Jiang}}, \bibinfo {author} {\bibfnamefont {X.}~\bibnamefont {Xu}}, \bibinfo {author} {\bibfnamefont {B.~A.}\ \bibnamefont {Bernevig}}, \bibinfo {author} {\bibfnamefont {T.}~\bibnamefont {Neupert}}, \ and\ \bibinfo {author} {\bibfnamefont {N.}~\bibnamefont {Regnault}},\ }\href@noop {} {\bibfield  {journal} {\bibinfo  {journal} {arXiv preprint arXiv:2407.02560}\ } (\bibinfo {year} {2024})}\BibitemShut {NoStop}%
\bibitem [{\citenamefont {Eugenio}\ and\ \citenamefont {Vafek}(2023)}]{eugenio2022twisted}%
  \BibitemOpen
  \bibfield  {author} {\bibinfo {author} {\bibfnamefont {P.~M.}\ \bibnamefont {Eugenio}}\ and\ \bibinfo {author} {\bibfnamefont {O.}~\bibnamefont {Vafek}},\ }\href@noop {} {\bibfield  {journal} {\bibinfo  {journal} {SciPost Physics}\ }\textbf {\bibinfo {volume} {15}},\ \bibinfo {pages} {081} (\bibinfo {year} {2023})}\BibitemShut {NoStop}%
\bibitem [{\citenamefont {Kharchev}\ and\ \citenamefont {Zabrodin}(2015)}]{kharchev2015theta}%
  \BibitemOpen
  \bibfield  {author} {\bibinfo {author} {\bibfnamefont {S.}~\bibnamefont {Kharchev}}\ and\ \bibinfo {author} {\bibfnamefont {A.}~\bibnamefont {Zabrodin}},\ }\href@noop {} {\bibfield  {journal} {\bibinfo  {journal} {Journal of Geometry and Physics}\ }\textbf {\bibinfo {volume} {94}},\ \bibinfo {pages} {19} (\bibinfo {year} {2015})}\BibitemShut {NoStop}%
\bibitem [{\citenamefont {Jain}(2007)}]{jain2007composite}%
  \BibitemOpen
  \bibfield  {author} {\bibinfo {author} {\bibfnamefont {J.~K.}\ \bibnamefont {Jain}},\ }\href@noop {} {\emph {\bibinfo {title} {Composite fermions}}}\ (\bibinfo  {publisher} {Cambridge University Press},\ \bibinfo {year} {2007})\BibitemShut {NoStop}%
\bibitem [{\citenamefont {Murthy}\ and\ \citenamefont {Shankar}(2012)}]{murthy2012hamiltonian}%
  \BibitemOpen
  \bibfield  {author} {\bibinfo {author} {\bibfnamefont {G.}~\bibnamefont {Murthy}}\ and\ \bibinfo {author} {\bibfnamefont {R.}~\bibnamefont {Shankar}},\ }\href@noop {} {\bibfield  {journal} {\bibinfo  {journal} {Physical Review B—Condensed Matter and Materials Physics}\ }\textbf {\bibinfo {volume} {86}},\ \bibinfo {pages} {195146} (\bibinfo {year} {2012})}\BibitemShut {NoStop}%
\bibitem [{\citenamefont {Goerbig}(2011)}]{goerbig2011electronic}%
  \BibitemOpen
  \bibfield  {author} {\bibinfo {author} {\bibfnamefont {M.}~\bibnamefont {Goerbig}},\ }\href@noop {} {\bibfield  {journal} {\bibinfo  {journal} {Reviews of Modern Physics}\ }\textbf {\bibinfo {volume} {83}},\ \bibinfo {pages} {1193} (\bibinfo {year} {2011})}\BibitemShut {NoStop}%
\bibitem [{\citenamefont {Sodemann}\ and\ \citenamefont {MacDonald}(2013)}]{sodemann2013landau}%
  \BibitemOpen
  \bibfield  {author} {\bibinfo {author} {\bibfnamefont {I.}~\bibnamefont {Sodemann}}\ and\ \bibinfo {author} {\bibfnamefont {A.}~\bibnamefont {MacDonald}},\ }\href@noop {} {\bibfield  {journal} {\bibinfo  {journal} {Physical Review B—Condensed Matter and Materials Physics}\ }\textbf {\bibinfo {volume} {87}},\ \bibinfo {pages} {245425} (\bibinfo {year} {2013})}\BibitemShut {NoStop}%
\end{thebibliography}%
\clearpage
\appendix
\setcounter{equation}{0}  %  this will re-count eq from 1
\setcounter{figure}{0}
\renewcommand{\theequation}{A\arabic{equation}}
\renewcommand{\thefigure}{A\arabic{figure}}

% \section{End Matter}
\subsection*{Appendix A. Many-body energy spectra and particle entanglement spectra for $\nu=1/3$ and $\nu=1/5$ FCI ground-states.}
\begin{figure}[h]
     \centering
\includegraphics[width=\linewidth]{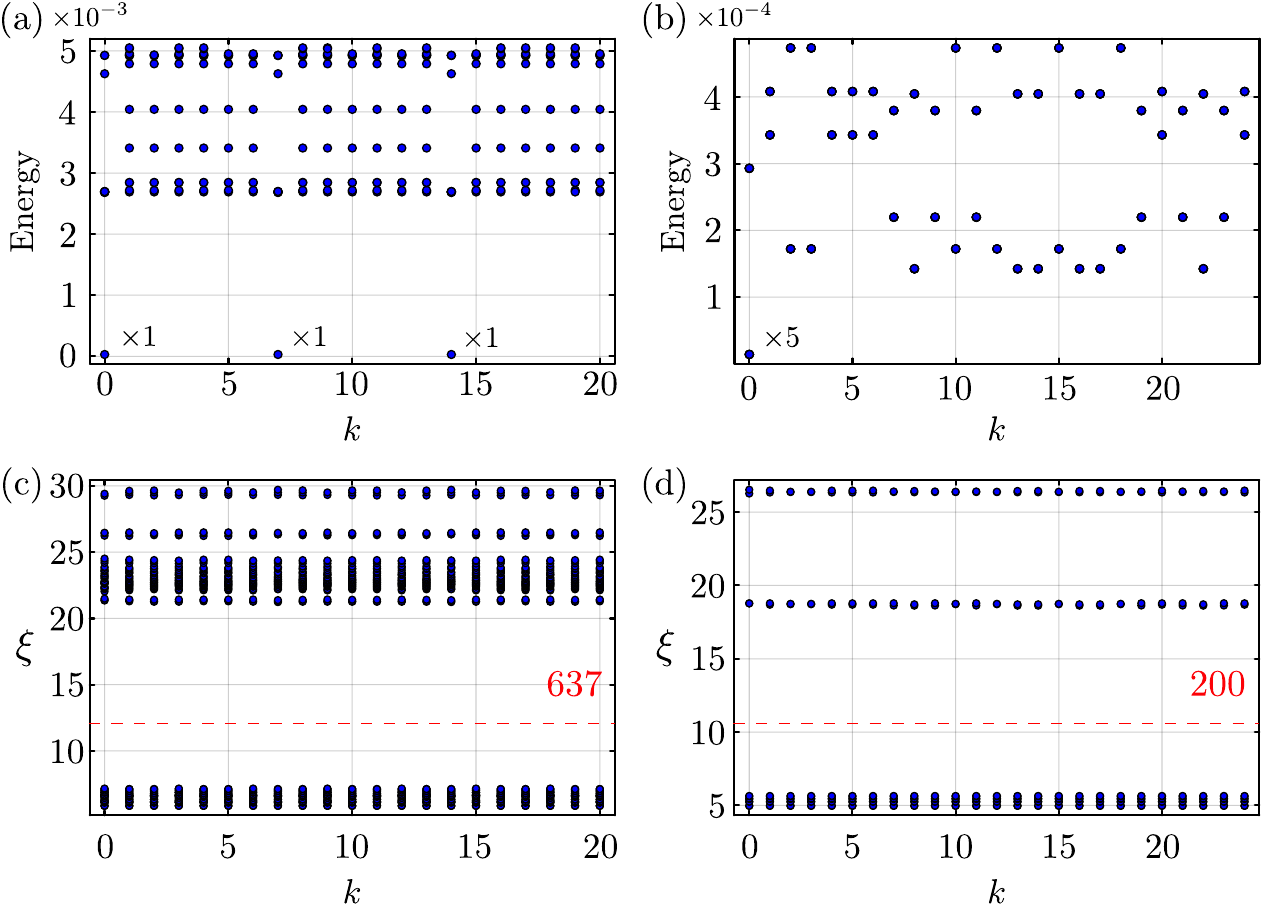}
     \caption{(a-b) Many-body energy spectra at filling $\nu=1/3$, $d_s/l_B=0.037$ and $\nu=1/5$, $d_s/l_B=0.23$ respectively. The energy is normalized by $U^{\rm LL}_{\mathrm{int}}$ as in Fig. 1.
     (c-d) Particle entanglement spectra of the ground states in (a-b) with particle cuts $N_A=3,N_B=4$ for $\nu=1/3$ and $N_A=2,N_B=3$ for $\nu=1/5$, respectively. The number of low lying states below the red dashed lines (as written in red) match the quasi-hole counting of Laughlin states at the respective filling fractions~\cite{Regnault2011fractional}. $\theta=\pi/4$ was used for all cases.
    }
     \label{fig:em1}
\end{figure}
\newpage
% \FloatBarrier

\subsection*{Appendix B. Many-body gap enhancement of Abelian states in a moir\'e model with significant single-particle quantum geometric fluctuations}
\begin{figure}[tbh]
     \centering
\includegraphics[width=\linewidth]{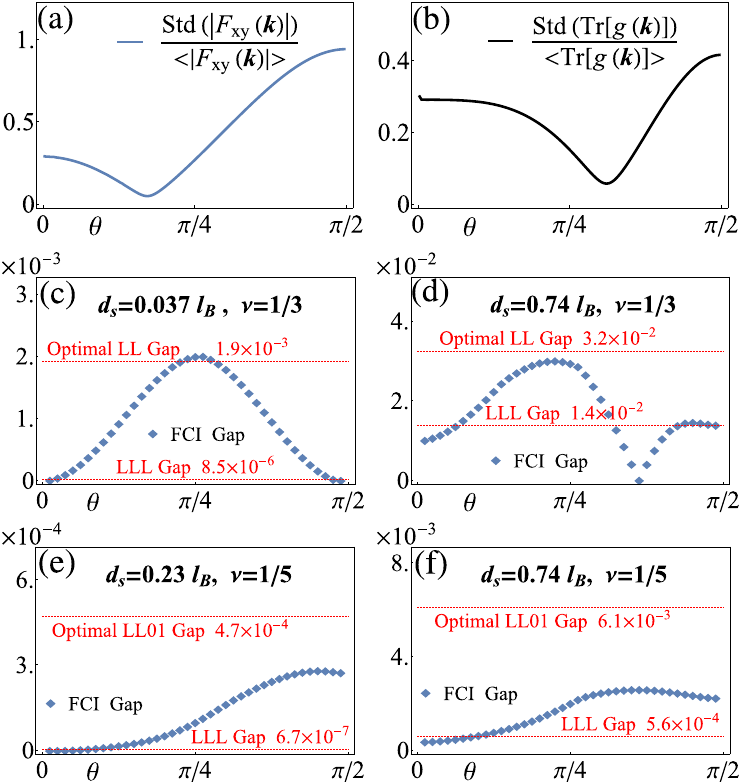}
     \caption{Many-body gap enhancement in a moir\'e model with significant single-particle quantum geometric fluctuations. The single particle Hamiltonian is the same as in Eq.~\eqref{eq:moireHamiltonian} with $\tilde{A}(\mathbf{r}) = -\alpha \sum_{n=1}^3 e^{i(1-n)\phi}\cos(\mathbf{G}_n\cdot\mathbf{r})/2$, where $\mathbf{G}_n$ and $\phi$ were defined below Eq.~\eqref{eq:moireHamiltonian}, and $\alpha = 0.62|G|^2$. Similar to Fig.~\ref{fig:moire}(a), there are four exact flatbands, two per sublattice. One of these two sublattice polarized bands in vortexable with wavefunction $\tilde{\Psi}_{\mathbf{k},1}(\mathbf{r})$, the other one higher vortexable $\tilde{\Psi}_{\mathbf{k},2}(\mathbf{r})$ as discussed in the main text. (a) and (b) show the fluctuations of Berry curvature and quantum metric for the higher vortexable band $\tilde{\Psi}_{\mathbf{k},2}(\mathbf{r})$, respectively. Clearly, for this choice of $\tilde{A}(\mathbf{r})$, the fluctuations in Berry curvature and quantum metric are much higher than the one in Fig.~\ref{fig:moire}. (c-f) Many-body gap of FCI states (blue dots) from ED as a function of $\theta$. As in Fig.~\ref{fig:moire}, the energy is measured in units of the Coulomb energy scale $U^{\rm LL}_{\rm int}$. The parameters used in ED for (c-f) are same as the parameter used in Fig.~\ref{fig:moire}(c-f), respectively. The lower and upper horizontal red dashed lines in (c–f) mark, respectively, the gap of the corresponding FQH states in the LLL and the maximal many-body gap achieved in the LL-hybridization model with the same interaction strength. Gap enhancement trends in this case remain the same as in Fig.~\ref{fig:moire}, although the maximum gap enhancements are significantly lower than those in Fig.~\ref{fig:moire}; the gap sizes do not follow the variation in $\text{tr}(g)$ and $F_{xy}$. In (d), a phase transition occurs at $\theta\approx 1.18$;  our analysis here focuses on the maximum gap of the FCI phase.
    }
     \label{fig:moire2}
\end{figure}

\subsection*{Appendix C. Many-body gap enhancement of non-Abelian Moore-Read state via LL hybridiazation}
% \FloatBarrier
\begin{figure}[tbh]
     \centering
\includegraphics[width=\linewidth]{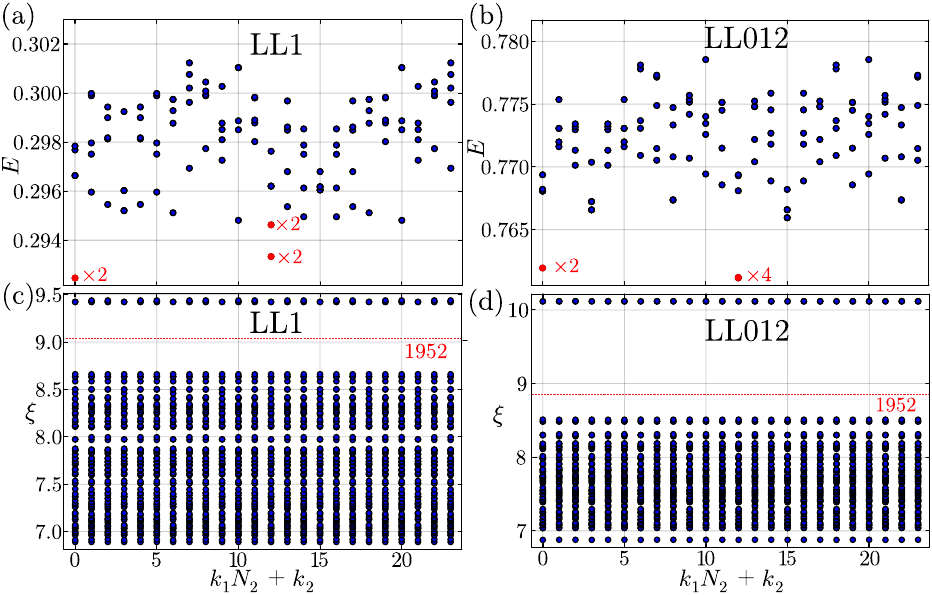}
     \caption{(a-b) Many-body energy spectra at filling $\nu=1/2$ and $d_s/l_B={0.5}$. Here we consider a triangular lattice of size $N_1\times N_2 = 4\times 6$. The red dots indicate Moore–Read states exhibiting sixfold quasi-degeneracy. In the $n = 1$ LL (a), the Moore–Read state does not display a clear gap. In panel (b), we use a three-component hybridized wavefunction constructed from the lowest three Landau levels, $\Psi=(\sqrt{w_0} \psi^{\text{LL}0},\sqrt{w_1} \psi^{\text{LL}1},\sqrt{w_2} \psi^{\text{LL}2})$, with $(w_0, w_1,w_2)={(0.273,0.273,0.454)}$.
     The gap size here is significantly increased to $0.004\, U_{\rm int}^{\rm LL}$.
     (c-d) Particle entanglement spectra of the ground states in (a-b) with particle cut $N_A=3,N_B=9$. The number of low lying states below the red dashed lines match the quasi-hole counting of the Moore-Read state~\cite{Regnault2011fractional,bernevig2012emergent}.
    }
     \label{fig:em1}
\end{figure}
\clearpage
% \pagebreak
\let\addcontentsline\oldaddcontentsline
\onecolumngrid
% \vspace{10cm}

%%%%%%%%%%%%%%%%%%%%%%%%%%%%%%%%%%%%%%
%%   Supplementary Information
%%%%%%%%%%%%%%%%%%%%%%%%%%%%%%%%%%%%%%
\makeatletter
\renewcommand \thesection{S-\@arabic\c@section}
\renewcommand\thetable{S\@arabic\c@table}
\renewcommand \thefigure{S\@arabic\c@figure}
\renewcommand \theequation{S\@arabic\c@equation}
\makeatother
\setcounter{equation}{0}  %  this will re-count eq from 1
\setcounter{figure}{0}  %  this will re-count eq from 1
\setcounter{section}{0}  %  this will re-count eq from 1
\counterwithin{figure}{section} 
{
    \center \bf \large 
    Supplemental Material\vspace*{0.1cm}\\ 
    \vspace*{0.0cm}
}
\maketitle
\tableofcontents

\section{Details of the flat band wavefunctions in the moir\'e system}
\begin{figure}[t]
  \centering
  \includegraphics[width= \textwidth,page=1]{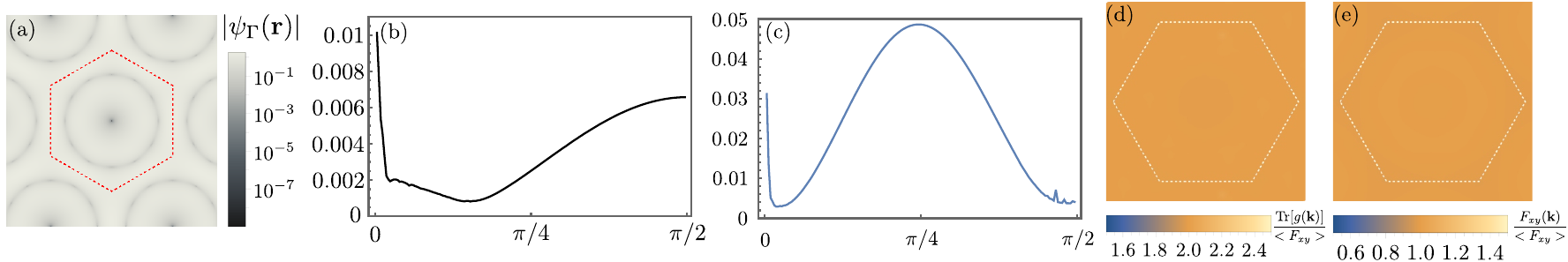}
  \caption{\textbf{Details of the flat band wavefunctions in the moir\'e system.} 
(a) Density plot of the absolute value of the sublattice polarized eigenfunction $\psi_\Gamma(\mathbf{r})$ of $\mathcal{H}_s(\mathbf{r})$ for $\tilde{A}(\mathbf{r})= -\frac{\alpha}{2} \sum_{n=1}^{3}(\frac{\eta}{2} e^{i(1-n)\phi} \cos\left(2\mathbf{G}_n \cdot \mathbf{r}\right)-\beta e^{i(2-n)\phi} \cos\left((\mathbf{G}_n-\mathbf{G}_{n+1}) \cdot \mathbf{r}\right) + e^{i(1-n)\phi} \cos\left(\mathbf{G}_n \cdot \mathbf{r} \right))$, where $\phi=2\pi /3$, $\mathbf{G}_{1} = \frac{4\pi}{\sqrt{3}a}(0,1)$ and $\mathbf{G}_{2,3} = \frac{4\pi}{\sqrt{3}a}(\mp\sqrt{3}/2,-1/2)$ are the reciprocal lattice vectors, $a$ is the moir\'e lattice constant, and $(\alpha,\beta,\eta,\gamma)\approx(4.38,0.5,-0.9,100)$. The plot shows a zero at the center of the moir\'e unit cell as well as a ring of zeros around the center. The red dashed hexagon mark the moir\'e unit cell. (b-c) Fluctuation of Berry curvature $F_{xy}(\mathbf{k})$ and $\text{Tr}(g(\mathbf{k}))$ in the moir\'e Brillouin zone as a function of $\theta$, where $\tan\theta = l_B\gamma/\sqrt{8}$, respectively. The fluctuations are normalized by the average value of the corresponding quantity in the moir\'e Brillouin zone. (d-e) Density plot of $\text{Tr}[g(\mathbf{k})]$ ($F_{xy}(\mathbf{k})$) normalized by the average Berry curvature $<F_{xy}(\mathbf{k})>$ for $\Psi_{\mathbf{k},2}$ at $(\alpha,\beta,\eta,\theta)\approx(4.38,0.5,-0.9,\pi/4)$, respectively.}
  \label{fig:moireSup}
\end{figure}

To understand the origin of exact flat bands in the moir\'e system described the Hamiltonian in {\color{red}Eq.~1} of the main text, we start from the following Hamiltonian
\begin{equation}
    \mathcal{H}_s(\mathbf{r}) = \begin{pmatrix}
        0 & \mathcal{D}_s^\dagger(\mathbf{r})\\\mathcal{D}_s(\mathbf{r}) & 0
    \end{pmatrix},\, \mathcal{D}_s(\mathbf{r}) = -4\overline{\partial_z}^2+\tilde{A}(\mathbf{r}),
\end{equation}
where $z=x+iy$ is the complex coordinate, overline stands for complex conjugation, $\tilde{A}(\mathbf{r}) = A_x(\mathbf{r})+iA_y(\mathbf{r})$ where $A_x=u_{xx}-u_{yy}$ and $A_y=u_{xy}$ are moir\'e periodic shear strain fields. This model describes a two-dimensional material with a chiral (or sublattice) symmetric quadratic band touching, subject to the a periodic moiré strain potential $\tilde{A}(\mathbf{r})$~\cite{wan2023topological}. If the strain field satisfies $\tilde{A}(\mathcal{C}_{6z} \mathbf{r}) = e^{-2\pi i/3}\tilde{A}(\mathbf{r})$ and $\tilde{A}(\mathcal{M}_x \mathbf{r}) = \overline{\tilde{A}(\mathbf{r})}$ (where $\mathcal{C}_{6z}$ and $\mathcal{M}_x$ are 6-fold rotation about out-of-plane axis $z$ and mirror reflection $x\rightarrow-x$, respectively), then the Hamiltonian has $p6mm$ symmetry; it satisfies $\mathcal{H}_s(\mathcal{C}_{6z}\mathbf{r}) = \rho(\mathcal{C}_{6z}) \mathcal{H}_s(\mathbf{r})\rho^\dagger(\mathcal{C}_{6z})$ with $\rho(\mathcal{C}_{6z}) = \text{diag}\{e^{-2\pi i/3},e^{2\pi i/3}\}$ and $\mathcal{H}_s(\mathcal{M}_{x}\mathbf{r}) = \sigma_x \mathcal{H}_s(\mathbf{r})\sigma_x$. Furthermore, by construction, the Hamiltonian has chiral symmetry $\sigma_z \mathcal{H}_s(\mathbf{r})\sigma_z = -\mathcal{H}_s(\mathbf{r})$ and time reversal symmetry $\sigma_x \mathcal{H}_s^*(\mathbf{r})\sigma_x = \mathcal{H}_s(\mathbf{r})$. It was shown in~\cite{wan2023topological,eugenio2022twisted,sarkar2023symmetry} that this type of Hamiltonians host exact flat bands upon tuning control parameters that change $\tilde{A}(\mathbf{r})$. An exact flat band of $\mathcal{H}_s(\mathbf{r})$ at energy $E= 0$ with wave function $\Psi_\mathbf{k}^{(s)}(\mathbf{r})$ satisfies $\mathcal{H}_s(\mathbf{r})\Psi_\mathbf{k}^{(s)}(\mathbf{r}) = \mathbf{0}$ for all $\mathbf{k}$. The construction of such a $\Psi_\mathbf{k}^{(s)}(\mathbf{r})$ is as follows. Note that due to $\mathcal{C}_{6z}$ and chiral symmetry, the two fold degeneracy of the quadratic band crossing at $\Gamma$ point remains at $E= 0$ for any $\tilde{A}(\mathbf{r})$ that keeps $\mathcal{C}_{6z}$ symmetry. This means that there are always two sublattice polarized wave functions $\Psi_{\Gamma,1}^{(s)}(\mathbf{r}) = \{\psi_{\Gamma}(\mathbf{r}),\mathbf{0}\}$ and $\Psi_{\Gamma}^{(s)}(\mathbf{r}) = \{\mathbf{0},\psi_{\Gamma}^*(\mathbf{r})\}$ satisfying $\mathcal{H}_s(\mathbf{r})\Psi_{\Gamma,i}^{(s)}(\mathbf{r})=\mathbf{0}$, or equivalently $\mathcal{D}_s(\mathbf{r})\psi_{\Gamma}(\mathbf{r})=0$. If there are exact flat bands, the wave functions can be written as $\{\psi_\mathbf{k}(\mathbf{r}),\mathbf{0}\}$ and $\{\mathbf{0},\psi_{-\mathbf{k}}^*(\mathbf{r})\}$. Since the kinetic part of $\mathcal{D}(\mathbf{r})$ contains antiholomorphic derivative, the trial wave function is naturally $\psi_{\mathbf{k}}(\mathbf{r}) = f_\mathbf{k}(z)\psi_{\Gamma}(\mathbf{r})$, where $f_\mathbf{k}(z)$ is holomorphic function satisfying $\overline{\partial_{z}}f_\mathbf{k}(z) = 0$. The function $f_\mathbf{k}(z)$ needs to satisfy Bloch periodicity (translation by moir\'e lattice vector $\mathbf{a}$ gives phase shift $e^{i\mathbf{k}\cdot\mathbf{a}}$). However, from Louiville's theorem, such a holomorphic function must have poles, making $\psi_\mathbf{k}(\mathbf{r})$ divergent, unless $\psi_{\Gamma}(\mathbf{r})$ has a zero that cancels the pole. Conversely, if $\psi_{\Gamma}(\mathbf{r})$ has zero at $\mathbf{r}_0$, a Bloch periodic holomorphic function
%= r_{01}\mathbf{a}_1+r_{02}\mathbf{a}_2$ ($\mathbf{a}_i$ are the moir\'e lattice vector)
% \begin{equation}
% \label{eq:fkz}
%     f_\mathbf{k}(z;\mathbf{r}_0)=\frac{\vartheta_{\frac{\mathbf{k} \cdot \mathbf{a}_1}{2\pi}-\frac{1}{2}-r_{02},\frac{1}{2}-r_{01}-\frac{\mathbf{k} \cdot \mathbf{a}_2}{2\pi}}(\frac{z}{a_1},\tau)}{\vartheta_{-\frac{1}{2}-r_{02},\frac{1}{2}-r_{01}}(\frac{z}{a_1},\tau)},
% \end{equation}
\begin{equation}
\label{eq:fkz}
\begin{split}
    &f_\mathbf{k}(z;\mathbf{r}_0)  = e^{i (\mathbf{k}\cdot\mathbf{a}_1) z/a_1}\frac{\vartheta\left(\frac{z-z_0}{a_1}-\frac{k}{G_2},\tau\right)}{\vartheta\left(\frac{z-z_0}{a_1},\tau\right)}= e^{i\mathbf{k}\cdot\mathbf{r}} \tilde{f}_\mathbf{k}(\mathbf{r};\mathbf{r}_0),\text{ where }\tilde{f}_\mathbf{k}(\mathbf{r};\mathbf{r}_0) = e^{-i(\mathbf{G}_2\cdot\mathbf{r})k/G_2}\frac{\vartheta\left(\frac{z-z_0}{a_1}-\frac{k}{G_2},\tau\right)}{\vartheta\left(\frac{z-z_0}{a_1},\tau\right)},
\end{split}
\end{equation}
with a pole at $\mathbf{r}_0$ can be constructed. Here $\vartheta (z,\tau) = -i\sum_{n=-\infty}^\infty(-1)^n e^{\pi i \tau(n+1/2)^2+\pi i(2n+1)z}$ is the Jacobi theta function of the first type~\cite{ledwith2020fractional}, $\mathbf{a}_i$ are lattice vectors, $\mathbf{G}_i$ are the corresponding reciprocal lattice vectors ($\mathbf{a}_i\cdot\mathbf{G}_j = 2\pi \delta_{ij}$), $a_i = (\mathbf{a}_i)_x+ i (\mathbf{a}_i)_y$, $G_i = (\mathbf{G}_i)_x+ i (\mathbf{G}_i)_y$, $z_0 = (\mathbf{r}_0)_x+ i (\mathbf{r}_0)_y$, $k = k_x + i k_y$, and $\tau = a_2/a_1$. Remarkably, at ``magic'' values of parameters of $\tilde{A}(\mathbf{r})$, the wave function $\psi_{\Gamma}(\mathbf{r})$ has a zero (at the center of the unit cell as shown in Fig.~\ref{fig:moireSup}(a)), which allows for such $f_\mathbf{k}(z;\mathbf{r}_0)$, and in turn gives rise to two exact flat bands. Note here that generically not all zeros of $\psi_\Gamma(\mathbf{r})$ can cancel the pole of $f_\mathbf{k}(z;\mathbf{r}_0)$ since near the pole, $f_\mathbf{k}(z;\mathbf{r}_0)\sim \frac{1}{z-z_0}$ has a special holomorphic structure, only the zeros that are at high symmetry points can generically cancel this type of poles; this was pointed out in~\cite{sarkar2023symmetry}. This is exactly why the ring of zeros of $\psi_\Gamma(\mathbf{r})$ in Fig.~\ref{fig:moireSup}(a) cannot be used to construct flat band wavefunctions (if they could be used to construct flat band wavefunctions, then there would be an infinite number of flat bands).
Furthermore, the periodic part $\tilde{f}_\mathbf{k}(\mathbf{r};\mathbf{r}_0)$ is a holomorphic function of $k$: $\overline{\partial_{k}}\tilde{f}_\mathbf{k}(\mathbf{r};\mathbf{r}_0) = \frac{1}{2}[(\partial_{k_x}+i\partial_{k_y})]\tilde{f}_\mathbf{k}(\mathbf{r};\mathbf{r}_0) = 0$~\cite{kharchev2015theta,ledwith2020fractional}; this property along with the presence of the zero in the wave function can be used to prove that  wave functions of this form carry Chern number $C = \pm 1$ (see~\cite{wang2021exact}). Moreover, since the periodic part $e^{-i\mathbf{k}\cdot\mathbf{r}}\psi_\mathbf{k}(\mathbf{r}) = \tilde{f}_\mathbf{k}(\mathbf{r};\mathbf{r}_0)\psi_\Gamma(\mathbf{r})$ is holomorphic in $k=k_x+ik_y$, this wave function satisfies ideal quantum geometry meaning i.e., trace of quantum metric $g(\mathbf{k})$ equals the absolute value Berry curvature $F_{xy}(\mathbf{k})$ at all momenta $\mathbf{k}$ (see~\cite{ledwith2020fractional} for a proof). Lastly, the wave function $\psi_\mathbf{k}(\mathbf{r})$ can be written as
\begin{equation}
\begin{split}
    \psi_\mathbf{k}(\mathbf{r}) &= f_\mathbf{k}(z;\mathbf{r}_0)\psi_\Gamma(\mathbf{r}) = \psi_\mathbf{k}^\text{LLL}(\mathbf{r})h(\mathbf{r}),\\ 
    &\text{ where }\psi_\mathbf{k}^\text{LLL}(\mathbf{r}) = e^{i (\mathbf{k}\cdot\mathbf{a}_1) z/a_1}\vartheta\left(\frac{z-z_0}{a_1}-\frac{k}{G_2},\tau\right) \exp\left(- \frac{(\hat{G}_2\cdot(\mathbf{r}-\mathbf{r}_0))^2}{2l_B^2}\right)\text{ and }h(\mathbf{r})=\frac{\psi_\Gamma(\mathbf{r})}{\psi_\Gamma^\text{LLL}(\mathbf{r})},
\end{split}
\end{equation}
where $\psi_\mathbf{k}^\text{LLL}(\mathbf{r})$ is the $n=0$ Landau Level (LL) wave function on a torus in Landau gauge $\mathbf{A}(\mathbf{r}) = B_0(\hat{G}_2\cdot(\mathbf{r}-\mathbf{r}_0))(\hat{z}\times\hat{G}_2)$, where $\hat{G}_2 = \mathbf{G}_2/|\mathbf{G}_2|$ and magnetic flux per moir\'e unit cell is one flux quantum $B_0 \hat{z}\cdot(\mathbf{a}_1\times\mathbf{a}_2)=2\pi \hbar/e$ such that the magnetic length $l_B$ satisfies $2\pi l_B^2 = \hat{z}\cdot(\mathbf{a}_1\times\mathbf{a}_2)$. It can be verified using the definition of Jacobi theta function that $\psi_\mathbf{k}^\text{LLL}(\mathbf{r})$ satisfies magnetic Bloch-periodicity
\begin{equation}
     \psi_\mathbf{k}^\text{LLL}(\mathbf{r}+\mathbf{a}_1) = -e^{i\mathbf{k}\cdot\mathbf{a}_1}\psi_\mathbf{k}^\text{LLL}(\mathbf{r}),\,\psi_\mathbf{k}^\text{LLL}(\mathbf{r}+\mathbf{a}_2) = -e^{i\mathbf{k}\cdot\mathbf{a}_2}e^{-2\pi i\frac{\mathbf{a}_1\cdot(\mathbf{r}-\mathbf{r}_0+\mathbf{a}_2/2)}{|\mathbf{a}_1|^2}}\psi_\mathbf{k}^\text{LLL}(\mathbf{r}).
\end{equation}
Due to this, along with the fact that $\psi_\Gamma(\mathbf{r}+\mathbf{a}_i) = \psi_\Gamma(\mathbf{r})$, we see that $\psi_\mathbf{k}(\mathbf{r}+\mathbf{a}_i) = e^{i\mathbf{k}\cdot\mathbf{a}_i}\psi_\mathbf{k}(\mathbf{r})$.

With the above knowledge, we next analyze the moir\'e Hamiltonian in the main text, reproduced here for convenience
\begin{equation}
    \mathcal{H}(\mathbf{r}) = \begin{pmatrix}0 & \mathcal{D}^\dagger(\mathbf{r})\\\mathcal{D}(\mathbf{r}) & 0\end{pmatrix},\, \mathcal{D}(\mathbf{r}) = \begin{pmatrix}
        -4\overline{\partial_z}^2+\tilde{A}(\mathbf{r}) & 2 i \gamma\overline{\partial_z}\\0 & -4\overline{\partial_z}^2+\tilde{A}(\mathbf{r})
    \end{pmatrix} = \begin{pmatrix}
        \mathcal{D}_s(\mathbf{r}) & 2 i \gamma\overline{\partial_z}\\0 & \mathcal{D}_s(\mathbf{r})
    \end{pmatrix}.
\end{equation}
Clearly if $\mathcal{D}_s(\mathbf{r})\psi_\mathbf{k}(\mathbf{r}) = 0$ for all $\mathbf{k}$, then $\mathcal{H}(\mathbf{r})$ must have an exact flat band at $E=0$ with wave function $\Psi_{\mathbf{k},1}(\mathbf{r})=\{\psi_\mathbf{k}(\mathbf{r}),0,0,0\}^T = \{\psi_\mathbf{k}^\text{LLL}(\mathbf{r}),0,0,0\}^Th(\mathbf{r})$. From our discussion on the properties of $\psi_\mathbf{k}(\mathbf{r})$, we know that this band must have Chern number $|C|=1$ and ideal quantum geometry. Next, we claim that $\mathcal{H}(\mathbf{r})$ has another flat band at $E=0$ with wave function 
\begin{equation}
\begin{split}
    &\Psi_{\mathbf{k},2}(\mathbf{r})= \{l_B\psi_\mathbf{k}^\text{LL1}(\mathbf{r})/\sqrt{8},\gamma^{-1}\psi_\mathbf{k}^\text{LLL}(\mathbf{r}),0,0\}^Th(\mathbf{r}) = \{l_B\psi_\mathbf{k}^\text{LL1}(\mathbf{r})/\sqrt{2},\gamma^{-1}\psi_\mathbf{k}^\text{LLL}(\mathbf{r}),0,0\}^T\frac{\psi_\Gamma(\mathbf{r})}{\psi_\Gamma^\text{LLL}(\mathbf{r})},\\
    &\psi_\mathbf{k}^\text{LL1}(\mathbf{r}) = -i\sqrt{2}l_B e^{i (\mathbf{k}\cdot\mathbf{a}_1) z/a_1} e^{\left(- \frac{(\hat{G}_2\cdot(\mathbf{r}-\mathbf{r}_0))^2}{2l_B^2}\right)}\left[\left(i \frac{\mathbf{k}\cdot\mathbf{a}_1}{a_1}-(\hat{G}_2\cdot(\mathbf{r}-\mathbf{r}_0))\frac{\overline{\hat{G}}_2}{l_B^2}\right)\vartheta\left(\frac{z-z_0}{a_1}-\frac{k}{G_2},\tau\right)+\partial_z\vartheta\left(\frac{z-z_0}{a_1}-\frac{k}{G_2},\tau\right) \right],
\end{split}
\end{equation}
where $\overline{\hat{G}}_2 = ((\mathbf{G}_i)_x - i (\mathbf{G}_i)_y)/|\mathbf{G}_2|$, $\psi_\mathbf{k}^\text{LL1}(\mathbf{r})$ is $n=1$ LL wave function in the same Landau gauge mentioned earlier. Indeed $\psi_\mathbf{k}^\text{LL1}(\mathbf{r})$ written above is just $\psi_\mathbf{k}^\text{LL1}(\mathbf{r}) = a^\dagger \psi_\mathbf{k}^\text{LLL}(\mathbf{r})$, where $a^\dagger = \frac{l_B}{\sqrt{2}\hbar}(\Pi_x -i\Pi_y)$ is the Landau level ladder operator with $\Pi_\alpha = \hbar(-i\partial_\alpha-eA_\alpha/\hbar)$. Since $a^\dagger$ commutes with magnetic translation operation, $\psi_\mathbf{k}^\text{LL1}(\mathbf{r})$ satisfies same magnetic Bloch periodicity as $\psi_\mathbf{k}^\text{LLL}(\mathbf{r})$; hence $\Psi_{\mathbf{k},2}(\mathbf{r})$ satisfies Bloch periodicity: $\Psi_{\mathbf{k},2}(\mathbf{r}+\mathbf{a}_i) = e^{i\mathbf{k}\cdot\mathbf{a}_i}\Psi_{\mathbf{k},2}(\mathbf{r})$. Next to prove $\mathcal{H}(\mathbf{r})\Psi_{\mathbf{k},2}(\mathbf{r}) = \mathbf{0}$, we must show $\mathcal{D}(\mathbf{r})\{l_B\psi_\mathbf{k}^\text{LL1}(\mathbf{r})/\sqrt{8},\gamma^{-1}\psi_\mathbf{k}^\text{LLL}(\mathbf{r})\}^Th(\mathbf{r}) = \mathbf{0}$. This can be verified via a straightforward explicit calculation.

Note that $\Psi_{\mathbf{k},1}(\mathbf{r})$ and $\Psi_{\mathbf{k},2}(\mathbf{r})$ are independent but not orthogonal to each other (they would be orthogonal if $h(\mathbf{r})=1$). We can construct the orthonormalized wavefunctions $\tilde{\Psi}_{\mathbf{k},1}(\mathbf{r})=\mathcal{N}_{\mathbf{k},1}\Psi_{\mathbf{k},1}(\mathbf{r})$ and $\tilde{\Psi}_{\mathbf{k},2}(\mathbf{r})=\mathcal{N}_{\mathbf{k},2}(\Psi_{\mathbf{k},2}(\mathbf{r}) - \langle\tilde{\Psi}_{\mathbf{k},1}|\Psi_{\mathbf{k},2} \rangle \tilde{\Psi}_{\mathbf{k},1}(\mathbf{r}))$, where $\mathcal{N}_{\mathbf{k},i}$ are normalization factors. But, we numerically find the overlap $\langle\tilde{\Psi}_{\mathbf{k},1}|\Psi_{\mathbf{k},2} \rangle$ is really small. Hence, $\tilde{\Psi}_{\mathbf{k},2}(\mathbf{r})\approx\mathcal{N}_{\mathbf{k},2}\Psi_{\mathbf{k},2}(\mathbf{r})$. In our numerics, we use the fully orthonormalized wavefunction $\tilde{\Psi}_{\mathbf{k},2}(\mathbf{r})=\mathcal{N}_{\mathbf{k},2}(\Psi_{\mathbf{k},2}(\mathbf{r}) - \langle\tilde{\Psi}_{\mathbf{k},1}|\Psi_{\mathbf{k},2} \rangle \tilde{\Psi}_{\mathbf{k},1}(\mathbf{r}))$. Since it is a superposition of $n=1$ and $n=0$ LL wavefunctions, its Chern number is $|C|=1$. 

Note further that the two wave functions $\Psi_{\mathbf{k},1}(\mathbf{r})$ and $\Psi_{\mathbf{k},2}(\mathbf{r})$ are sublattice polarized (lowest two components of each of these two four component vectors are zero). Due to time reversal symmetry, which flips sublattice, their are two other exact flat bands at $E=0$ with wave functions $\Psi_{\mathbf{k},3}(\mathbf{r}) =\{0,0,\overline{\psi_{-\mathbf{k}}^\text{LLL}(\mathbf{r})},0\}^T\overline{h(\mathbf{r})}$ and $\Psi_{\mathbf{k},4}(\mathbf{r}) = \{0,0,l_B\overline{\psi_{-\mathbf{k}}^\text{LL1}(\mathbf{r})}/\sqrt{8},\gamma^{-1}\overline{\psi_{-\mathbf{k}}^\text{LLL}(\mathbf{r})}\}^T\overline{h(\mathbf{r})}$ polarized on the other sublattice.

The wave functions $\tilde{\Psi}_{\mathbf{k},1}$ and $\tilde{\Psi}_{\mathbf{k},2}$ have the same structure as the flat band wavefunctions found in~\cite{fujimoto2024higher}; they called the wave functions of the form $\tilde{\Psi}_{\mathbf{k},2}$ ``first vortexable''.

For the particular choice of strain field $\tilde{A}(\mathbf{r})$, we numerically find that the Berry curvature and trace of quantum metric distributions of $\tilde{\Psi}_{\mathbf{k},1}$ and $\tilde{\Psi}_{\mathbf{k},2}$ are really flat (as a function of $\mathbf{k}$) for all values of $\gamma$ (as is evident from the plots of fluctuations in $F_{xy}$ and $\text{tr}(g)$ in Figs.~\ref{fig:moireSup}(b,c)).

\section{Exact diagonalization clusters and $k$-point labeling}

\begin{figure}[tbh]
  \centering
  \includegraphics[width=0.8 \textwidth,page=1]{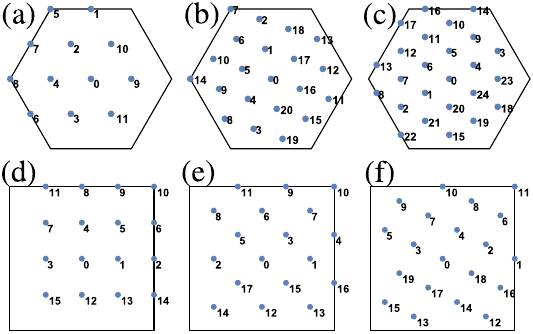}
  \caption{\textbf{ED $k$-space clusters and labeling.} 
(a–c) show triangular lattice ED clusters with reciprocal basis vectors 
$\mathbf{G}_1=\frac{2\sqrt{\pi}}{3^{1/4}}(0,\,1)$ and $\mathbf{G}_2=\frac{2\sqrt{\pi}}{3^{1/4}}\left(-\sqrt{3}/2,\,-1/2\right)$, corresponding to (a) triangle 12, (b) triangle 21, and (c) triangle 25 in our notation.  
(d–f) show square lattice clusters with reciprocal basis vectors 
$\mathbf{G}_1=\sqrt{2\pi}\,(1,\,0)$ and $\mathbf{G}_2=\sqrt{2\pi}\,(0,\,1)$, corresponding to (d) square 16, (e) square 18, and (f) square 20. In all panels, each selected momentum in the Brillouin zone is represented as a point and labeled by an integer $k_n$ ranging from $0$ to $N_s\!-\!1$, where $N_s$ is the cluster size.}
  \label{fig:EDclusters}
\end{figure}
For exact diagonalization (ED) calculation, we employ two families of finite clusters: triangular and square lattice geometries, as illustrated in Fig.~\ref{fig:EDclusters}. To ensure a consistent comparison, we fix the magnetic length to unity, $l_B=1$, and construct reciprocal lattice vectors accordingly.

Panels~(a-c) show the triangular lattice clusters, whose Brillouin zone is a regular hexagon generated by the reciprocal basis vectors
\begin{equation*}
    \mathbf{G}_1 = |G|\,(0, 1), \quad 
    \mathbf{G}_2 = |G|\!\left(-\tfrac{\sqrt{3}}{2}, -\tfrac{1}{2}\right), \quad |G| = \frac{2\sqrt{\pi}}{3^{1/4}\,l_B}.
\end{equation*}
Panels~(d-f) display the square lattice clusters, generated by the reciprocal basis
\begin{equation*}
    \mathbf{G}_1 = |G|\,(1, 0), \quad 
    \mathbf{G}_2 = |G|\,(0, 1), \quad |G| = \frac{\sqrt{2\pi}}{l_B}.
\end{equation*}
In this convention, both lattice geometries are constructed such that the magnetic length is fixed, ensuring that all energies can be expressed in terms of the same coulomb scale $U_{\rm int}=e^2/\varepsilon l_B$, and length can be expressed in terms of $l_B$.
In both families, each sampling crystal momentum $\mathbf{k}$ in the finite Brillouin zone is plotted as a point and assigned an integer label
\[
k_n=0,1,\ldots,N_s-1,
\]
where $N_s$ is the number of $k$ points (cluster size). The integer printed next to each dot in the figure is precisely this label $k_n$.

In our enumeration, the triangular clusters shown in Fig.~\ref{fig:EDclusters}(a–c) are denoted \emph{triangle 12}, \emph{triangle 21}, and \emph{triangle 25}.  
All have aspect ratio $1$ and include the $\Gamma$ point at the Brillouin-zone center.  
Clusters 12 and 21 also contain the $K$ and $K'$ points of the hexagonal Brillouin zone, while cluster 25 does not.  
Similarly, the square clusters shown in Fig.~\ref{fig:EDclusters}(d–f) are denoted \emph{square 16}, \emph{square 18}, and \emph{square 20}.  
All have aspect ratio $1$, and each contains both the $\Gamma$ point and the $M$ points.  
In addition, clusters 16 and 20 also contain the $X$ point.  
These clusters are chosen based on commensuration between the cluster tiling and the reciprocal lattice symmetry, which in turn determines the set of symmetries that can appear in the ED spectra.

\section{Coulomb scale in moir\'e and Landau level systems}

This section summarizes how we determine the Coulomb energy scale in moir\'e systems and Landau level systems and how the two results are benchmarked with each other. 

We use superscripts $(a)$ and $(n)$ to distinguish analytical expressions obtained from theory and numerical formulas implemented in the ED code. The characteristic length scale for the moir\'e system is the moir\'e lattice constant $a$, and for the LL system it is the magnetic length $l_B$. Then, denote $A_{\rm u.c.}$ as the unit cell area and $N_s$ the number of unit cells, so total system area is $A=N_sA_{\rm u.c.}$. In moir\'e systems, $A_{\rm u.c.}=\sqrt{3}/2\;a^2$ for triangular lattices and $A_{\rm u.c.}=a^2$ for square lattices. In the LL system, the unit cell area is $A_{\rm u.c.}=2\pi{l_B}^2$. For both systems, the analytical form of the normalized interaction is written as
\begin{equation}
\frac{V^{(a)}(q)}{A}
=\frac{2\pi e^2}{\varepsilon\, q\, A}\,\tanh(q\, d_s),
\label{eq:analytical_Vq}
\end{equation}
where $d_s$ is the screening length.

Since the characteristic length scale for the moir\'e system is $a$, whereas for the LL system it is $l_B$, we define the  Coulomb energy scale for the moir\'e system and the LL system separately as:
\begin{equation}
    U^{\rm m}_{\rm int}=\frac{e^2}{\varepsilon a},\qquad U^{\rm LL}_{\rm int}=\frac{e^2}{\varepsilon {l_B}}.
\end{equation}
They are directly related by
\begin{equation}
    U^{\rm LL}_{\rm int}=U^{\rm m}_{\rm int}\,\frac{a}{{l_B}},
\end{equation}
providing the conversion between the moir\'e and LL energy scales.
With this definition, one can express the analytical normalized interaction (for moir\'e systems) as
\begin{equation}
    \frac{V^{(a)}(q)}{A}
=\frac{4\pi\, U^{\rm m}_{\rm int}}{N_s\sqrt{3}\,q\,a}\tanh(q\, d_s),
\end{equation}
for the triangular lattice case (we use the triangular lattice for our moire model; an analogous form for the square lattice can also be derived). 

We take the following form of interaction in the numerical calculation for the moir\'e system,
\begin{equation}
    \frac{V^{(n)}(q)}{A}
=\frac{V^{\rm m}_{0}}{N_s}\frac{\tanh(\tilde q\,\tilde d_s)}{\tilde q},
\end{equation}
where $\tilde q=q/|G|\ \ \text{and}\ \ 
\tilde d_s=d_s|G|$ are the dimensionless momentum and length, $|G|$ is the magnitude of the moir\'e reciprocal lattice vector (we choose $|G|=1$ in the moir\'e ED calculation). $V^{\rm m}_{0}$ is the numerical parameter we choose representing the strength of the interaction.
Comparing with the analytical expression~\ref{eq:analytical_Vq}, which we rewrite as
\[
\frac{V^{(a)}(q)}{A}
=\frac{2\pi e^2}{\varepsilon\, q\, A}\tanh(q\, d_s)=\frac{2\pi e^2}{\varepsilon A |G|}
\frac{\tanh(\tilde q\,\tilde d_s)}{\tilde q},
\]
we get
\begin{equation}
\frac{V^{\rm m}_{0}}{N_s}=\frac{2\pi e^2}{\varepsilon A |G|}.
\end{equation}
Note that $A=N_s A_{\rm u.c.}$ and $A_{\rm u.c.}|G|=2\pi a$,
which gives
\begin{equation}
    U^{\rm m}_{\rm int}=\frac{e^2}{\varepsilon{a}}=V^{\rm m}_{0}.
\end{equation}
Therefore, in our convention, the Coulomb scale for the moir\'e system is the same as the numerical parameter $V^{\rm m}_{0}$ we chose in the ED calculation.

In the LL system numerical implementation, the interaction used is
\begin{equation}
    \frac{V^{(n)}(q)}{A}
=\frac{V^{\rm LL}_{0}}{N_s}\frac{\tanh(\tilde q\,\tilde d_s)}{\tilde q},
\end{equation}
where $\tilde q=q{l_B}\ \ \text{and}\ \ 
\tilde d_s=d_s/{l_B}$ are the dimensionless momentum and length are (we choose $l_B=1$ in the LL ED calculation). $V^{\rm LL}_{0}$ is the parameter for interaction strength. Comparing this with the analytical form~\ref{eq:analytical_Vq}, which is rewritten as:
\[
\frac{V^{(a)}(q)}{A}
=\frac{2\pi e^2{l_B}}{\varepsilon A}
\frac{\tanh(\tilde q\,\tilde d_s)}{\tilde q},
\]
we find
\begin{equation}
    \frac{V^{\rm LL}_{0}}{N_s}=\frac{2\pi e^2{l_B}}{\varepsilon A},
\end{equation}
and noting that $A=2\pi N_s{l_B}^2$ gives
\begin{equation}
    U^{\rm LL}_{\rm int}=\frac{e^2}{\varepsilon{l_B}}=V^{\rm LL}_{0}.
\end{equation}
Thus, the numerical interaction strength parameter $V^{\rm LL}_{0}$ used in the code also directly corresponds to the Coulomb scale we chose for the LL system $U^{\rm LL}_{\rm int}$.

In summary, the Coulomb scales are related as
\begin{equation}
U^{\rm m}_{\rm int}=\frac{e^2}{\varepsilon a}=V^{\rm m}_{0},\qquad
U^{\rm LL}_{\rm int}=\frac{e^2}{\varepsilon {l_B}}=V^{\rm LL}_{0}, \qquad U^{\rm LL}_{\rm int}=U^{\rm m}_{\rm int}\frac{a}{{l_B}},
\end{equation}
showing that our analytical and numerical conventions are fully consistent, 
which allows a direct and quantitative comparison between our moir\'e ED results 
and the Landau level ED results, as shown in Fig.~1 of the main text. For convenience, we always used the LL Coulomb scale $U^{\rm LL}_{\rm int}$ through out the main text.
 
\clearpage
\section{Derivation of the Landau level form factor}
We denote the $n$-th Landau level single-particle states on a torus as $|n\mathbf{k}\rangle$ at momentum $\mathbf{k}$, and the lowest Landau level (LLL) state $|0\mathbf{k}\rangle$ as $|\mathbf{k}\rangle$. The form factor for scattering with wavevector $\mathbf{q}$ is
\begin{equation}
    \lambda^{n_1,n_2}_{\mathbf{k}',\mathbf{k}}(\mathbf{q}) 
    = \langle \mathbf{k}' n_{1} | e^{i\mathbf{q}\cdot\mathbf{r}} | \mathbf{k} n_{2} \rangle ,
\end{equation}
Due to Bloch periodicity, nonzero matrix elements have to satisfy $\mathbf{k}' = [\mathbf{k}+\mathbf{q}]$, i.e.\ $\mathbf{k}'-\mathbf{k}-\mathbf{q}=\mathbf{G}$ for some reciprocal lattice vector $\mathbf{G}$. Substituting $\mathbf{q} = \mathbf{k}'-\mathbf{k}-\mathbf{G}$ gives
\begin{equation}
    \lambda^{n_1,n_2}_{\mathbf{k}',\mathbf{k}}(\mathbf{q}) 
    = \langle \mathbf{k}' n_{1} | e^{i(\mathbf{k}'-\mathbf{k}-\mathbf{G})\cdot\mathbf{r}} | \mathbf{k} n_{2} \rangle .
\end{equation}

Splitting $\mathbf r = \mathbf R + \boldsymbol{\eta}$ (guiding center + cyclotron), and using $[\mathbf{R},\boldsymbol{\eta}]=0$, we factorize the matrix element into the cyclotron part and the guiding center part:
\begin{equation}
    \begin{aligned}
        \text{BCH formula}&\Rightarrow e^{i \mathbf{q} \cdot \mathbf{r}}=e^{i \mathbf{q} \cdot(\mathbf{R} + \boldsymbol{\eta})}=e^{i \mathbf{q} \cdot \mathbf{R}} e^{i \mathbf{q} \cdot \boldsymbol{\eta}} . \\
\lambda_{\mathbf{k'},\mathbf{k}}^{n_1 n_{2}}(\mathbf{q})= \left\langle\mathbf{k'} n_{1}\right| e^{i \mathbf{q} \cdot \mathbf{R}}\left|\mathbf{k} n_{2}\right\rangle&\left\langle\mathbf{k'} n_{1}\right| e^{i \mathbf{q} \cdot \boldsymbol{\eta}}\left|\mathbf{k} n_{2}\right\rangle
=\langle\mathbf{k'}| e^{i \mathbf{q} \cdot \mathbf{R}}|\mathbf{k}\rangle\left\langle n_{1}\right| e^{i \mathbf{q} \cdot \boldsymbol{\eta}}\left|n_{2}\right\rangle .
    \end{aligned}
\end{equation}

We first evaluate the cyclotron matrix element $\langle n_{1} | e^{i \mathbf{q} \cdot \boldsymbol{\eta}} | n_{2} \rangle$ by introducing ladder operators
\begin{equation}
a=\tfrac{1}{\sqrt{2}l_B}(\eta_{x}+i\eta_{y}), \quad 
a^{\dagger}=\tfrac{1}{\sqrt{2}l_B}(\eta_{x}-i\eta_{y}), \quad [a,a^{\dagger}]=1,
\end{equation}
so that
\begin{equation}
\mathbf{q}\cdot\boldsymbol{\eta} 
= q_{x}\eta_{x}+q_{y}\eta_{y}
= \tfrac{l}{\sqrt{2}}(\bar{z}a+z a^{\dagger}), 
\quad z=q_{x}+iq_{y}.
\end{equation}
Defining $\alpha=i l_B z/\sqrt{2}$, we can write
\begin{equation}
e^{i\mathbf{q}\cdot\boldsymbol{\eta}}
= e^{\alpha a^{\dagger}-\alpha^{*}a}
= e^{-\tfrac{1}{2}|\alpha|^{2}} e^{\alpha a^{\dagger}} e^{-\alpha^{*} a}.
\end{equation}
The matrix element becomes
\begin{equation}
e^{-|\alpha|^{2}/2}\,\langle n_{1}| e^{\alpha a^{\dagger}} e^{-\alpha^{*} a}|n_{2}\rangle=e^{-q^2l_B^2/4}\,\langle n_{1}| e^{\alpha a^{\dagger}} e^{-\alpha^{*} a}|n_{2}\rangle.
\end{equation}
Expanding the exponentials,
\begin{equation}
e^{\alpha a^{\dagger}}=\sum_{k=0}^{\infty}\frac{\alpha^{k}}{k!}(a^{\dagger})^{k}, 
\quad 
e^{-\alpha^{*}a}=\sum_{l=0}^{\infty}\frac{(-\alpha^{*})^{l}}{l!}a^{l},
\end{equation}
yields
\begin{equation}
\langle n_{1}| e^{\alpha a^{\dagger}} e^{-\alpha^{*} a}|n_{2}\rangle
=\sum_{k,l}\frac{\alpha^{k}}{k!}\frac{(-\alpha^{*})^{l}}{l!}\langle n_{1}|(a^{\dagger})^{k}a^{l}|n_{2}\rangle.
\end{equation}
Then, using the standard relation of creation and annihilation operators
\begin{equation}
a^{l}|n_{2}\rangle=\sqrt{\tfrac{n_{2}!}{(n_{2}-l)!}}\,|n_{2}-l\rangle, 
\quad 
(a^{\dagger})^{k}|m\rangle=\sqrt{\tfrac{(m+k)!}{m!}}\,|m+k\rangle,
\end{equation}
and in the case when $n_1\geq n_2$, we obtain
\begin{equation}
\langle n_{1}|(a^{\dagger})^{k}a^{l}|n_{2}\rangle
=\sqrt{\tfrac{n_{1}!n_{2}!}{[(n_{2}-l)!]^{2}}}\,\delta_{n_{1}-n_{2},k-l}.
\end{equation}
Only terms with $k-l=n_{1}-n_{2}\equiv d$ contribute, and it requires $l\leq n_2$, so the sum reduces to
\begin{equation}
\langle n_{1}| e^{\alpha a^{\dagger}} e^{-\alpha^{*} a}|n_{2}\rangle
=\sum_{l=0}^{n_{2}}\frac{\alpha^{l+d}(-\alpha^{*})^{l}}{(l+d)!\,l!}
\sqrt{\frac{n_{1}!n_{2}!}{[(n_{2}-l)!]^{2}}}.
\end{equation}
Recognizing the definition of the associated Laguerre polynomial
\begin{equation}
L^{d}_{n_{2}}(x)=\sum_{p=0}^{n_{2}}
\frac{(n_{2}+d)!}{(n_{2}-p)!(p+d)!}\,\frac{(-x)^{p}}{p!},
\end{equation}
we simplify to
\begin{equation}
\langle n_{1}| e^{\alpha a^{\dagger}} e^{-\alpha^{*} a}|n_{2}\rangle
=\alpha^{d}\sqrt{\tfrac{n_{2}!}{n_{1}!}}\,L^{d}_{n_{<}}(|\alpha|^{2}),
\end{equation}
where $n_{<}=\min(n_{1},n_{2})$. The case when $n_1\leq n_2$ can be worked out similarly.

Finally, the the cyclotron radius matrix element takes the compact form ~\cite{jain2007composite, murthy2012hamiltonian}
\begin{equation}\label{eq:CyclotronMixedLL}
\Lambda^{n_1n_2}(\mathbf{q})=\langle n_{1} | e^{i \mathbf{q} \cdot \boldsymbol{\eta}} | n_{2} \rangle
=e^{-q^2l_B^2/4}\sqrt{\tfrac{n_{<}!}{n_{>}!}}
L^{|n_{1}-n_{2}|}_{n_{<}}\!\!\left(\tfrac{q^2l_B^2}{2}\right)
\times
\begin{cases}
\bigl(\tfrac{i l z}{\sqrt{2}}\bigr)^{n_{1}-n_{2}}, & n_{1}\geq n_{2}, \\[6pt]
\bigl(\tfrac{i l \bar{z}}{\sqrt{2}}\bigr)^{n_{2}-n_{1}}, & n_{1}\leq n_{2}.
\end{cases}
\end{equation}

To evaluate the guiding center part of the form factor, we use the Baker--Campbell--Hausdorff formula 
$e^{A+B}=e^{A}e^{B}e^{-\frac{1}{2}[A,B]}$, together with the commutation relation $[R_{x},R_{y}]=-i l_B^{2}$. 
For momentum transfer $\mathbf{q}=\mathbf{k}'-\mathbf{k}-\mathbf{G}$, one finds
\begin{equation}
\begin{aligned}
e^{\,i(\mathbf{k}'-\mathbf{k}-\mathbf{G})\cdot\mathbf{R}}
&= e^{-i\mathbf{G}\cdot\mathbf{R}}\,
   e^{\,i(\mathbf{k}'-\mathbf{k})\cdot\mathbf{R}}\,
   e^{-\tfrac{1}{2}[-i\mathbf{G}\cdot\mathbf{R},\,i(\mathbf{k}'-\mathbf{k})\cdot\mathbf{R}]} \\
&= e^{-i\mathbf{G}\cdot\mathbf{R}}\,
   e^{\,i(\mathbf{k}'-\mathbf{k})\cdot\mathbf{R}}\,
   \exp\!\Bigl(\tfrac{i}{2}\,l_B^2\,\mathbf{G}\times(\mathbf{k}'-\mathbf{k})\Bigr),
\end{aligned}
\end{equation}
where the phase factor arises directly from the noncommutative nature of guiding center coordinates.

Thus, the matrix element reads
\begin{equation}
\langle \mathbf{k}'|\,e^{\,i(\mathbf{k}'-\mathbf{k}-\mathbf{G})\cdot\mathbf{R}}\,|\mathbf{k}\rangle
= \langle \mathbf{k}'|\,e^{-i\mathbf{G}\cdot\mathbf{R}}\,
   e^{\,i(\mathbf{k}'-\mathbf{k})\cdot\mathbf{R}}\,|\mathbf{k}'\rangle\,
   \exp\!\Bigl(\tfrac{i}{2}\,l_B^2\,\mathbf{G}\times(\mathbf{k}'-\mathbf{k})\Bigr).
\end{equation}

The evaluation relies on magnetic translation properties of the LLL Bloch states. 
For a general vector $\mathbf{q}$ and reciprocal-lattice vector $\mathbf{G}$ ~\cite{wang2021exact},
\begin{equation}
\begin{aligned}
e^{\,i\mathbf{q}\cdot\mathbf{R}}|\mathbf{k}\rangle
  &= \exp\!\Bigl(\tfrac{i}{2}\,l_B^2\,\mathbf{q}\times\mathbf{k}\Bigr)\,
     |\mathbf{k}+\mathbf{q}\rangle, \\[4pt]
e^{\,i\mathbf{G}\cdot\mathbf{R}}|\mathbf{k}\rangle
  &= -\,\exp\!\bigl(i\,l_B^2\,\mathbf{G}\times\mathbf{k}\bigr)\,
     |\mathbf{k}\rangle, \\[4pt]
|\mathbf{k}+\mathbf{G}\rangle
  &= -\,\exp\!\Bigl(\tfrac{i}{2}\,l_B^2\,\mathbf{G}\times\mathbf{k}\Bigr)\,
     |\mathbf{k}\rangle.
\end{aligned}
\label{eq:magnetic_translation}
\end{equation}
These relations show that the states shifted by reciprocal lattice vectors differ by phase factors determined by the magnetic flux through the unit cell.

Applying the first magnetic-translation relation, we obtain
\begin{equation}
\langle\mathbf{k}'|\,e^{\,i(\mathbf{k}'-\mathbf{k}-\mathbf{G})\cdot\mathbf{R}}\,|\mathbf{k}\rangle
= \langle\mathbf{k}'|\,e^{-i\mathbf{G}\cdot\mathbf{R}}\,|\mathbf{k}'\rangle\,
   \exp\!\Bigl(\tfrac{i}{2}\,l_B^2(\mathbf{k}'-\mathbf{k})\times\mathbf{k}\Bigr)\,
   \exp\!\Bigl(\tfrac{i}{2}\,l_B^2\,\mathbf{G}\times(\mathbf{k}'-\mathbf{k})\Bigr).
\end{equation}

We now decompose the reciprocal lattice vector as $\mathbf{G}=n\mathbf{G}_{1}+m\mathbf{G}_{2}$ with integers $n,m$. 
Using the algebra of magnetic translations again, one finds
\begin{equation}
e^{-\,i\mathbf{G}\cdot\mathbf{R}}
= e^{-\,in\mathbf{G}_{1}\cdot\mathbf{R}}\,
  e^{-\,im\mathbf{G}_{2}\cdot\mathbf{R}}\,
  \exp\!\Bigl(\tfrac{i}{2}\,l_B^2\,mn\,\mathbf{G}_{1}\times\mathbf{G}_{2}\Bigr).
\end{equation}
The additional phase encodes the noncommutativity of successive translations in a magnetic field.

To simplify this factor, we use the lattice geometry of the magnetic Brillouin zone:
\begin{equation}
\mathbf{a}_{1}\times\mathbf{a}_{2}=2\pi l_B^2, 
\qquad 
\mathbf{a}_{i}\cdot\mathbf{G}_{j}=2\pi\delta_{ij}, \quad \mathbf{G}_{1}\times\mathbf{G}_{2}=2\pi/l_B^2,
\end{equation}
so that
\begin{equation}
e^{-\,i\mathbf{G}\cdot\mathbf{R}}
= e^{-\,in\mathbf{G}_{1}\cdot\mathbf{R}}\,
  e^{-\,im\mathbf{G}_{2}\cdot\mathbf{R}}\,
  (-1)^{mn}.
\end{equation}
Finally, the guiding center contribution can be evaluated explicitly. From the magnetic translation properties \ref{eq:magnetic_translation}, we obtain
\begin{equation}
\begin{aligned}
\langle\mathbf{k}'|\,e^{i(\mathbf{k}'-\mathbf{k}-\mathbf{G})\cdot\mathbf{R}}\,|\mathbf{k}\rangle
&=\langle\mathbf{k}'|\,e^{-i n \mathbf{G}_{1}\cdot\mathbf{R}}\,
                      e^{-i m \mathbf{G}_{2}\cdot\mathbf{R}}\,|\mathbf{k}'\rangle\,
   (-1)^{mn}\,
   e^{\tfrac{i}{2}l_B^2\mathbf{k}'\times\mathbf{k}}\,
   e^{\tfrac{i}{2}l_B^2\mathbf{G}\times(\mathbf{k}'-\mathbf{k})} \\
&=(-1)^{n} e^{-i n l_B^2\mathbf{G}_{1}\times\mathbf{k}'}\;
  (-1)^{m} e^{-i m l_B^2\mathbf{G}_{2}\times\mathbf{k}'}\;
  (-1)^{mn}\,
  e^{\tfrac{i}{2}l_B^2\mathbf{k}'\times\mathbf{k}}\,
  e^{\tfrac{i}{2}l_B^2\mathbf{G}\times(\mathbf{k}'-\mathbf{k})}.
\end{aligned}
\end{equation}
This can be compactly written as ~\cite{wang2021exact}
\begin{equation}
\langle\mathbf{k}'|\,e^{i(\mathbf{k}'-\mathbf{k}-\mathbf{G})\cdot\mathbf{R}}\,|\mathbf{k}\rangle
= \eta_{\mathbf{G}}\,
  \exp\!\Bigl[-\tfrac{i}{2}l_B^2\,\mathbf{G}\times(\mathbf{k}'+\mathbf{k})
              +\tfrac{i}{2}l_B^2\,\mathbf{k}'\times\mathbf{k}\Bigr],
\end{equation}
where the overall sign factor
\begin{equation}
\eta_{\mathbf{G}}=(-1)^{n+m+mn}
=\begin{cases}
  +1, & n,m \ \text{both even (so $\mathbf{G}/2$ is a reciprocal vector)}, \\[4pt]
  -1, & \text{otherwise}.
\end{cases}
\end{equation}

Collecting both the cyclotron part and the guiding center part, the most general Landau level form factor is
\begin{equation}
\lambda^{n_{1},n_{2}}_{\mathbf{k}',\mathbf{k}}(\mathbf{q})
=\langle \mathbf{k}' n_{1}|\,e^{i\mathbf{q}\cdot\mathbf{r}}\,|\mathbf{k} n_{2}\rangle=\lambda^{\mathrm{LLL}}_{\mathbf{k}',\mathbf{k}}(\mathbf{q})\sqrt{\frac{n_{<}!}{n_{>}!}}\;
  L_{n_{<}}^{\,|n_{1}-n_{2}|}\!\Bigl(\tfrac{q^2l_B^2}{2}\Bigr)
  \times
  \begin{cases}
  \bigl(\tfrac{ilz}{\sqrt{2}}\bigr)^{n_{1}-n_{2}}, & n_{1}\ge n_{2}, \\[6pt]
  \bigl(\tfrac{il\bar z}{\sqrt{2}}\bigr)^{n_{2}-n_{1}}, & n_{2}\ge n_{1},
  \end{cases},
\end{equation}
with $z=q_{x}+iq_{y}$, $n_{<}=\min(n_{1},n_{2})$, and $n_{>}=\max(n_{1},n_{2})$. The lowest Landau level (LLL) form factor is
\begin{equation}
\lambda^{\mathrm{LLL}}_{\mathbf{k}',\mathbf{k}}(\mathbf{q})
=\eta_{\mathbf{G}}\,
  \exp\!\Bigl[-\tfrac{q^2l_B^2}{4}
              -\tfrac{i}{2}l_B^2\,\mathbf{G}\times(\mathbf{k}'+\mathbf{k})
              +\tfrac{i}{2}l_B^2\,\mathbf{k}'\times\mathbf{k}\Bigr]\,
  \delta_{\mathbf{q},\,\mathbf{k}'-\mathbf{k}-\mathbf{G}}.
\end{equation}

Thus, the form factor factorizes into a universal LLL part fixed by the guiding center algebra, while all information about Landau level hybridization enters solely through the cyclotron part governed by the Laguerre polynomials.

\section{Analytic framework: Haldane pseudopotentials and Laguerre polynomials}

We will start from the derivation of the Haldane pseudopotential for the $n$-th LL and then generalize it to the LL-hybridization case.

\noindent\textbf{Calculation of Haldane pseudopotential}
For two identical particles projected to the same Landau level $n$, define the Haldane pseudopotential as the matrix element between the relative angular momentum states
\begin{equation}
V^{(n)}_{m}\;\equiv\;\langle n,m|\;V(r_{\rm rel})\;|n,m\rangle,\qquad
r_{\rm rel}=|\mathbf r_1-\mathbf r_2|,
\end{equation}
where $m=0,1,2,\cdots$ is the relative guiding center angular momentum.  
With the Fourier decomposition $V(r_{\rm rel})=\int\!\frac{d^2\mathbf{q}}{(2\pi)^2}\,V(q)\,e^{i\mathbf q\cdot \mathbf r_{\rm rel}},$ we have
\begin{equation}
V^{(n)}_{m}=\int\!\frac{d^2\mathbf{q}}{(2\pi)^2}\,V(q)\;\langle n,m|\,e^{i\mathbf q\cdot \mathbf r_{\rm rel}}\,|n,m\rangle.
\label{eq:Vmn-start}
\end{equation}
To calculate the matrix element, decompose each particle’s position as $\mathbf r=\mathbf R+\boldsymbol\eta$, where $\mathbf R$ is the guiding center and $\boldsymbol\eta$ is the cyclotron coordinate; for two particles the relative coordinate splits as $\mathbf r_{\rm rel}=\mathbf R_{\rm rel}+\boldsymbol\eta_{\rm rel}, \text{with}\  [\mathbf R_{\rm rel},\boldsymbol\eta_{\rm rel}]=0,$ hence $e^{i\mathbf q\cdot \mathbf r_{\rm rel}}=e^{i\mathbf q\cdot \mathbf R_{\rm rel}}\,e^{i\mathbf q\cdot \boldsymbol\eta_{\rm rel}}.$

Landau level projection fixes the cyclotron state of each particle to $|n\rangle$, so the matrix element factorizes into:
\begin{equation}
\langle n,m|\,e^{i\mathbf q\cdot \mathbf r_{\rm rel}}\,|n,m\rangle
=\langle n|e^{i\mathbf q\cdot \boldsymbol\eta_1}|n\rangle
\langle n|e^{-i\mathbf q\cdot \boldsymbol\eta_2}|n\rangle
\times\langle m|e^{i\mathbf q\cdot \mathbf R_{\rm rel}}|m\rangle.
\end{equation}
For each particle in the $n$-th LL, the LL projection gives the form factor $F_n(q)\equiv \langle n|e^{i\mathbf q\cdot \boldsymbol\eta}|n\rangle
= L_m\!\Big(\tfrac{q^2 l_B^2}{2}\Big)\,e^{-q^2 l_B^2/4}.$ Now we define the relative guiding center ladder operator
\begin{equation}
  b_r=\frac{R_{{\rm rel},x}-iR_{{\rm rel},y}}{2l_B},\qquad |m\rangle=\frac{(b_r^\dagger)^m}{\sqrt{m!}}\,|0\rangle.  
\end{equation}
Using $[R_{1x},R_{1y}]=-i l_B^2$, $[R_{2x},R_{2y}]=-i l_B^2$, and $\mathbf{R}_{\rm rel}=\mathbf{R}_1-\mathbf{R}_2$, one finds $[R_{{\rm rel},x},R_{{\rm rel},y}]=[R_{1x},R_{1y}]+[R_{2x},R_{2y}]=-2i l_B^2,$ so that $[b_r,b_r^\dagger]=\frac{1}{4l_B^2}\,[R_{{\rm rel},x}-iR_{{\rm rel},y},\,R_{{\rm rel},x}+iR_{{\rm rel},y}]=1.$ We use the exact same trick when projecting onto Landau level, writing $e^{i\mathbf q\cdot \mathbf R_{\rm rel}}=\exp(\lambda b_r^\dagger-\lambda^* b_r)$ with
$\lambda=il_B(q_x+i q_y)$, and obtain
\begin{equation}
\langle m|e^{i\mathbf q\cdot \mathbf R_{\rm rel}}|m\rangle
=\langle m|e^{\lambda b_r^\dagger-\lambda^* b_r}|m\rangle=e^{-|\lambda|^2/2}\,L_m(|\lambda|^2)=e^{-q^2 l_B^2/2}\,L_m(q^2 l_B^2).
\end{equation}
Inserting the projection form factor $F_n(q)$, we get the Haldane pseudopotential for the $n$-th LL~\cite{jain2007composite,goerbig2011electronic}
\begin{equation}
V_m^{(n)}=\int\!\frac{d^2\mathbf{q}}{(2\pi)^2}\;V(q)\;
\Big[L_n\!\Big(\tfrac{q^2 l_B^2}{2}\Big)\Big]^2\,
L_m(q^2 l_B^2)\,e^{-q^2 l_B^2}.
\label{eq:Vmn-final}
\end{equation}

\begin{table}[h]
\centering
\caption{Pseudopotential $c_m$ in Coulomb scale $U_{\rm int}^{\rm LL}$  for models LL01, LL02, and LL03 at the short-range interaction limit.}
\label{tab:cm_models010203_shortrange}
\begin{tabular}{c|ccc}
\hline\hline
 & LL01 & LL02 & LL03 \\
\hline
$c_{0}$ & $\tfrac{1}{16}\!\left(11 + 4\cos(2\theta) + \cos(4\theta)\right)$ 
        & $\tfrac{1}{64}\!\left(37 + 20\cos(2\theta) + 7\cos(4\theta)\right)$ 
        & $\tfrac{1}{128}\!\left(67 + 44\cos(2\theta) + 17\cos(4\theta)\right)$ \\
$c_{1}$ & $\tfrac{1}{4}\,\sin^{2}(2\theta)$ 
        & $\tfrac{1}{4}\,\sin^{2}(2\theta)$ 
        & $\tfrac{3}{16}\,\sin^{2}(2\theta)$ \\
$c_{2}$ & $\tfrac{1}{2}\,\sin^{4}\!\theta$ 
        & $\tfrac{1}{8}\,(3+\cos(2\theta))\,\sin^{2}\!\theta$ 
        & $\tfrac{3}{16}\,\bigl(\sin^{4}\!\theta + \sin^{2}(2\theta)\bigr)$ \\
$c_{3}$ & $0$ 
        & $0$ 
        & $\tfrac{1}{16}\,\sin^{2}(2\theta)$ \\
$c_{4}$ & $0$ 
        & $\tfrac{3}{8}\,\sin^{4}\!\theta$ 
        & $\tfrac{3}{16}\,\sin^{4}\!\theta$ \\
$c_{5}$ & $0$ 
        & $0$ 
        & $0$\\
\hline\hline
\end{tabular}
\end{table}

\begin{table}[h]
\centering
\caption{Pseudopotential $c_m$ in Coulomb scale $U_{\rm int}^{\rm LL}$ for models LL12, LL13, and LL23 at the short-range interaction limit.}
\label{tab:cm_models121323_shortrange}
\begin{tabular}{c|ccc}
\hline\hline
 & LL12 & LL13 & LL23 \\
\hline
$c_{0}$ & $\tfrac{1}{64}\!\left(27 + 4\cos(2\theta) + \cos(4\theta)\right)$ 
        & $\tfrac{1}{128}\!\left(47 + 12\cos(2\theta) + 5\cos(4\theta)\right)$ 
        & $\tfrac{1}{128}\!\left(43 + 4\cos(2\theta) + \cos(4\theta)\right)$ \\
$c_{1}$ & $\tfrac{1}{16}\,\sin^{2}(2\theta)$ 
        & $\tfrac{1}{8}\,\sin^{2}(2\theta)$ 
        & $\tfrac{1}{32}\,\sin^{2}(2\theta)$ \\
$c_{2}$ & $\tfrac{1}{16}\!\left(5 + 2\cos(2\theta) + \cos(4\theta)\right)$ 
        & $\tfrac{1}{16}\!\left(8\cos^{4}\!\theta + 3\sin^{4}\!\theta\right)$ 
        & $\tfrac{1}{128}\!\left(25 + 4\cos(2\theta) + 3\cos(4\theta)\right)$ \\
$c_{3}$ & $\tfrac{3}{16}\,\sin^{2}(2\theta)$ 
        & $\tfrac{1}{8}\,\sin^{2}(2\theta)$ 
        & $\tfrac{1}{16}\,\sin^{2}(2\theta)$ \\
$c_{4}$ & $\tfrac{3}{8}\,\sin^{4}\!\theta$ 
        & $\tfrac{1}{32}\,\bigl(11 + 5\cos(2\theta)\bigr)\,\sin^{2}\!\theta$ 
        & $\tfrac{1}{128}\!\left(29 + 12\cos(2\theta) + 7\cos(4\theta)\right)$ \\
$c_{5}$ & $0$ 
        & $0$ 
        & $\tfrac{5}{32}\,\sin^{2}(2\theta)$ \\
\hline\hline
\end{tabular}
\end{table}

To generalize the pseudopotential construction to the Landau level hybridization case, it is convenient to introduce the basis functions
\begin{equation}
v_m(q)\;\equiv\;\langle m|e^{i\mathbf q\cdot \mathbf R_{\rm rel}}|m\rangle
= e^{-q^2 l_B^2/2}\,L_m(q^2 l_B^2),
\label{eq:vm_pseudopotential}
\end{equation}
which encode the guiding center structure of a two-particle state with relative angular momentum $m$, ($m=0,1,2,\cdots$).  
Within this setup, the pseudopotential for a general interaction takes the compact form
\begin{equation}
c_m=\int\!\frac{d^2\mathbf q}{(2\pi)^2}\;V_{\rm eff}(q)\,v_m(q),
\label{eq:general_pseudopotential}
\end{equation}
where $V_{\rm eff}(q)=V(q)F(q)^2$ is the effective interaction after Landau level projection and $F(q)$ denotes the corresponding form factor. In this language, the effective interaction admits a decomposition in terms of the guiding-center basis functions $\{v_m\}$:
\begin{equation}
V_{\rm eff}(q)=4\pi l_B^2\sum_{m=0}^{\infty} c_m\,v_m(q).
\label{eq:expansion}
\end{equation}
which follows from the normalization condition
\begin{equation}
\int\!\frac{d^2\mathbf q}{(2\pi)^2}\;\,v_m(q)\,v_{m'}(q)=\frac{1}{4\pi l_B^2}\,\delta_{mm'},
\label{eq:expansion}
\end{equation}
Thus, the set $\{v_m(q)\}$ provides a complete basis for expanding any rotationally symmetric effective interaction $V_{\rm eff}(q)$.

\noindent \textbf{Pseudopotential for hybridized LL in this framework.}
In our Landau level hybridization model with hybridization angle $\theta\in[0,\pi/2]$, the cyclotron part of the form factor entering $V_{\rm eff}$ is
\begin{equation}
F(q;\theta)=\big(\cos \theta\  \langle\psi^{\text{LL}n_1}|,\sin \theta\  \langle\psi^{\text{LL}n_2}|\big)\;e^{i\mathbf q\cdot \boldsymbol\eta}\begin{pmatrix}\; \cos \theta\  |\psi^{\text{LL}n_1}\rangle \\ \sin \theta\  |\psi^{\text{LL}n_2}\rangle \end{pmatrix}=\Big[\cos^2\!\theta\,L_{n_1}\!\Big(\tfrac{q^2 l_B^2}{2}\Big)
+\sin^2\!\theta\,L_{n_2}\!\Big(\tfrac{q^2 l_B^2}{2}\Big)\Big]\;e^{-q^2 l_B^2/4},
\end{equation}
We choose the screened Coulomb interaction $V(q)=2\pi e^2\tanh(d_sq)/\varepsilon q$ wtih screening distance $d_s$, so that
\begin{equation}
V_{\rm eff}(q)=V(q)F(q;\theta)^2=\frac{2\pi e^2 \tanh(q d_s)}{\varepsilon q}\;
\Big[\cos^2\!\theta\,L_{n_1}\!\Big(\tfrac{q^2 l_B^2}{2}\Big)
+\sin^2\!\theta\,L_{n_2}\!\Big(\tfrac{q^2 l_B^2}{2}\Big)\Big]^2\,
e^{-q^2 l_B^2/2}.
\label{eq:Veff-hybridization}
\end{equation}
Plugging \eqref{eq:Veff-hybridization} into the general pseudopotential formula \eqref{eq:general_pseudopotential} and integrating over the angle gives the hybridized LL pseudopotentials:
\begin{equation}
\begin{split}
c_m(\theta;d_s)&=\frac{e^2}{\varepsilon l_B}\int_0^\infty\!ql_B\,d(ql_B)\;\frac{\tanh(ql_B\, d_s/l_B)}{q l_B}\;
\Big[\cos^2\!\theta\,L_{n_1}\!\Big(\tfrac{q^2 l_B^2}{2}\Big)
+\sin^2\!\theta\,L_{n_2}\!\Big(\tfrac{q^2 l_B^2}{2}\Big)\Big]^2
L_m(q^2 l_B^2)\,e^{-q^2 l_B^2}\\
\end{split}
\label{eq:cm-hybridization}
\end{equation}
Note that all formulas above remain valid for a general flat band with an arbitrary (isotropic) form factor $F(q)$. 

\begin{figure}[t]
  \centering
  \includegraphics[width=\linewidth]{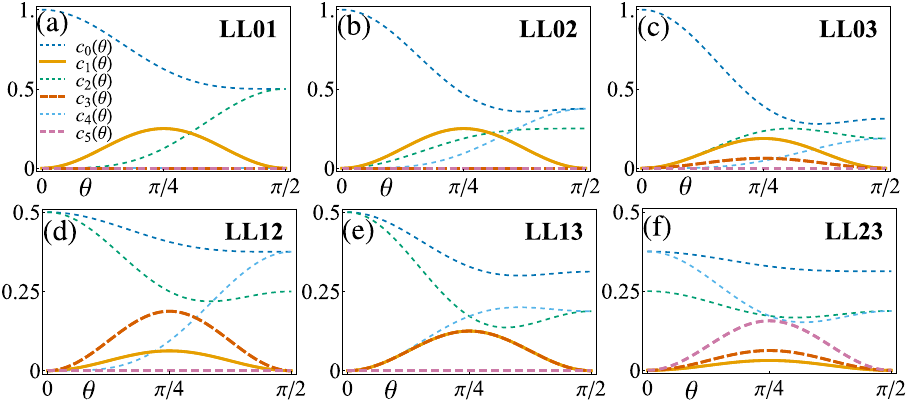}
  \caption{\textbf{Pseudopotential decomposition along Landau level hybridization paths in the short-range interaction limit $V(q)=4\pi l_Be^2/\varepsilon$.}
  Each panel shows the pseudopotentials $c_m(\theta)$ ($m=0,\dots,5$) in terms of the Coulomb scale $U_{\rm int}^{\rm LL}=e^2/(\varepsilon l_B)$, where $\theta$ continuously interpolates the form factor between two LLs.
  (a) LL01 model: $c_1(\theta)$ exhibits a single dome with a maximum at $\theta=\pi/4$; quantitatively $c_1(\theta)/U_{\rm int}^{\rm LL}=\sin^2(2\theta)/4$, accounting for the strong enhancement of the many-body gap at filling $\nu=1/3$. 
  (b) LL02 model: $c_1(\theta)$ follows the same analytic form $c_1(\theta)/U_{\rm int}^{\rm LL}=\sin^2(2\theta)/4$ as the LL01 model, explaining why the $\nu=1/3$ gap closely resembles that of LL01.
  (c) LL03 model: the $c_1$ dome is visibly smaller, and weight is partially redistributed into $c_3$, consistent with a weaker gap improvement at $\nu=1/3$, and the appearance of the $c_3$ dome would lift up the $1/3$ filling ground state energy from 0 and enhance the many-body gap for $1/5$ filling.
  (d) LL12 model: the odd $m$ channel shifts upward in $m$, $c_3(\theta)$ becomes the most dominant piece with a maximum at $\theta=\pi/4$, correlating with a pronounced enhancement of the $\nu=1/5$ gap, while $c_1$ remaining small throughout. 
  (e) LL13 model: similar to (d) but with a slightly reduced $c_3$ peak, implying a more moderate $\nu=1/5$ enhancement, and note that $c_1$ has exactly the same function form in $\theta$ as $c_3$. 
  (f) LL23 model: pseudopotential weight moves further toward higher-$m$ channels (notably $c_5$) with only a small $c_3$ and $c_1$, suppressing $\nu=1/3$ gap improvement.}
  \label{fig:pseudo_short}
\end{figure}

\begin{figure}[t]
  \centering
  \includegraphics[width=\linewidth]{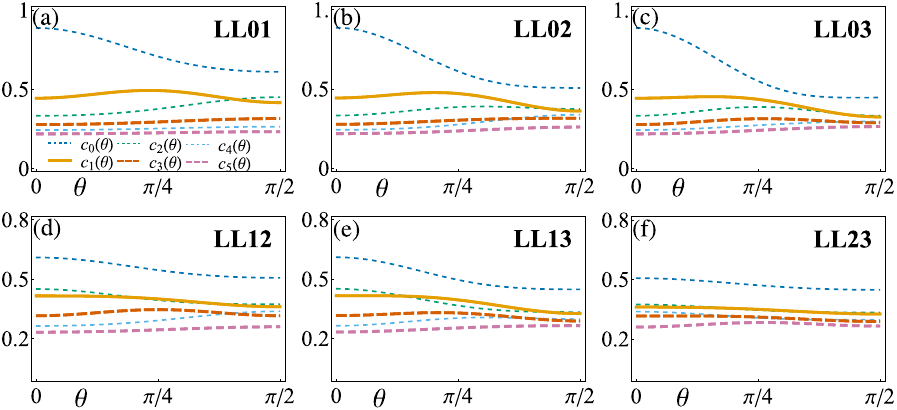}
  \caption{\textbf{Pseudopotential decomposition along Landau level hybridization paths in the long-range interaction limit $V(q)=2\pi e^2/\varepsilon q$.}
Shown are the pseudopotentials $c_m(\theta)$ ($m=0,\dots,5$) in terms of the Coulomb scale $U_{\rm int}^{\rm LL}$ for (a) LL01, (b) LL02, (c) LL03, (d) LL12, (e) LL13, and (f) LL23.
In the long-range regime, infinitely many channels are nonzero, in contrast to the short-range case where only a few small-$m$ terms dominate. Across all panels the $\theta$ dependence is weak, and no sharp enhancement of the relevant odd $m$ channels is observed. The only clear exception is a modest but
systematic increase of $c_3(\theta)$ for LL01 and LL02, which is consistent with the
enhanced many-body gap we observe at $\nu=1/5$. For higher-level hybridizations (LL03, LL13,
LL23) the weight shifts toward higher relative angular momentum channels, suppressing the $\nu=1/3$ Laughlin state.}
  \label{fig:pseudo_long}
\end{figure}

In the short-range limit $d_s\!\to\!0$ and long-range limit $d_s\!\to\!\infty$, the integral \eqref{eq:cm-hybridization} simplifies and can be evaluated in closed form by using standard Laguerre identities. A simple verification of our calculation result is that for LL01, $c_m$ agrees with the well-known result of LLL pseudopotential under Coulomb interaction $V(q)=2\pi e^2/\varepsilon q$~\cite{jain2007composite,sodemann2013landau}: $c_m(0)=V_m^{(0)} = \frac{e^2}{\varepsilon l_B}\,\frac{\Gamma\!\big(m+\tfrac12\big)}{2\, m!}$. The full comparison between different LL hybridization models (LL01, LL02, LL03, LL12, LL13, LL23) are shown in Fig.~\ref{fig:pseudo_short} and Fig.~\ref{fig:pseudo_long} for the short-range interaction limit and the long-range interaction limit respectively. (We choose $V(q)=4\pi l_Be^2/\varepsilon$ in the short-range limit and $V(q)=2\pi e^2/\varepsilon q$ in the long-range limit. Since the absolute value of the pseudopotentials is not important, we only care about the relative change.)

\noindent\textbf{The short-range interaction limit.}
In the short-range interaction limit, the closed expressions for the leading coefficients are summarized in the above tables \ref{tab:cm_models010203_shortrange} and \ref{tab:cm_models121323_shortrange}. For lower Landau levels LL01, hybridization produces a single dominant odd $m$ channel $c_1$: $c_1$ is maximized at $\theta=\pi/4$. LL02 behaves similarly to LL01, sharing the same $c_1(\theta)/U_{\rm int}^{\rm LL}=\sin^2(2\theta)/4$, which accounts for their nearly identical gap enhancement behavior at filling $\nu=1/3$. In higher-level combinations (LL03, LL13, LL23), however, the pseudopotential weight is redistributed into higher-$m$ channels (notably $m=3,5$). The $c_1$ dome is suppressed, weakening the benefit for $\nu=1/3$. In LL23, the growth of $c_5$ together with a small $c_3$ suggests that even the $\nu=1/5$ state gains only limited enhancement.

These results demonstrate that even for the same bare interaction $V(q)$, the projected (effective) flat-band interaction $V_{\rm eff}$ can vary significantly. Different LL hybridizations favor distinct pseudopotential channels: hybridization of lower Landau levels (LL01, LL02) enhances channels with low relative angular momentum ($c_1$), whereas higher-level hybridization (LL12, LL13, LL23) shifts weight into channels with larger relative angular momentum ($c_3, c_5$).

\noindent\textbf{The long-range interaction limit.}
In the long-range limit $V(q)\!\propto\!1/q$, the pseudopotential projection spreads interaction weight over infinitely many $m$ channels rather than concentrating it in the lowest few channels. As a result, tuning $\theta$ has only a very gentle effect: $c_1$ lacks a pronounced dome and decreases with $\theta$ for higher Landau levels hybridization. Most higher-$m$ coefficients change only
gradually. This dilution of small $m$ weight explains why “form-factor engineering’’ is much less effective in the long-range interaction case: the many-body gap does not exhibit a considerable rise because the odd channel pseudopotentials do not exhibit any notable enhancement.

\noindent\textbf{Derivation of odd channel pseudopotential summation in the short-range interaction limit.} 
% For convenience, we choose $e^2/\varepsilon=1$, and $l_B=1$ below.
If we take a closer look at Table \ref{tab:cm_models010203_shortrange} and \ref{tab:cm_models121323_shortrange}, we might be able to find an interesting conclusion: the odd channel pseudopotential for any LL hybridization model would sum up to the same function $\sin^2 2\theta/4$. We will prove this holds for all LL hybridization models and generalize this result to many LL hybridizations.

In the short-range limit, we take $V(q)=4\pi l_Be^2/\varepsilon$, so that the pseudopotentials are
\begin{equation}
c_m=\frac{e^2}{\varepsilon l_B}\int_{0}^{\infty}2\,\left(\cos^2\theta\,L_{n_1}\!\Big(\tfrac{q^2l_B^2}{2}\Big)+\sin^2\theta\,L_{n_2}\!\Big(\tfrac{q^2l_B^2}{2}\Big)\right)^{\!2} L_m(q^2l_B^2)\,e^{-q^2l_B^2}\,ql_B\,d(ql_B) =\frac{e^2}{\varepsilon l_B}\;\tilde{c}_m.
\end{equation}
where we defined the normalized (dimensionless) pseudopotentials as
\begin{equation}
    \tilde{c}_m=c_m/U_{\rm int}^{\rm LL}=c_m/\frac{e^2}{\varepsilon l_B}=\int_{0}^{\infty}\,g(x)\,L_m(x)\,e^{-x}\,dx,
\end{equation}
and we changed the variables to $x=q^2l_B^2$, $g(x)=\Big(\cos^2\theta\,L_{n_1}\!\big(\tfrac{x}{2}\big)+\sin^2\theta\,L_{n_2}\!\big(\tfrac{x}{2}\big)\Big)^{\!2}$. Then $\{ \tilde{c}_m\}$ are the Laguerre-Fourier coefficients of $g$. We define the generating function for the normalized pseudopotential as
\begin{equation}
    C(t)=\sum_{m=0}^{\infty} \tilde{c}_m\,t^m =\int_0^\infty e^{-x}\,g(x)\,\Big(\sum_{m=0}^{\infty}L_m(x)\,t^m\Big)\,dx=\frac{1}{1-t}\int_0^\infty g(x)\,\exp\!\Big(-\frac{x}{1-t}\Big)\,dx.
\end{equation}
where we used Laguerre polynomial generating function $\sum_{m=0}^{\infty}L_m(x)\,t^m=\frac{1}{1-t}\exp\!\Big(-\frac{t}{1-t}\,x\Big),\ |t|<1.$ The odd sums are extracted via $\sum_{\text{odd }m}\tilde{c}_m=\left(C(1^-)-C(-1)\right)/2$,
where
\begin{equation}
    C(1^-)=\lim_{t\to 1^-}C(t)=g(0)=\big(\cos^2\theta\,L_{n_1}(0)+\sin^2\theta\,L_{n_2}(0)\big)^2=1,
\end{equation}
and for $n_1\neq n_2$, we can apply the orthogonality of Laguerre polynomials $\int_0^\infty e^{-y}L_m(y)L_{n}(y)\,dy=\delta_{m,n}$ and get
\begin{equation}
    C(-1)=\frac{1}{2}\int_0^\infty g(x)\,e^{-x/2}\,dx=\int_0^\infty \big(\cos^2\theta L_{n_1}(y)+\sin^2\theta L_{n_2}(y)\big)^2 e^{-y}\,dy=\cos^4\theta+\sin^4\theta.
\end{equation}
Finally we get the sum over odd $m$,
\begin{equation}
    \sum_{\text{odd }m\ge 0} \tilde{c}_m=\frac{C(1^-)-C(-1)}{2} =\cos^2\theta\, \sin^2\theta=\frac{1}{4}\sin^22\theta.
\end{equation}

\noindent\textbf{General hybridization of many Landau levels.}
We can generalize this sum rule result to the hybridization between $N$ Landau levels. Let $\{n_j\}_{j=1}^N$ be Landau level indices ($0\leq n_1<n_2<\ldots<n_N$), and let $\{w_{n_j}\}_{j=1}^N$ be weights ($w_{n_j}>0$) for each Landau level. Define the hybridization function $g(x)$ and the normalized pseudopotentials $\tilde{c}_m$ as
\begin{equation}
    g(x) \;=\; \Big(\sum_{j} w_{n_j}\, L_{n_j}(x/2)\Big)^2 ,
\qquad
\tilde{c}_m \;=\; \int_{0}^{\infty} g(x)\,L_m(x)\,e^{-x}\,dx
\quad (m=0,1,2,\dots),
\end{equation}
Now set the generating function the same as before
\begin{equation}
    C(t)\;:=\;\sum_{m=0}^{\infty} \tilde{c}_m\,t^m
=\int_0^\infty e^{-x}\,g(x)\,\Big(\sum_{m=0}^{\infty}L_m(x)\,t^m\Big)\,dx
=\frac{1}{1-t}\int_0^\infty g(x)\exp\!\Big(-\frac{x}{1-t}\Big)\,dx ,
\end{equation}
similar to the previous case, we get
\begin{equation}
    C(1^-)\;=\;g(0)\;=\;\Big(\sum_{j} w_{n_j}\Big)^2,\quad \text{and} \ \ C(-1)\;=\;\frac12\int_0^\infty g(x)\,e^{-x/2}\,dx=\sum_{j} w_{n_j}^2 .
\end{equation}
Extracting the odd and even coefficent summations by $\sum_{\text{odd }m} \tilde{c}_m=\big(C(1^-)-C(-1)\big)/2$, and $\sum_{\text{even }m} \tilde{c}_m=\big(C(1^-)+C(-1)\big)/2,$
we obtain the odd and even sum identities for normalized pseudopotentials
\begin{equation}
    \sum_{\text{odd }m\ge 0} \tilde{c}_m
=\frac{\Big(\sum_{j} w_{n_j}\Big)^2 - \sum_{j} w_{n_j}^2}{2}
=\sum_{j<k} w_{n_j} w_{n_k},\quad\sum_{\text{even }m\ge 0} \tilde{c}_m
=\frac{\Big(\sum_{j} w_{n_j}\Big)^2 + \sum_{j} w_{n_j}^2}{2}
\end{equation}
Note that the total sum satisfies $\sum_{m\ge 0} \tilde{c}_m = C(1^-)=\big(\sum_{j} w_{n_j}\big)^2$, if one imposes normalization $\sum_j w_{n_j}=1$, then we have $\sum_{m\ge 0} \tilde{c}_m=1$.
From the general sum formula for odd channel coefficients, we can realize that the odd channels only receive contribution from the LL hybridization effect under the short-range interaction limit.

\section{Comparison between two Landau level hybridization models}

\begin{figure}[t]
    \centering
    \includegraphics[width=\linewidth]{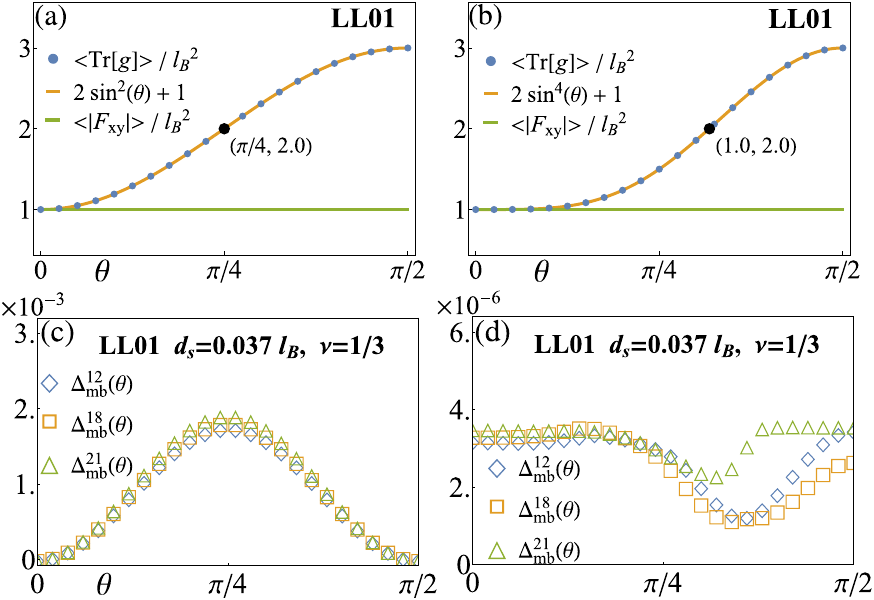}
    
\caption{\textbf{Comparison of quantum geometry and many-body gaps between two toy models.}
Panels (a) and (c) display the quantum geometry and many-body gap at $\nu = 1/3$ for model 1. The corresponding quantities for model 2 are shown in panels (b) and (d). (a) Model\,1: average trace of normalized quantum metric $\langle\mathrm{Tr}[g]\rangle/l_B^2$ (blue dots) and the average of normalized Berry curvature magnitude $\langle | F_{xy}|\rangle/l_B^2$ (green line).  Here the trace of the quantum metric follows $\langle\mathrm{Tr}[g]\rangle/l_B^2=1+2\sin^2\theta$. The black dot marks the special point $(\theta,\langle\mathrm{Tr}[g]\rangle/l_B^2)=(\pi/4,2.0)$.
(b) Model\,2: $\langle\mathrm{Tr}[g]\rangle/l_B^2=1+2\sin^4\theta$; the black dot indicates the
point $(\theta,\langle\mathrm{Tr}[g]\rangle/l_B^2)=(1.0,2.0)$ which has the same mean trace as
the black point in (a).  (c) many-body gaps under the Coulomb scale $U_{\rm int}^{\rm LL}$ for model\,1: note the enormous gap enhancement around $\pi/4$. (d) many-body gaps under the Coulomb scale for model\,2, which shows no comparable enhancement. The ED clusters used here are trianglule 12, square 18 and triangle 21.}
\label{fig:model12QG}
\end{figure}

Now we would like to consider other LL hybridization models and point out the importance of the role Haldane pseudopotentials play in these models.

\noindent\textbf{Model 1:}
If we start from a two-component wave function $\begin{pmatrix} \cos \theta\  |\psi^{\text{LL}n_1}\rangle \\ \sin \theta\  |\psi^{\text{LL}n_2}\rangle \end{pmatrix}$, we should get the cyclotron part of the form factor
\begin{equation}
    F_1(q;\theta)=\cos ^{2} \theta\,\Lambda^{n_1n_1}(q)+\sin ^{2} \theta\,\Lambda^{n_2n_2}(q)
\end{equation}
where $\Lambda^{n_1n_2}(\mathbf{q})$ is defined in Eq.~\eqref{eq:CyclotronMixedLL}. When $n_1=n_2=n$, it reduces to $\Lambda^{nn}(q)=L_n(q^2l_B^2/2)e^{-q^2l_B^2/4}$.
This is the LL hybridization model we used in the paper.

\noindent\textbf{Model 2:} Consider another model with two Landau levels hybridized together: $|\psi\rangle = \cos\theta\,|\psi^{\text{LL}n_1}\rangle + \sin\theta\,|\psi^{\text{LL}n_2}\rangle$. Then when we compute the cyclotron part of the form factor, we should get two extra terms
\begin{equation}
    F_2(\mathbf{q};\theta)=\cos ^{2} \theta\,\Lambda^{n_1n_1}(q)+\sin ^{2} \theta\,\Lambda^{n_2n_2}(q)+\frac{1}{2} \sin 2\theta\,\Lambda^{n_1n_2}(\mathbf{q})+\frac{1}{2} \sin 2\theta\,\Lambda^{n_2n_1}(\mathbf{q})
\end{equation}
As we can see from the form factor, this model will break the rotational symmetry in general. 

We have two interesting observations that apply to both models: 
(i) the mean trace of the many-body quantum metric obtained from ED (blue dots) exactly matches that of the single-particle quantum metric (orange line); and (ii) the Berry curvature (green) remains perfectly constant, reflecting the complete flatness of the Berry curvature in the Landau levels.

Fig.~\ref{fig:model12QG} panels (a)–(d) illustrate a striking, yet intentionally simple comparison: Despite having the same
average trace value at the two black-marker points $(\pi/4,2.0)$ in (a) and $(1.0,2.0)$ in (b), the many-body gap responses are dramatically different. Model 1 exhibits a huge gap enhancement (panel c, peak $\sim0.2\%$ of the Coulomb scale) whereas model\,2 shows no comparable increase (panel d, only $\sim0.0003\%$ of the Coulomb scale at the matched $\langle \mathrm{Tr}[g]\rangle$). These plots therefore emphasize a simple empirical point: the mean quantum geometry measure $\langle\mathrm{Tr}[g]\rangle$ does not by itself determine the gap
enhancement. The two models share the same mean trace and Berry curvature
magnitude at the marked points, yet produce very different many-body gaps,
implying that other details of the form factor (e.g., pseudopotential decompositions) control whether hybridization yields a large gap.

\section{Effect of the Coulomb screening distance on many-body gap enhancement for LL01 hybridization model}
\begin{figure}[h]
  \centering
  \includegraphics[width=\linewidth]{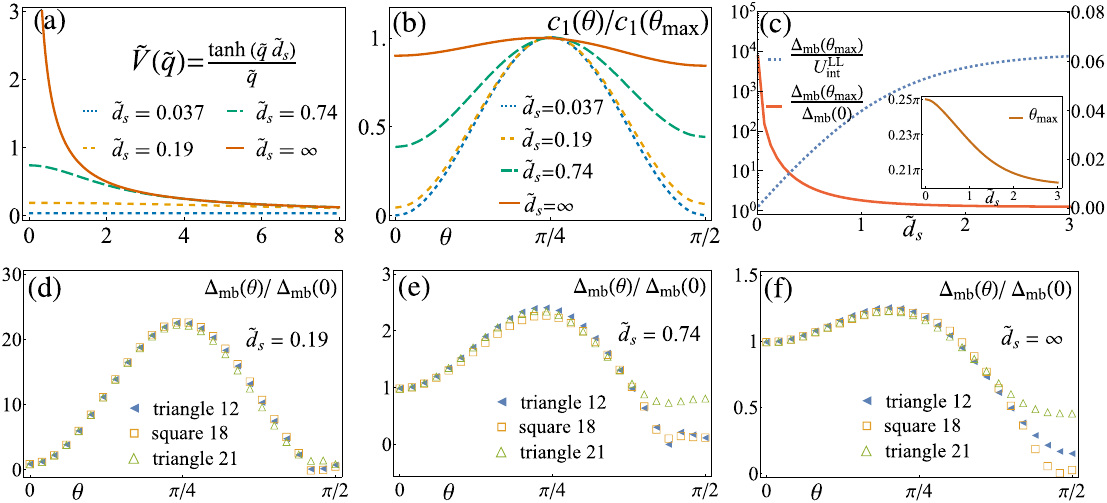}
  \caption{\textbf{Screened interaction and the enhancement of the many-body gap for LL01 hybridization.}
(a) Normalized momentum space interaction
$\tilde{V}(\tilde{q})=V(q)/(AU_{\rm int}^{\rm LL})=\tanh(\tilde{q} \tilde{d_s})/\tilde{q}$ with dimensionless screening length $\tilde{d_s}=d_s/l_B$ and dimensionless momentum $\tilde{q}=ql_B$. (b) Scaled pseudopotential $c_1(\theta)/c_1(\theta_{\rm max})$ of the LL01 hybridization model for different $\tilde{d_s}$. Such scaling will show the relative enhancement ratio of the pseudopotentials as we vary $\theta$.
For very small $\tilde{d_s}$, $c_1(\theta)$ develops a sharp peak near $\theta\!\approx\!\pi/4$; as $\tilde{d_s}$ grows, the $\theta$ dependence flattens and the peak becomes much less pronounced. (c) Combined plot showing (red solid line) the maximal gap enhancement ratio 
$\Delta_{\rm mb}(\theta_{\rm max}) / \Delta_{\rm mb}(0)$ and (blue dashed line) the corresponding maximal gap magnitude $\Delta_{\rm mb}(\theta_{\rm max}) /U^{\rm LL}_{\rm int}$, both as functions of the dimensionless screening length $\tilde{d_s}$. The inset displays the position of the maximal gap, $\theta_{\rm max}$, as a function of $\tilde{d_s}$. The results show that while the absolute many-body gap increases with the screening length, the enhancement ratio decreases rapidly. Moreover, the optimal hybridization weight gradually shifts toward LL0 as $\tilde{d_s}$ increases. (d–f) many-body gap ratio $\Delta_{\mathrm{mb}}(\theta)/\Delta_{\mathrm{mb}}(0)$ versus $\theta$ for different screening lengths. At $\tilde{d_s}=0.19$ (d), the ratio increases by more than 20 times.
For $\tilde{d_s}=0.74$ (e), a clear but much weaker (less than 3-fold) enhancement remains. In the unscreened limit $\tilde{d_s}=\infty$ (f), the ratio shows only marginal enhancement (an increase of 20\%). Note that for longer-range interactions, the ED results exhibit stronger finite-size effects near possible phase transitions near LL1. This explains the deviations between clusters as $\theta \to \pi/2$. In this work, we therefore only restrict our attention to the regimes where the clusters are consistent, i.e., deep within the FQH phase. }
  \label{fig:dsc_gap}
\end{figure}

\section{Pseudopotential effects on the ground-state energy and optimization of the many-body gap}
\begin{figure}[h]
    \centering
\includegraphics[width=\linewidth]{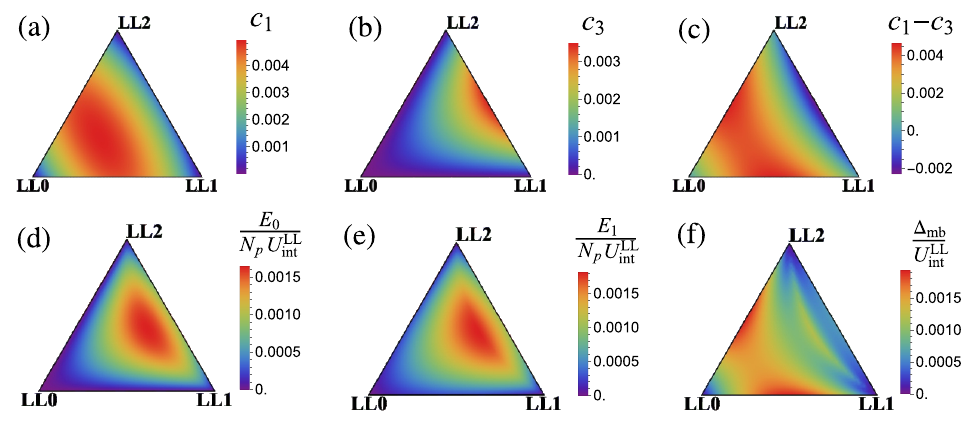}
    \caption{\textbf{LL012 hybridization at $d_s/l_B = 0.037$ (close to the short-range limit): pseudopotentials, energies, and gap on the LL012 hybridization simplex.}
Each point on the simplex corresponds to a weight vector 
$\mathbf w=(w_0,w_1,w_2)$ with $w_0+w_1+w_2=1$, giving the hybridization weight of LL0, LL1, and LL2. The three vertices are the pure LLs
(LL0, LL1, LL2), and the edges represent two-level hybridization.
Panels~(a)–(c) show the two leading odd Haldane pseudopotentials and their difference:
(a) $c_1$ peaks close to the short-range optimal weight distribution
$w_0\!\approx\!7/15$, $w_1\!=\!w_2\!\approx\!4/15$; (b) $c_3$ is maximized near the middle of the LL12 hybridization. (c) The difference $c_1-c_3$ is maximized near the middle of the two edges LL01 and LL02 (and away from the LL12 edge). Panels~(d)–(f) report the ED results for the triangle 21 cluster: (d) the ground state energy per particle (in units of Coulomb scale) $E_0/(N_p U^{\rm LL}_{\rm int})$ remains small along the LL01 and LL02 edges, which is consistent with weak $c_3$ there, but rises in the interior where three-level mixing is significant and $c_3$ is enhanced;
(e) the first excited state energy $E_1/(N_p U^{\rm LL}_{\rm int})$ follows a similar trend, shifting upward with increased LL12 admixture; 
(f) the many-body gap $\Delta_{\rm mb}/U^{\rm LL}_{\rm int}$ is strongly
correlated with panel~(c): it is largest near the maximum point of $c_1-c_3$ and diminishes toward the LL12 edge where $c_3$ dominates. Taken together, the maps reveal a clear physical mechanism: at filling $\nu=1/3$, the stability and magnitude of the incompressible state are
controlled primarily by the competition between $c_1$ (which favors the
Laughlin zero of order three) and $c_3$ (which penalizes configurations that still have residual short-range overlap between particles). Optimizing LL012 hybridization toward the region of
large $c_1-c_3$ maximizes the gap, whereas pushing weight into the
LL12 sector elevates $c_3$, increases the absolute ground state
energy, and suppresses the gap. That is the reason why the gap is maximized near the middle of the LL01 and LL02, not in the interior of LL012 hybridization, where $c_1$ is actually maximized.}
    \label{fig:3ll_hybridization}
\end{figure}

We consider the pseudopotential decomposition for the many LL hybridization model,
\begin{equation}
    c_m \;=\;\int\!\frac{d^2\mathbf q}{(2\pi)^2}\;V_{\rm eff}(q)\,v_m(q),
    \quad V_{\rm eff}(q)=V(q) \Big(\sum_{j} w_{n_j}\, L_{n_j}(q^2l_B^2/2)\Big)^2 ,
\end{equation}
with weights for each $n_j$th LL  satisfying $w_{n_j} > 0$ and $\sum_j w_{n_j}=1$. $c_m$ are the Haldane pseudopotentials of the projected interaction. We are interested in how to distribute the weights $\{w_{n_j}\}$ among available indices $\{n_j\}$ so as to maximize either a single coefficient $c_m$, or more physically relevant differences such as $c_{2m-1}-c_{2m+1}$ that control neutral excitation gaps at filling $\nu=1/(2m+1)$. Below, we will provide a generic approach to the optimization.

Expanding $V_{\rm eff}(q)$, we obtain
\begin{equation}
c_m \;=\; \sum_{i,j} w_{n_i} w_{n_j} \, A_m(n_i,n_j), 
\qquad 
A_m(n_i,n_j) := \frac{1}{2\pi}\int_0^\infty V(q)L_{n_i}(q^2l_B^2/2)\,L_{n_j}(q^2l_B^2/2)\,L_m(q^2l_B^2)\,e^{-q^2l_B^2}q\,dq .
\end{equation}
Thus the pseudopotentials are quadratic forms in the weights $\{w_{n_i}\}$. Now suppose we choose our hybridization model to be a set of $k$ LLs $S=\{n_1,\dots,n_k\}$ with weights $\mathbf{w}_S=\{w_{n_1},\dots,w_{n_k}\}.$ To maximize $c_m(S) = \mathbf{w}_S^\top A^{(m)} \mathbf{w}_S$ subject to $w_{n_j\in S} > 0$ and $\sum_{n_j\in S} w_{n_j}=1$, we introduce a Lagrange multiplier for the normalization condition. 
\begin{equation}
    \mathcal{L}(\mathbf{w}_S,\lambda_S)=\mathbf{w}_S^\top A_S^{(m)} \mathbf{w}_S-\lambda_S\Big(\mathbf{1}^\top \mathbf{w}_S-1\Big),
\end{equation}
where $\mathbf{1}$ represents a constant vector with the same length as $\mathbf{w}_S$. The stationarity condition requires
\begin{equation*}
    A^{(m)}_S\,\mathbf{w}_S = \frac{\lambda_S}{2}\,\mathbf{1}, 
\qquad \mathbf{1}^\top \mathbf{w}_S = 1, 
\qquad w_{n_j\in S}>0 .
\end{equation*}
Hence the optimal weights and the corresponding maximal value are
\begin{equation}
    \mathbf{w}_S = \frac{\big(A^{(m)}_S\big)^{-1}\mathbf{1}}{\mathbf{1}^\top \big(A^{(m)}_S\big)^{-1}\mathbf{1}},\quad \max c_m(S) = \frac{1}{\mathbf{1}^\top \big(A^{(m)}_S\big)^{-1}\mathbf{1}}
\end{equation}
Thus the problem reduces to selecting a model set $S$, computing the corresponding weight vector $\mathbf{w}_S$ with $w_{n_j}>0$ for all $n_j \in S$, and then comparing the resulting maximal values $c_m(S)$ across different choices of $S$. Below, we will give a few examples of finding the optimal weight distribution in this LL hybridization model. For simplicity, we focus on the short-range and long-range interaction limit, but note the above method applies to any general interactions.

\noindent\textbf{Maximizing $c_m$ in the short-range interaction limit.} As an example, we first maximize $c_1$ through listing the $4\times 4$ principal blocks of the infinite matrix $A_1$ normalized by $U_{\rm int}^{\rm LL}$ with rows/columns labeled by $n_i,n_j\in\{0,1,2,3\}$:
\begin{equation}
\renewcommand{\arraystretch}{1.4}
A_1 / U_{\rm int}^{\rm LL} =
\begin{array}{c|ccccc}
 & 0 & 1 & 2 & 3 & \cdots \\
\hline
0 & 0 & \tfrac{1}{2} & \tfrac{1}{2} & \tfrac{3}{8} & \cdots \\
1 & \tfrac{1}{2} & 0 & \tfrac{1}{8} & \tfrac{1}{4} & \cdots \\
2 & \tfrac{1}{2} & \tfrac{1}{8} & 0 & \tfrac{1}{16} & \cdots \\
3 & \tfrac{3}{8} & \tfrac{1}{4} & \tfrac{1}{16} & 0 & \cdots \\
\vdots & \vdots & \vdots & \vdots & \vdots & \ddots
\end{array}
\end{equation}

First consider $S=\{0,1\}$. The optimal weights are 
$w_0 = w_1 = 1/2$, giving 
\(\max c_m(\{0,1\})/U_{\rm int}^{\rm LL} = 1/4\)
For $S=\{0,1,2\}$, the optimal distribution is 
$w_0 = 7/15$, $w_1 = w_2 = 4/15$, yielding 
\(\max c_m(\{0,1,2\})/U_{\rm int}^{\rm LL} = 4/15\), which exceeds the $\{0,1\}$ case. 
Based on our numerical results, in the short-range interaction limit, the optimal hybridizations and weight distributions that maximize the odd pseudopotentials $c_{2m-1}$ are
\begin{equation}
    S =
\begin{cases}
\{0,1,2\}, & m=1, \\
\{m-1,m\}, & m \ge 2 ,
\end{cases}
\qquad
\mathbf{w}_S =
\begin{cases}
\{7/15,\,4/15,\,4/15\}, & m=1, \\
\{1/2,\,1/2\}, & m \ge 2 ,
\end{cases}
\end{equation}
with corresponding maximal values
\begin{equation}
    \max c_{2m-1}/U_{\rm int}^{\rm LL} =
\begin{cases}
4/15, & m=1, \\
2^{-2m}\binom{2m-1}{m-1}, & m \ge 2 .
\end{cases}
\end{equation}

\noindent\textbf{Maximizing $c_{2m-1}-c_{2m+1}$ in the short-range interaction limit.}
For Laughlin fillings $\nu=1/(2m+1)$, the neutral gap will be determined primarily by
\begin{equation}
    \Phi_m := c_{2m-1}-c_{2m+1}
= \sum_{i,j} w_{n_i} w_{n_j}\,B_m(n_i,n_j), \qquad B_m=A_{2m-1}-A_{2m+1}.
\end{equation}
We can maximize $\Phi_m$ following exactly the same procedure as before by computnig the $B_S^{(m)}$ matrix. Numerical evidence suggests that the maximum value of $\Phi_m$ is attained at $N=2$ hybridization by equal weights on the maximizing pair $S=\{m,m-1\}$:
\begin{equation}
    w_{m-1}=w_m=\frac12,\qquad \max\Phi_m/U_{\rm int}^{\rm LL}=2^{-2m}\binom{2m-1}{m-1}.
\end{equation}
e.g. for $m=1$ ($\nu=1/3$), optimum is $\{0,1\}$ or $\{0,2\}$ with $\Phi_1^{\max}/U_{\rm int}^{\rm LL}=1/4$; for $m=2$ ($\nu=1/5$), optimum is $\{1,2\}$ with $\Phi_2^{\max}/U_{\rm int}^{\rm LL}=3/16$. This explains the ED spectrum: maximizing $c_1$ raises the first excitation, but the largest neutral gap corresponds to maximizing $c_{2m-1}-c_{2m+1}$ (e.g.\ $c_1-c_3$ for $\nu=1/3$, $c_3-c_5$ for $\nu=1/5$).

\noindent\textbf{Maximizing $c_{1}$ and $c_{1}-c_{3}$ in the long-range interaction limit (unscreened Coulomb interaction).} First, we compute the normalized $A_1$ and $B_1$ matricies with Coulomb interaction $V(q)=2\pi e^2/\varepsilon q$.
\begin{equation}
\renewcommand{\arraystretch}{1.4}
A_1/U_{\rm int}^{\rm LL} =
\begin{array}{c|cccc}
 & 0 & 1 & 2 & \cdots \\
\hline
0 & \tfrac{\sqrt{\pi}}{4} & \tfrac{5\sqrt{\pi}}{16} & \tfrac{39\sqrt{\pi}}{128} & \cdots \\
1 & \tfrac{5\sqrt{\pi}}{16} & \tfrac{15\sqrt{\pi}}{64} & \tfrac{119\sqrt{\pi}}{512} & \cdots \\
2 & \tfrac{39\sqrt{\pi}}{128} & \tfrac{119\sqrt{\pi}}{512} & \tfrac{833\sqrt{\pi}}{4096} & \cdots \\
\vdots & \vdots & \vdots & \vdots & \ddots
\end{array}
\qquad
B_1/U_{\rm int}^{\rm LL} =
\begin{array}{c|cccc}
 & 0 & 1 & 2 & \cdots \\
\hline
0 & \tfrac{3\sqrt{\pi}}{32} & \tfrac{19\sqrt{\pi}}{128} & \tfrac{133\sqrt{\pi}}{1024} & \cdots \\
1 & \tfrac{19\sqrt{\pi}}{128} & \tfrac{29\sqrt{\pi}}{512} & \tfrac{81\sqrt{\pi}}{4096} & \cdots \\
2 & \tfrac{133\sqrt{\pi}}{1024} & \tfrac{81\sqrt{\pi}}{4096} & \tfrac{835\sqrt{\pi}}{32768} & \cdots \\
\vdots & \vdots & \vdots & \vdots & \ddots
\end{array}
\end{equation}
Numerical results show that both $c_{1}$ and $c_{1}-c_{3}$ are maximized via LL01 hybridization, but with different weights. 
\begin{equation}
\begin{aligned}
        \max c_1/U_{\rm int}^{\rm LL}&=\frac{5\sqrt{\pi}}{18}, \qquad w_0=\frac{5}{9}, \;w_1=\frac{4}{9}\\
        \max (c_1-c_3)/U_{\rm int}^{\rm LL}&=\frac{137\sqrt{\pi}}{1200}, \qquad w_0=\frac{47}{75},\; w_1=\frac{28}{75}.
\end{aligned}
\end{equation}
Note that the true value of the $\theta_{\rm max}$ is around $0.2\pi$ in the long-range interaction limit from Fig.~\ref{fig:dsc_gap} (f). While from the above optimization process of $c_{1}-c_{3}$, we obtain $\theta_{\rm max}=\arccos{\sqrt{47/75}}\approx 0.21\pi$, which is close to the ED result.

\end{document}